\documentclass{emulateapj}
\usepackage{lscape}
\usepackage{amsmath}
\usepackage{morefloats}

\newcommand {\ergs} {erg s$^{-1}$}
\newcommand {\ergcms} {erg cm$^{-2}$ s$^{-1}$}

\shorttitle{ GOODS XLF GALAXIES}
\shortauthors{Ptak et al.}
\slugcomment{June 11, 2007}
\begin{document}

\title{X-ray Luminosity Functions of Normal Galaxies in the GOODS}
 
\author {Andrew Ptak\altaffilmark{1}, 
  Bahram Mobasher\altaffilmark{2}, Ann Hornschemeier\altaffilmark{1,4},
  Franz Bauer,\altaffilmark{5}, and   Colin Norman\altaffilmark{1,2,3}
 } 

\altaffiltext{1}{The Johns Hopkins University, Homewood Campus,
  Baltimore, MD 21218}

\altaffiltext{2}{Space Telescope Science Institute, 3700 San Martin Drive, Baltimore, MD 21218}

\altaffiltext{3}{European Southern Observatory, Karl-Schwarzschild-Strasse 2, Garching, D-85748, Germany}

\altaffiltext{4}{NASA/GSFC Laboratory for X-ray Astrophysics, Code 662, Greenbelt, MD 20771}
\altaffiltext{5}{Chandra Fellow, Columbia University, 550 W. 112th Street,
New York, NY 10027}

\begin{abstract}
We present soft (0.5-2 keV) X-ray luminosity functions (XLFs) in the Great
Observatories Origins Deep Survey (GOODS) fields, derived for galaxies at
$z \sim 0.25$ and 0.75.  SED fitting was used to estimate photometric
redshifts and separate galaxy types, resulting in a sample of 40
early-type galaxies and 46 late-type galaxies.  We estimate
k-corrections for both the X-ray/optical and X-ray/NIR flux ratios,
which facilitates the separation of AGN from the normal/starburst
galaxies. We fit the XLFs with a power-law model using both
traditional and Markov-Chain Monte Carlo (MCMC) procedures.  A key
advantage of the MCMC approach is that it explicitly takes into
account upper-limits and allows errors on ``derived'' quantities such
as luminosity densities to be computed directly (i.e., without
potentially questionable assumptions concerning the propagation of
errors). The slopes of the early-type galaxy XLFs tend to be slightly
flatter than the late-type galaxy XLFs although the effect is
significant at only the 90\% and 97\% levels for z$\sim 0.25$ and 0.75.
The XLFs differ between $z<0.5$ and $z>0.5$, at
 $>99\%$ significance levels for  
early-type, late-type and all (early and late-type) galaxies.
We also fit Schechter and log-normal models to the XLFs, fitting the
low and high redshift XLFs for a given sample simultaneously assuming
only pure luminosity evolution.  In the case of log-normal fits, the
results of MCMC fitting of the local FIR luminosity function were used
as priors for the faint and bright-end slopes (similar to ``fixing'' these
parameters at the FIR values except here the FIR uncertainty is
included).  The best-fit values of the change in $\log L^*$ with
redshift were $\Delta \log L^* = 0.23 \pm 0.16$ dex (for early-type
galaxies) and $0.34 \pm 0.12$ dex (for late-type galaxies),
corresponding to $(1+z)^{1.6}$ and $(1+z)^{2.3}$. These results were
insensitive to whether the Schechter or log-normal function was adopted. 
\end{abstract}


\section{Introduction}
  
Recently \cite[hereafter N04]{Norman2004} presented the first X-ray luminosity
functions (XLFs) of
normal/starburst galaxies at cosmologically-interesting redshifts (at
z $\sim$ 0.3 and  
0.7).  These XLFs were derived from the Chandra Deep Field North (2
Ms) and South (1 Ms) surveys (hereafter the ``CDF" fields).  
It was found that the normal galaxy XLF was consistent in
normalization and shape 
to the far-infrared LF, assuming pure luminosity evolution of
$ \sim (1+z)^3$.  While the errors were large due to the
limited numbers of galaxies in each luminosity bin, the estimated
star-formation rate (SFR) derived from 
the XLFs was found to be consistent with other SFR measures such as
the H$\alpha$ luminosity.  As discussed in N04, an
interesting aspect of the XLF is that the X-ray emission from
star-forming galaxies often has a large component due to X-ray
binaries, particularly at hard X-ray energies above 2 keV.  Therefore,
the XLF is in part a probe of the {\it binary} stellar mass functions,
and hence indirectly a measure of the SFR.
It is quite possible that binaries 
play a critical role in the evolution of the majority of 
stellar systems,
and X-ray emission provides one of very few
{\it direct} probes of such phenomena.

The Great Observatories Origins Deep Survey
\citep[GOODS;][]{MauroGOODSletter}  is a 
multi-wavelength survey of a sub-area of the CDF fields. This survey
entails deep imaging by three of NASA's Great Observatories: the Hubble Space Telescope, the
Chandra X-ray Observatory, and the Spitzer
Space Telescope, as well as extensive photometric and spectroscopic
observation by ground based facilities.  These data
allow the extension of the N04 results in several new directions.
First, improved redshift determinations are now available for many of
the sources.  Second, the multi-band data have been used to model the
spectral-energy distributions (SEDs) of the sources and estimate their
spectral types.  Therefore XLFs can now be generated as a 
function of galaxy spectral type, which is the primary goal of this paper.
Finally, we have improved the selection of 
AGN vs. normal/starburst galaxies over N04 by applying
k-corrections to both X-ray/optical and X-ray/NIR flux ratios.  Such
k-corrections are  critical to correctly separating AGN from
normal galaxies \citep[][]{BauerLogN}. 

A motivation for deriving galaxy type-selected XLFs  is the fact
that there are multiple contributors 
to the X-ray emission of galaxies, namely low-mass X-ray binaries
(LMXRB), high-mass X-ray binaries (HMXRB), hot ISM, AGN, and to a lesser extent, individual supernovae (SN) and massive stars, 
with the relative contribution of these expected to be dependent on
galaxy type \citep[see ][for a review]{Ptak2001LLAGN}. 
 Specifically, the X-ray emission of early-type galaxies is known to
 be dominated by LMXRB emission, and in
the case of massive, gas-rich ellipticals, hot ISM (at the virial temperature
of the galaxy).  Late-type spiral galaxies should have significant
contributions from both LMXRB and HMXRB
populations, with the former being associated with the older (t $\ga
10^{8-9}$ year) populations.   Late-type spiral galaxies often exhibit
hot ISM due to heating associated with recent episodes of star
formation.  Finally, starburst galaxies should have the largest
contribution from hot ISM (including potentially a superwind outflow)
and HMXRB \citep[see ][]{Persic2003}. 
We would therefore naively expect somewhat
different evolution in the X-ray
luminosity density of these various galaxy types, with the LMXRB
contribution
following the global SFR history of the Universe with a delay of the order
of the evolutionary time scale of low-mass stars, i.e., $\sim 10^{9}$
years \citep{Ghosh01}
~and the HMXRB and hot ISM contribution
tracking the SFR history instantaneously (relative to a Hubble time).

In this paper we are primarily concerned with sources with $L_X <
10^{42}$~\ergs, and hence any AGN present would be a low-luminosity AGN
(LLAGN).
LLAGN are found in all galaxy types, with LINERs preferentially
being found in early-type galaxies \citep[see ][ and references
  therein]{Ho2003}. 
As in N04, we assess the impact of AGN by classifying sources 
based on a Bayesian statistical analysis. In the Appendix we also separately address
the properties of the X-ray flux ratios (i.e., ratio of X-ray to
optical and near-IR flux) which provide important criteria for source
classification.  Ascertaining the presence of any luminosity or redshift
dependence in the X-ray/optical or X-ray/NIR flux ratios would also indicate
possible evolution of the relative contributions of different sources
of X-ray flux in galaxies, i.e., LMXRB, HMXRB, hot gas, and AGN \citep[for examples
of study of the evolution of X-ray/optical flux ratios, please see][]{Ptak2001,Hornstacking}.

This paper is organized as follows.  In Section \ref{s:anal} we
describe our sample selection and data analysis.  The results of our
analysis are given in Section \ref{s:results}, and we discuss the
results in Section \ref{s:disc}.  In the Appendix we give the details
of our galaxy/AGN source classification procedure, basic statistics
concerning the sample, and a discussion of the X-ray/optical and
X-ray/NIR flux ratios.  We assume the WMAP
cosmology of $H_0$ = 70 km s$^{-1}$
Mpc$^{-1}$, $\Omega_m = 0.3$ and $\Omega_{\Lambda}=0.7$.

\section{Methodology \label{s:anal}}
\subsection{X-ray Sample Selection and Redshifts}
The X-ray data used in our analysis are taken from \cite{davocatalog}, 
where the positions and X-ray fluxes for sources in both the CDF-N and
CDF-S are tabulated.  Bauer et al. (2007) have carried
out detailed matching of the GOODS ACS data with the
\cite{davocatalog} X-ray catalog, assigning matching probabilities
based on optical faintness of potential counterparts, resulting in 263
(185) sources in 
the CDF-N (CDF-S) having a single ACS counterpart. We found 44 (28) CDF-N 
(CDF-S) sources with no optical counterpart, while 3 (4) with multiple
counterparts were not included in our analysis.  We also included
the off-nuclear X-ray sources \citep[14 in the CDF-N and 6 in the
CDF-S; e.g., ][]{HornULX}.
We then used
the ACS coordinates in the Bauer catalog to match with the
publicly available GOODS spectroscopic redshift 
catalogs\footnote{http://www.stsci.edu/science/goods} and
\citet{MobasherGOODSphotz}
~photometric redshift catalog to obtain both
redshifts and spectral types for all sources.  Stars were excluded.  
The main parameters for each galaxy are
the soft-band (0.5-2.0 keV) flux \citep[from][]{davocatalog}, 
the hardness ratio (defined as $\frac{H-S}{H+S}$ where H is the
2-8 keV vignetting-corrected count rate and S is the 0.5-2.0 keV
vignetting-corrected count rate; from Alexander et al. 2003), the
redshifts, and the optical spectral types.  We broadly
divided the spectral types into the groups early-type, late-type, and
starburst/irregular galaxies, however we only extract XLFs for early-type and
late-type galaxies since the numbers of irregular galaxies were very small.  

\subsubsection{Photometric Redshifts}
The photometric redshifts for GOODS fields are estimated using
template fitting technique (Mobasher et al 1996; Chen et al 1999;
Arnouts et al., 1999; Benitez 2001; Bolzonella et al 2000, Dahlen et
al. 2007).
The rest-frame Spectral Energy Distribution (SED)
for galaxies of different types are convolved with
filter response functions of the filters used in photometric observations
of galaxies.
The convolved SEDs, shifted in redshift space, were then fitted
to observed SEDs of individual galaxies by minimizing the $\chi^2$
function

$$\chi^2 = \Sigma_{i=1}^n [(F^i_{obs} - \alpha
F^i_{template})/\sigma^i)]^2$$

\noindent where the summation, $i$, is over the passbands (i.e. number of
photometric points) and $n$ is the total number of passbands. $F^i_{obs}$
and
$F^i_{template}$ are, respectively, the observed and template
fluxes at any given passband. $\sigma^i$ is the uncertainty in the
observed
flux and $\alpha$ is the normalization. The redshift and SED (i.e.
spectral
type) corresponding to the minimum $\chi^2$ value for a given galaxy were
then assigned to that galaxy. We used priors based on luminosity functions
 (LFs).
The main effect of a LF prior is to discriminate between cases in which
the redshift probability distribution function, which identifies the most
likely redshift,
shows two or multiple peaks (i.e. more than a single optimum redshift) due
to
confusion between the Lyman break and the 4000 \AA~ break features.
The absolute magnitudes of the object at the redshift peaks can then
discriminate between these possibilities. Absorption due to
intergalactic HI is included using the parametrization in Madau (1995).

We use template spectral energy distributions (SEDs) for normal
galaxies consisting of E, Sbc, Scd and Im from Colman et al (1980) and
two starburst templates from Kinney et al (1996)-(SB2 and SB3).
 To increase the
spectral resolution, we construct intermediate-type templates by using
the weighted mean of the adjacent templates and interpolate between them.
This is done by defining five intermediate-type templates between the main
spectral types used. Therefore, we use a total of 31 SED templates in
this study. Photometric redshifts were then measured using the 
the observed photometry in GOODS-N (UBVR{\it iz}JK) and GOODS-S (UBVR{\it i}JHK).
%
\subsubsection{Spectroscopic Redshifts}

\begin{figure*}
\plottwo{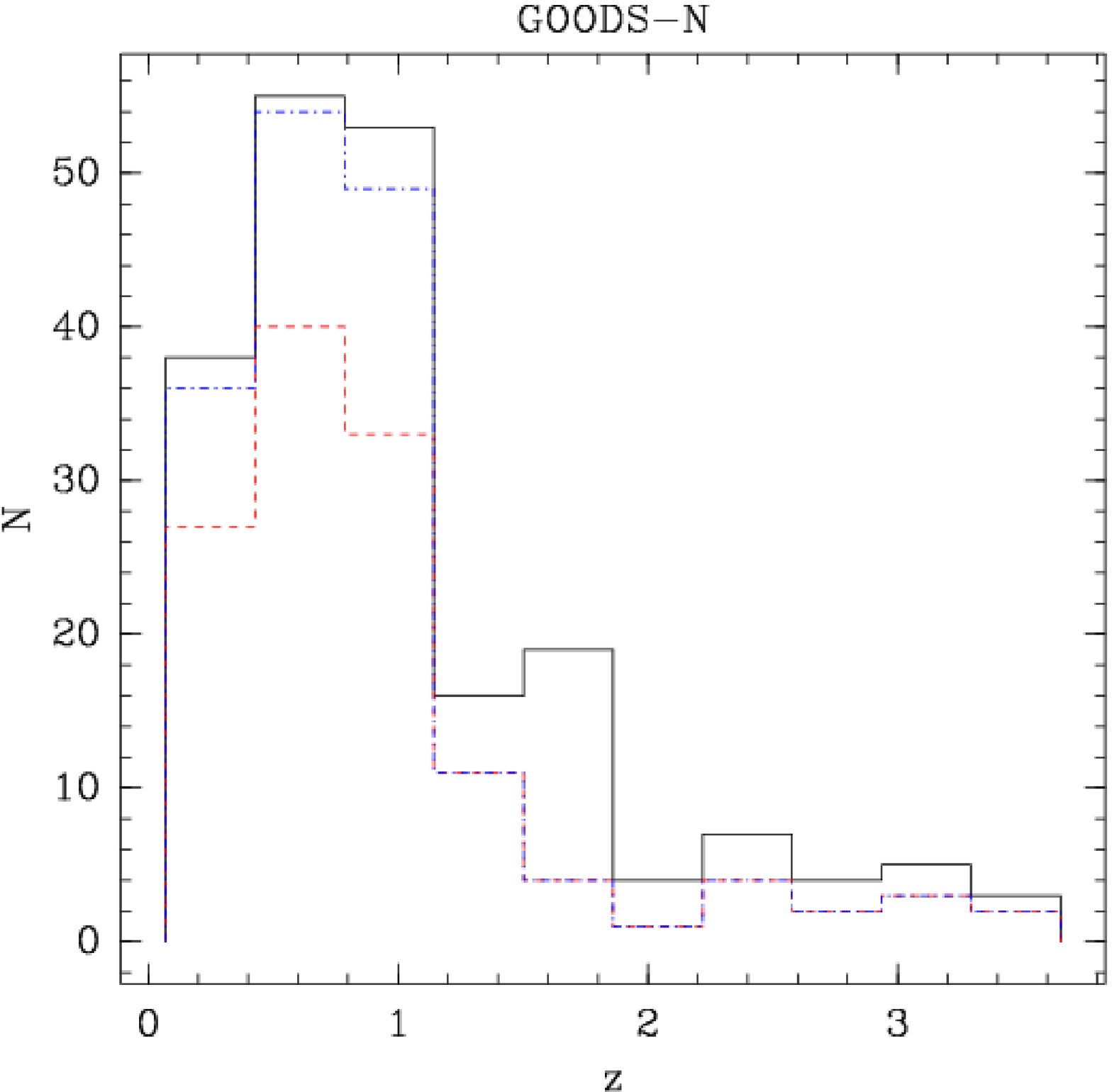}{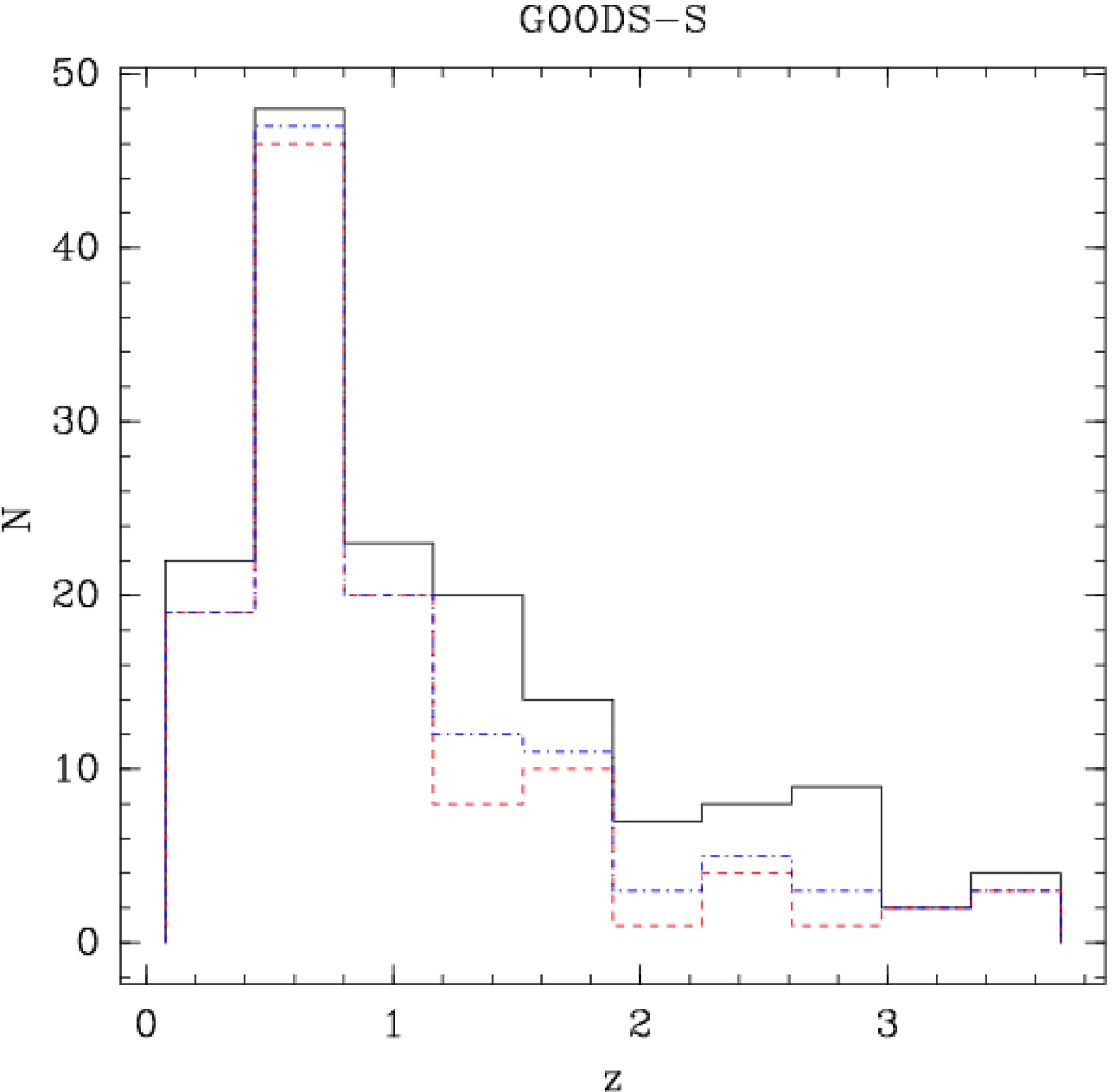}
\caption{The redshift distributions for the CDF-N (left) and CDF-S
  (right) fields. The spectroscopic redshift distributions are shown
  in blue with dot-dashed lines. In cases where the spectroscopic
  redshift quality was unknown or flagged as low and the photometric
  redshift error was small (see text for details), the photometric
  redshift was adopted.   This distribution of sources for which a
  spectroscopic redshift was adopted is shown in red with dashed
  lines.  \label{f_zdist}}
\end{figure*}

For a subset of the galaxies, we use the available spectroscopic 
redshifts. 
The majority of
the spectroscopic redshifts were derived from Keck DEIMOS data for the
CDF-N \citep[see ][ and references therein]{Wirth2004} and 
VLT FORS2 or VIMOS data for the CDF-S 
\citep{Szokoly2004,Vanzella2006}
.
The limiting magnitude for spectroscopic redshift determination was
typically $R_{AB} \sim 24$.  If a
spectroscopic redshift was not available then a photometric redshift
was used, with the photometric redshift
limiting magnitude typically being $R_{AB} \sim 25$
\citep{MobasherGOODSphotz}.  In cases where a 
quality assessment was available for the spectroscopic redshift and
was considered to be poor, and a photometric redshift with error
$\delta z <0.2$ was available \footnote{$\delta z$ is defined as the 68\%
uncertainty on the photometric redshift derived from the posterior
probability}, the photometric redshift was used.  In cases where there
was no spectroscopic redshift quality given and a photometric 
redshift with error $\delta z <0.1$ was available, the photometric redshift was
used.
This resulted in 204 (129) and 157 (116) total (spectroscopic) redshift
determinations  
for the CDF-N and CDF-S sources, respectively (i.e., 73\% and 82\% of the
GOODS X-ray sources with a unique optical counterpart have a redshift
estimate). The redshift distributions are shown in Figure \ref{f_zdist}. 

\begin{figure*}
\plottwo{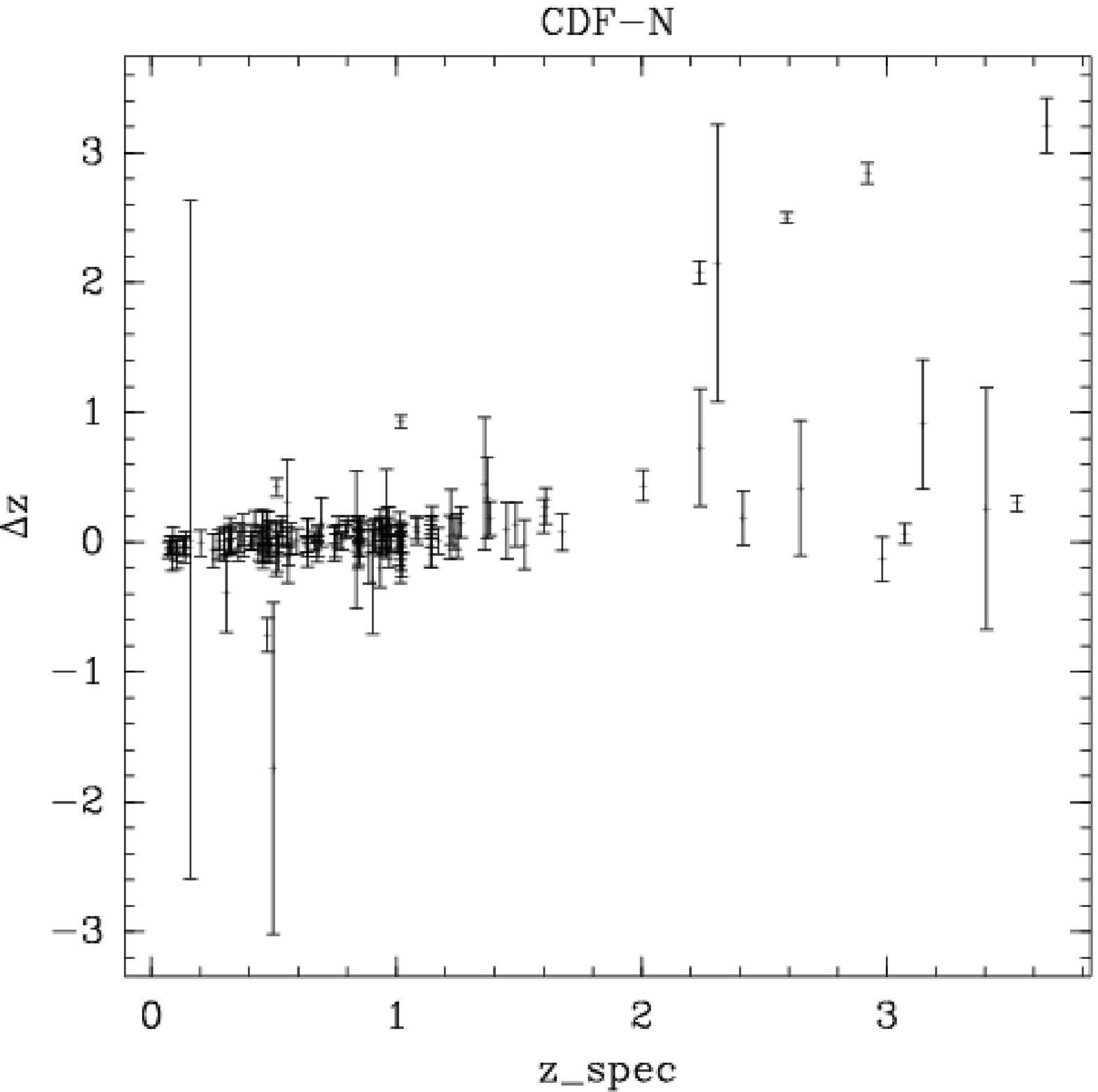}{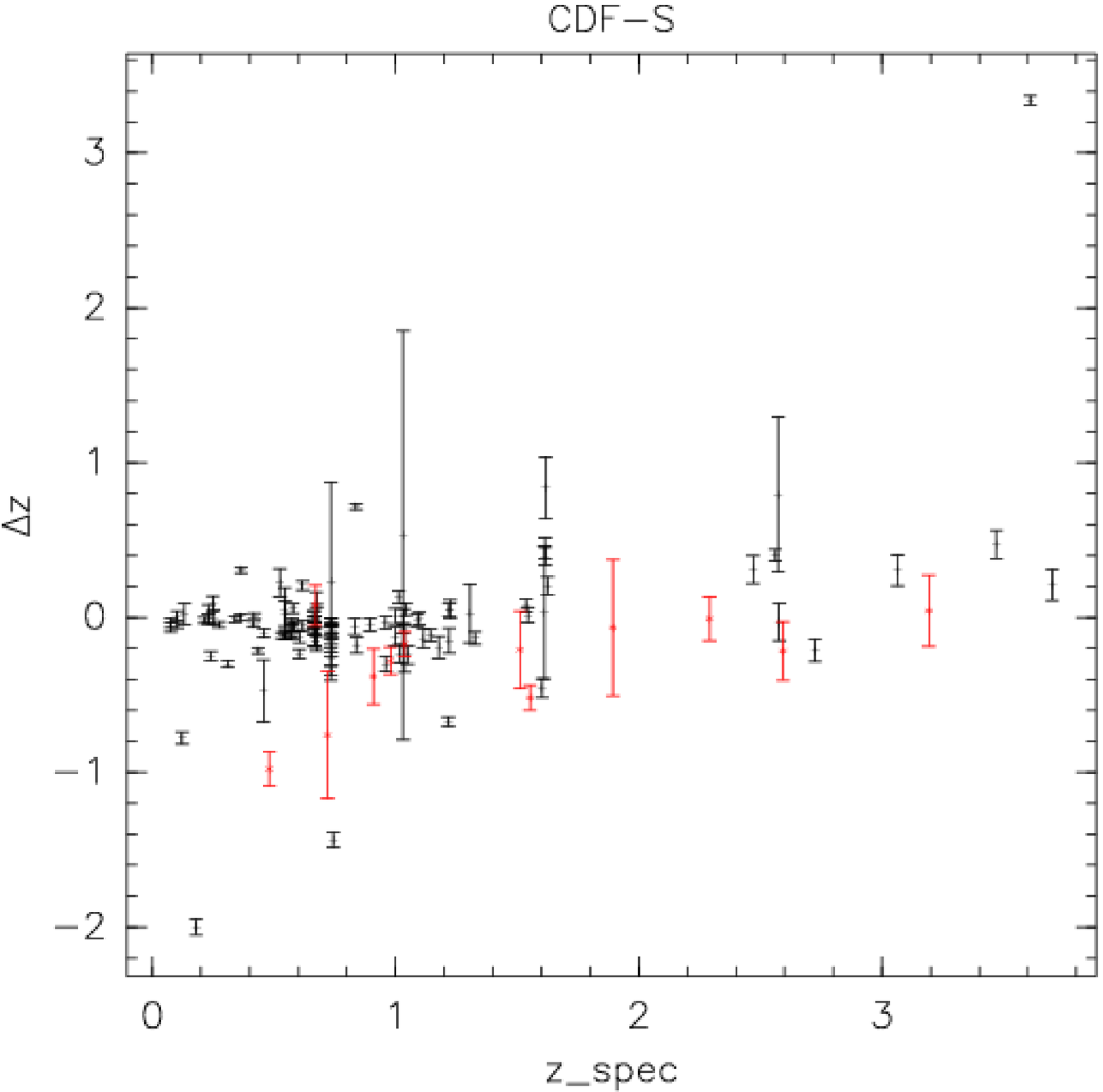}
\caption{$\Delta z = z_{spec} - z_{phot}$ plotted as a function of
  $z_{spec}$ for CDF-N (left) and CDF-S (right). In the CDF-S plot,
  the sources with poor spectroscopic redshift determinations (often
  based on a single line identification) are shown in red and marked with
  crosses) The errors
  plotted for the points are based solely on the photometric redshift
  error.
\label{f_z_dz}} 
\end{figure*}

In cases where both
a photometric and spectroscopic
redshift were available, the mean absolute redshift difference was 
0.09 (CDF-N) and 0.13 (CDF-S) after removing a small number of
  outliers with $\Delta z \equiv |z_{\rm spec}-z_{\rm phot}| > 0.5$. 
In Figure \ref{f_z_dz} we plot the difference between the photometric
and spectroscopic redshift for sources with both measurements.  The
errors plotted in the figure are based solely on the photometric
redshift errors.
This figure shows that our
typical redshift uncertainty is  $\Delta z \sim
0.1-0.2$ for $z \la 1.5$, and increases significantly for $z > 1.5$.  We
conservatively limit our analysis to $z \le 1.2$.  
This leaves a total of
148 sources in the CDF-N and 95 sources in the CDF-S regions, of
which 138 and 81 have soft-band X-ray detections.  Note that 131
(CDF-N) and 75 (CDF-S) of the X-ray sources have $L_X < 1 \times
10^{43} \rm \ ergs \ s^{-1}$, the highest X-ray luminosity considered
in the X-ray luminosity functions discussed here.

\subsection{K-Corrections}
The soft-band X-ray fluxes, listed in Alexander et al. (2003),
are based on the observed count rates and a count rate to flux
conversion computed from an effective photon index (based on band
ratios).  This photon index ($\Gamma$) was used to ``k-correct'' the
observed fluxes, i.e.,  $F_{X, kcor} = 
(1+z)^{\Gamma-2}F_{X, obs}$, and $L_X = 4\pi d_L^2F_{X, kcor}$ where
$d_L$ is the luminosity distance. 
In Appendix A we present the sources,
the adopted redshifts and X-ray luminosities.

\begin{figure*}
\plottwo{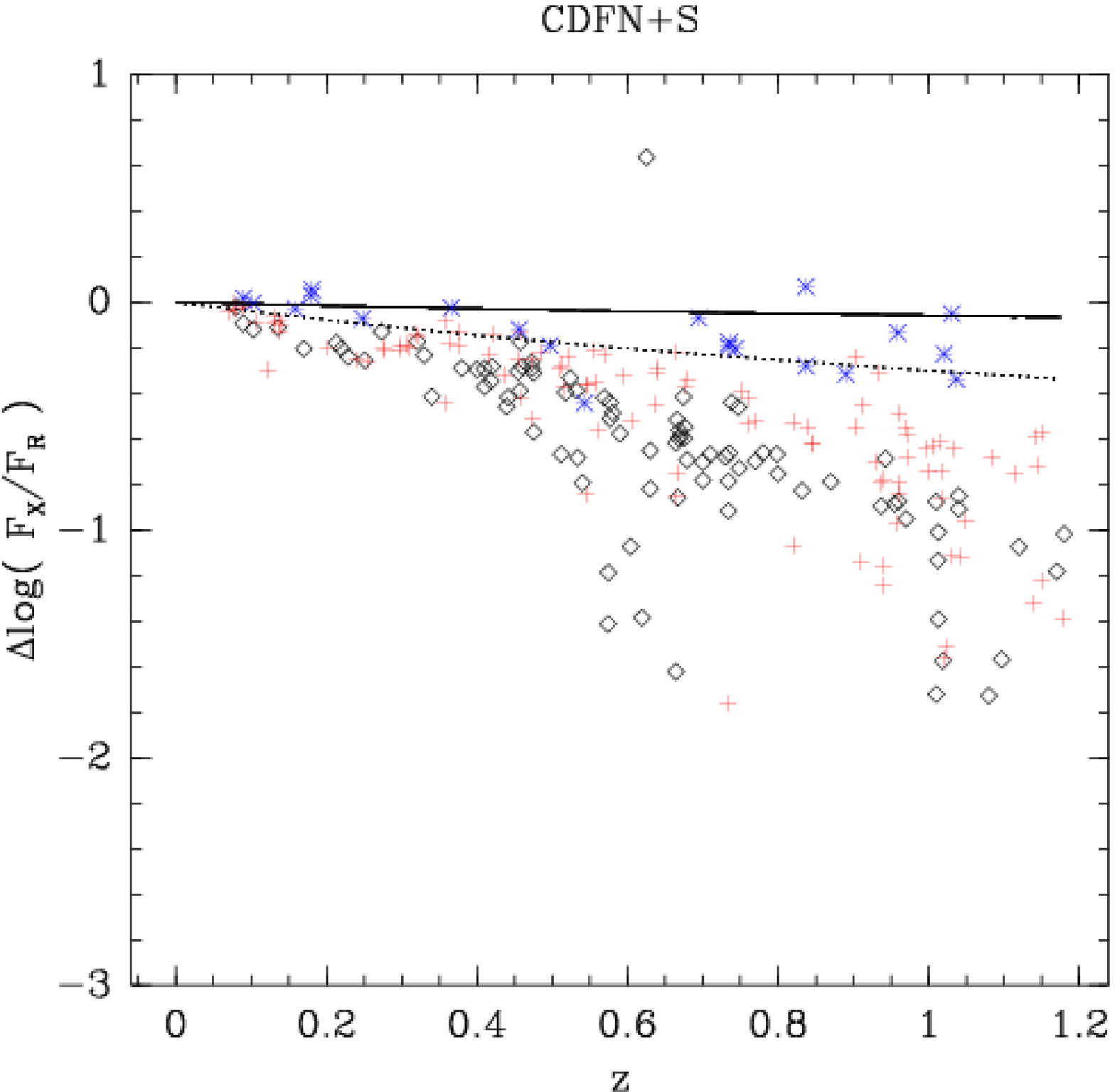}{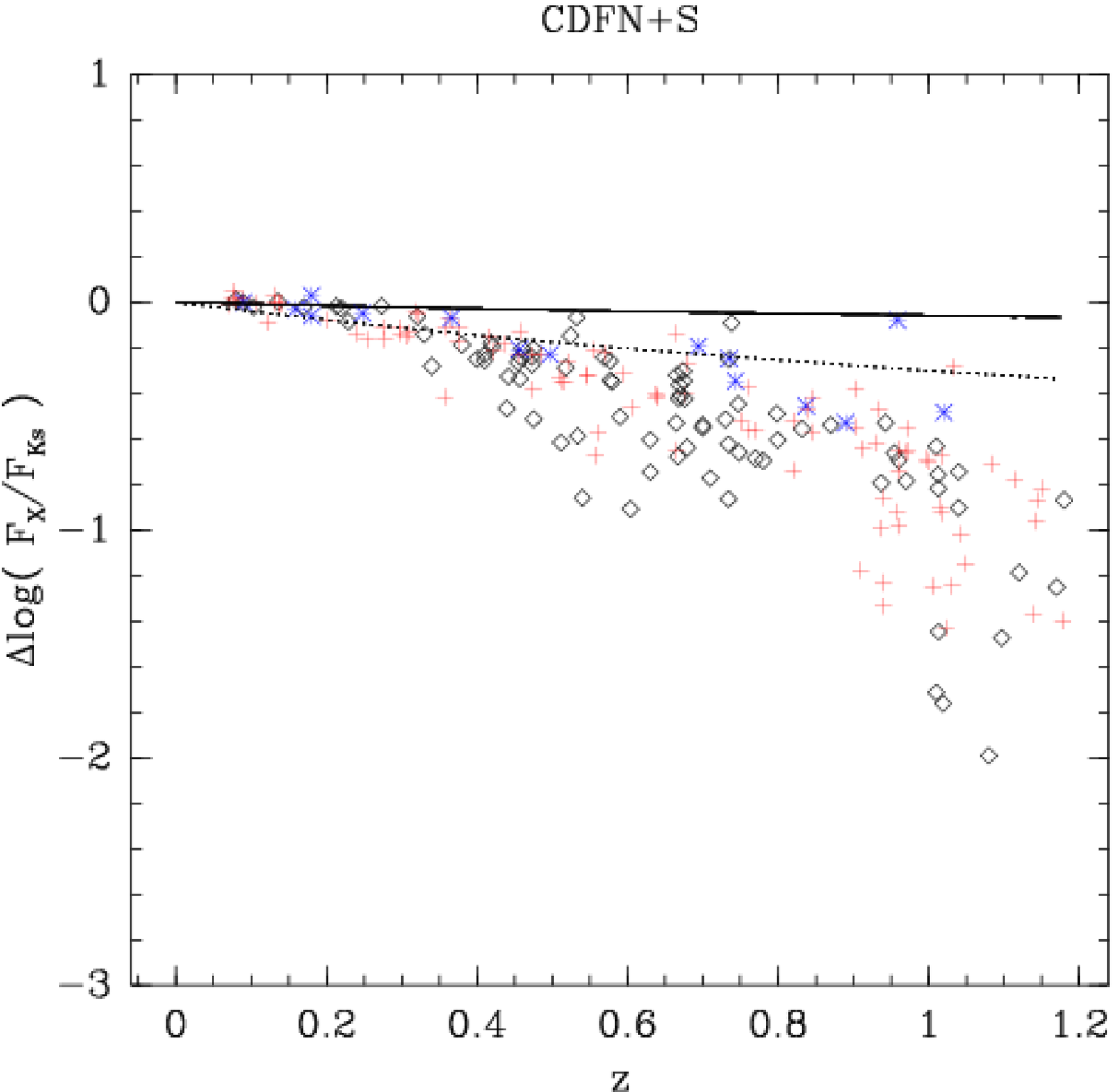}
\caption{The k-corrections applied to the X-ray flux ratios for the R (left)
  and $\rm K_s$ (right) bands, for both of the GOODS fields. The 
  points show 
$\Delta\log[F_X/F_{opt}] = \log [(F_X/F_{opt})_{z=0}] - \log
  [(F_X/F_{opt})_{z}]$, 
where $\log [(F_X/F_{opt})_{z=0}]$ is the flux ratio after applying
k-corrections and $\log [(F_X/F_{opt})_{z}]$ is the ``observed''
flux ratio. Early-type sources are marked with (black) diamonds,
  late-type sources are marked with (red) pluses, and 
  irregular/starburst sources are marked with (blue)
  asterisks. The solid and dotted lines show
  the X-ray k-correction for $\Gamma=1.8$ and $\Gamma=1.0$,
  respectively. 
\label{f:kcor}}
\end{figure*}

For a given band Q, we define $F_Q$ to be $\nu F_\nu$ at the central
frequency of the band $\nu_Q$. 
Since the GOODS magnitudes are in the AB system, $\log(F_X/F_Q)$ is
given by $\log F_X + 0.4(Q +  48.6) - \log\nu_Q$,  e.g., in the case of R
band, $\log(F_X/F_R) = \log F_X + 0.4R_{AB} + 4.8$.  
The optical magnitudes were k-corrected by interpolating between the
magnitudes whose central wavelengths bracket $(1+z)\lambda$,  where
$\lambda$ is the central wavelength of the filter of interest (e.g.,
4400\AA\ for B-band).  The bandpass correction term (1+z) was also
included in the k-correction 
\citep{BlantonKCorrect}. In Figure
\ref{f:kcor} we show the k-corrections applied to R and $\rm K_{s}$ band
flux ratios.  Note that in the case of $\rm K_s$ band magnitudes, the
interpolation discussed above necessarily amounts to an extrapolation
since $\rm K_s$ is the longest wavelength band in the dataset and every
source considered here has $z>0$. 
The corrections are plotted separately for each spectral type, where it can be
seen that the largest k-corrections occur for B band fluxes from
early-type galaxies, which is not surprising considering their red
color.  Also, as mentioned in Bauer et al. (2004), the k-corrections
to the flux ratios are dominated by the optical correction rather than
the X-ray correction.

\subsection{Luminosity Function Estimation}
We construct binned luminosity functions using the estimator described in
\citet{page2000}.  Briefly, this assumes negligible variation in 
both the luminosity function and its evolution across each
$\Delta L \Delta V(z)$ bin, where $V(z)$ is the co-moving volume:
\begin{equation}
N \simeq \phi(L,
z)\int^{L_{max}}_{L_{min}}\int^{z_{max}}_{z_{min}}C(L,z)V(z)dzdL
\end{equation}
$N$ is the number of sources in the XLF bin bounded by
$L_{min}$, $L_{max}$, $z_{min}$ and $z_{max}$, and $C(L, z)$ is a completeness
correction. Note that $z_{max}$ is a function of luminosity since the
limit is chosen at the highest observable redshift given the limiting
flux (or, equivalently, $C(L,z) = 0$ where $F(L,z) < F_{lim}$). 
The error for each XLF bin is accordingly derived from the
Poisson error on the number of galaxies in each bin (note this assumes
negligible statistical and systematic error in the volume integral terms),
and for bins with no galaxies we used N=1.841 as the 1$\sigma$ upper limit \citep{Gehrels86}.
Small luminosity function bins 
are preferred, however larger bins minimize Poisson noise and reduce
the effect of luminosity uncertainties within a bin (due to redshift
uncertainties). 
For example, for the uncertainty in luminosity resulting from redshift
error to be on the order of the bin size, bin sizes of $\Delta \log L
= 0.5$ are required for sources with $z \ga 0.4$ and redshift errors
of 0.1 or less. 
Given the small number of sources, we divide our sample
into just two redshift bins, $z < 0.5$ and $0.5 < z < 1.2$.   We used
the soft-band GOODS sky coverage shown in \citet{Triester2004}.  

\subsection{Completeness Correction}
\begin{figure}
\plotone{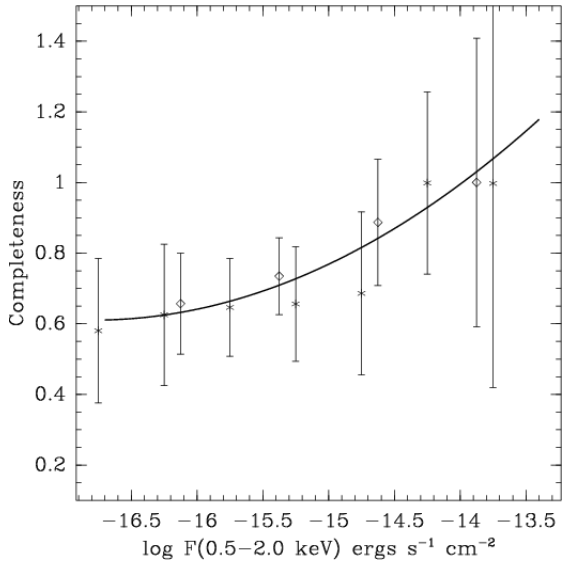}
\caption{Completeness corrections as a function of flux calculated for
  the GOODS-N (crosses) and GOODS-S (diamonds) regions.  
  The line shows a quadratic fit to the combined completeness
  corrections (i.e., to be applied to either data set).
  \label{complfig}
}
\end{figure}

The completeness correction $C(L, z)$
is given by the product of the X-ray detection completeness, the
probability of a given X-ray source having an optical/NIR
counterpart, and the probability of a counterpart having a redshift.
The latter two terms reduce to the probability of an X-ray
source having a redshift.  For the X-ray detection completeness we
used the results of simulations originally performed for the full CDF
areas discussed in Bauer et al. (2004), where the effective solid
angle for each source was computed (i.e., the maximum solid angle over
which the source could have been detected).  The ratio of the effective solid
angle to the (total) CDF solid angle at the flux of the source is then
an estimate of the completeness for sources at a similar flux and
position.  We then computed the completeness as a function of
flux by multiplying the mean per-source completeness 
(i.e., effectively averaging over off-axis angle) and the fraction of
GOODS x-ray sources with a redshift in the given flux range (the bin
sizes were 0.5 dex for the GOODS-N sources and 0.75 dex for the
GOODS-S sources), shown in Figure \ref{complfig}. In both the GOODS-N
and GOODS-S the completenesses ranged from $\sim 60\%$ at
$F_{0.5-2.0\rm\ keV} = 2 \times 10^{-17}$ \ergcms in the North and
$F_{0.5-2.0\rm\ keV} = 7 \times 10^{-17}$ in 
the South to 100\% for $F_{0.5-2.0\rm \ keV} > 10^{-14}$\ergcms,
although with errors on the order of 20\% in each flux bin.  We fit
the completeness with a quadratic function, both to avoid the impact
of statistical fluctuations and to have an analytic expression for use in
calculating Eq. 1.  The best-fitting relation was 
$C(F_X) = 14.56 + 1.66 \log F_X + 0.0495 (\log F_X)^2$ and is shown in
Figure \ref{complfig}.

\subsection{Galaxy Classification}
N04 classified sources as AGN1 (broad-line
AGN), AGN2 (narrow-line AGN), and ``galaxy'' (i.e., normal/starburst galaxy)
using a prior based on the observed distributions of X-ray luminosity
($\log L_X$), the 
X-ray-to-optical flux ratio ($\log F_X + R$), and X-ray hardness
ratio derived from a subset of the sources 
where the optical classification was secure.  These distributions were
used as priors along with the likelihoods of these measured values in
computing the posterior probabilities.  A source was then classified
as a galaxy if the galaxy probability exceeded the AGN1 and AGN2
probabilities.

Here we employ a similar algorithm, discussed in Appendix A .
However, there are a number of differences with the approach in 
N04, including use of the X-ray/K-band flux  ratio,
inclusion of k-corrections in the computation of X-ray/optical and
X-ray/NIR flux ratios, and a more conservative galaxy selection
threshold. Note that our priors are based on the spectroscopic data in
Szokoly et al. (2004) which covers the entire CDF-S area.  In Szokoly
et al. only R and K band values are listed, and accordingly
k-corrections could not be interpolated as was done with the GOODS
data (see \S 2.2).  However, we estimated the spectral type of the
sources based on their R-K color, and applied a mean k-correction
derived for each type.

In the case of X-ray hardness ratio, the means and standard deviations
of the prior samples were $-0.19 \pm 0.46$ for galaxies, $0.16 \pm
0.37$ for AGN2 and $-0.51 \pm 0.05$ for AGN1, very similar to
those used in N04.  
For comparison, 
\citet{Peterson2006}
artificially redshifted a
sample of local galaxies to $z=0.3$ and found that hardness
ratios of AGN2 was split evenly around HR=0.  They noted that soft
emission from starbursts could be resulting in HR$<$0 in
AGN2/starburst composites (the case for 4 out of the 7 AGN2 with HR$<$0).
Our final normal/starburst galaxy sample consisted of 64 CDF-N sources
and 23 CDF-S sources. We also consider an ``optimistic''
normal/starburst galaxy sample which includes all sources that were
{\it not} classified as an AGN.  This sample therefore includes
sources with ambiguous classification, similar to the original
selection in N04.  The optimistic normal/starburst
sample contains 98 CDF-N and 54 CDF-S sources.

\subsection{XLF Model Fitting}
In order to quantify differences among the various XLFs discussed
here, we fit the XLFs with several models (note that all luminosity
functions discussed here are binned).  The simplest model is a 
linear function in $\log 
\phi$-$\log L_X$ space, i.e., $\log \phi(L_X) = a\log {L_X} +
b$. We performed linear fits with 
two conventional approaches, namely using the survival analysis linear
regression (hereafter Method 1) discussed in \citet{Isobe1986}, and 
least-squares fitting (using the {\sc qdp} program) after excluding the upper limits (Method 2).  
The linear regression approach has the disadvantage of not explicitly
including any error information (i.e., the algorithm only takes as
input the value of each detection or limit), and the confidence level
used in computing an upper limit is arbitrary.  

\subsubsection{Markov-Chain Monte Carlo Fitting}
Bayesian parameter estimation has been gaining popularity in various
fields of astronomy, most notably in cosmology \citep[see, e.g.,][]{sperg07}. 
A key advantage of Bayesian parameter estimation is that common
statistical issues, such as the treatment of upper limits in fitting
and propagation of errors, are inherently handled properly.  In
addition the probability distributions for parameters of interest are
computed and can be shown rather than simply a summary such as the
68\% confidence interval.  This latter point is particularly relevant
when there are relative minima in the fitting statistic, in which case
the meaning of traditional error bar is not well defined.

The basis of Bayesian parameter estimation is the computation of the
posterior probability distribution:
\begin{equation}
p(\theta|D) = \frac{p(\theta)p(D|\theta)}{p(D)}
\end{equation}
where $\theta$ is the vector of model parameters (e.g., $\theta =
\{a,b\}$ for the linear model), $D$ represents the
data, $p(\theta)$ is the prior probability distribution for the
parameters, $p(D|\theta)$ is the likelihood function, and $p(D)$ is a
normalization constant (i.e., since it does not depend on the
parameter values). $p(D)$ is given by $\int p(\theta)p(D|\theta)d\theta$. 
In practice, computing $p(\theta|D)$ is 
computationally difficult since it requires an $n$-dimensional
integral when fitting an $n$ parameter model.  The Markov-chain Monte
Carlo (MCMC) procedure circumvents this by performing a directed random
walk through the parameter space \citep{vanDyk01, ford06}.  Here we
assume Gaussian priors for each parameter (with mean and standard
deviation $\mu_{\theta_i}$ and $\sigma_{\theta_i}$ for parameter
$\theta_i$), and chose very large $\sigma_{\theta_i}$ values for the
case of a flat (uninformative) prior.  The likelihood function
for the number of counts in an XLF bin is the Poisson distribution,
giving
\begin{multline}
p(\theta|D) \propto \prod_{i=1}^{n} G(\theta_i|\mu_{\theta_i},
\sigma_{\theta_i}) \times \\
\times \prod_{j=1}^{m} Pois[N_j|\phi(\theta_1,..,\theta_n, L_{X,j},
z_j) V(L_{X,j}, z_j)]
\end{multline}
where there are $n$ model parameters and $m$ XLF bins.
$G(\theta_i|\mu_{\theta_i}, \sigma_{\theta_i})$ is the Gaussian 
prior for parameter $\theta_i$. $N_j$ is the number of sources and
$V(L_{X,j}, z_j)$ is the co-moving volume corresponding to the jth
bin (see Eq. 1).
$\phi(\theta_1,..,\theta_n, L_{X,j}, z_j)$ is the model XLF evaluated
at $L_{X,j}$ and $z_j$.  The $Pois[]$ terms give the likelihood of
detecting $N_j$ galaxies in XLF bin $j$ given an expectation of $\phi V$.
In addition
to the simple linear model we also fitted Schechter and log-normal
functions to the individual XLFs.  We used the Metropolis-Hastings MCMC algorithm, 
in which a ``proposal'' distribution\footnote{We used the Gaussian
distribution as the proposal distribution as is common practice.} is
used to guide the variation of 
the parameters \citep[see][]{ford05, russel07}.  In this procedure
random offsets for each 
parameter are drawn from the proposal distribution, and accordingly
the step sizes (i.e., the Gaussian $\sigma$s) are preferably on the
order of the final error for that parameter.  A step is accepted if
the probability of the model given the new parameter values is higher,
and also at random intervals when the probability is lower (i.e.,
occassionally the fit is allowed to proceed ``downhill'' to avoid
relative minima).  For a given run, three chains were produced
with a length of at least $2 \times 10^6$ iterations,
and the parameter step sizes were adjusted during the first chain to
achieve acceptance rates in the range 0.35-0.4.
Only the last chain in a given run was used for analysis.  At least 10
runs were performed for a given fit and we computed the convergence
$R$ statistic from 
\citet{gelm04}.  $R$ values $\la 1.2$ indicate convergence, and in
every case the $R$ values were $< 1.01$.  In practice, the linear
model parameters $a$ and $b$ were highly correlated, which leads to
inefficient MCMC fitting.  As discussed in \citet{gelm04}, we
addressed this by instead fitting for $a_1$, $a_2$ and $b$, where $a =
a_1 + a_2b$. The initial values of $a_1$ and $a_2$ were determined
from a linear regression fit to the output of a short MCMC run, and
the value of $a_2$ was held approximately fixed by using a tight
prior ($a_1$ and $a_2$ are ``nuisance'' parameters and are not
discussed further).  The initial parameter values for a given run was
chosen randomly and the range of the allowed values was $\pm2 
\times$ the initial step size from the prior mean.  The dispersion between the
best-fitting parameter values from the runs in a given fit was always
$<10\%$ of the final error in the parameters, showing that the
chains converged independently of the starting values.  For each
parameter of interest the chain values were binned into normalized
histograms, which represent the marginalized posterior probability distribution
of the parameter.   The peak
value (i.e., the mode) was taken as the best-fit value and the 68\%
confidence interval was derived from the probability density by
steeping from the peak in the direction of the smallest decrease until
the integrated area equaled 68\%, as discussed in \citet{Kraft91}.  In
order to assess how constraining the prior is, we computed the ratio of
the prior $\sigma$ values to the standard deviation of the MCMC
parameter values (i.e., the 68\% error in the case of a Gaussian
posterior probability distribution).  In tables showing the fit
results we marked parameter values and errors as having a
tightly-constraining prior when this ratio is $<$1.1 and as a
moderately-constraining prior when it is between 1.1 and 2.0.  

\subsubsection{Fit Quality}
Unfortunately the MCMC procedure does not directly provide a model
probability estimate since this requires the normalization of Eq. 2 to
be computed.  Since efficient model selection is complicated (and
the subject of active research, see, e.g., Trotta 2007), we defer this
and here we compute $\chi^2$ for the models after excluding upper
limits.  This also has the advantage that it can be consistently
applied to all of the model fitting methods discussed here in the case
of the linear model. 

\section{Results \label{s:results}}
\subsection{Linear Model Fits}
\begin{figure*}
\plottwo{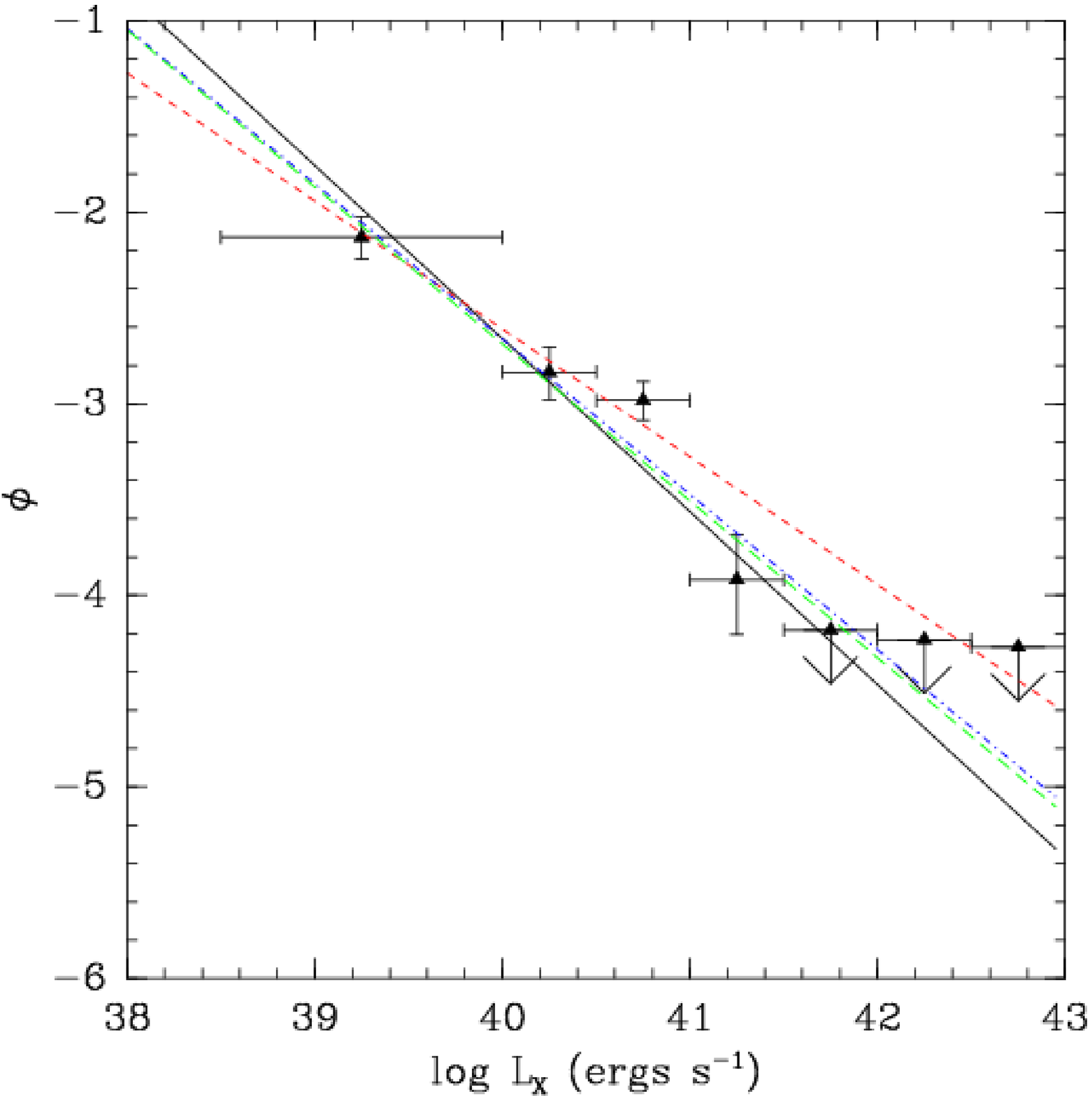}{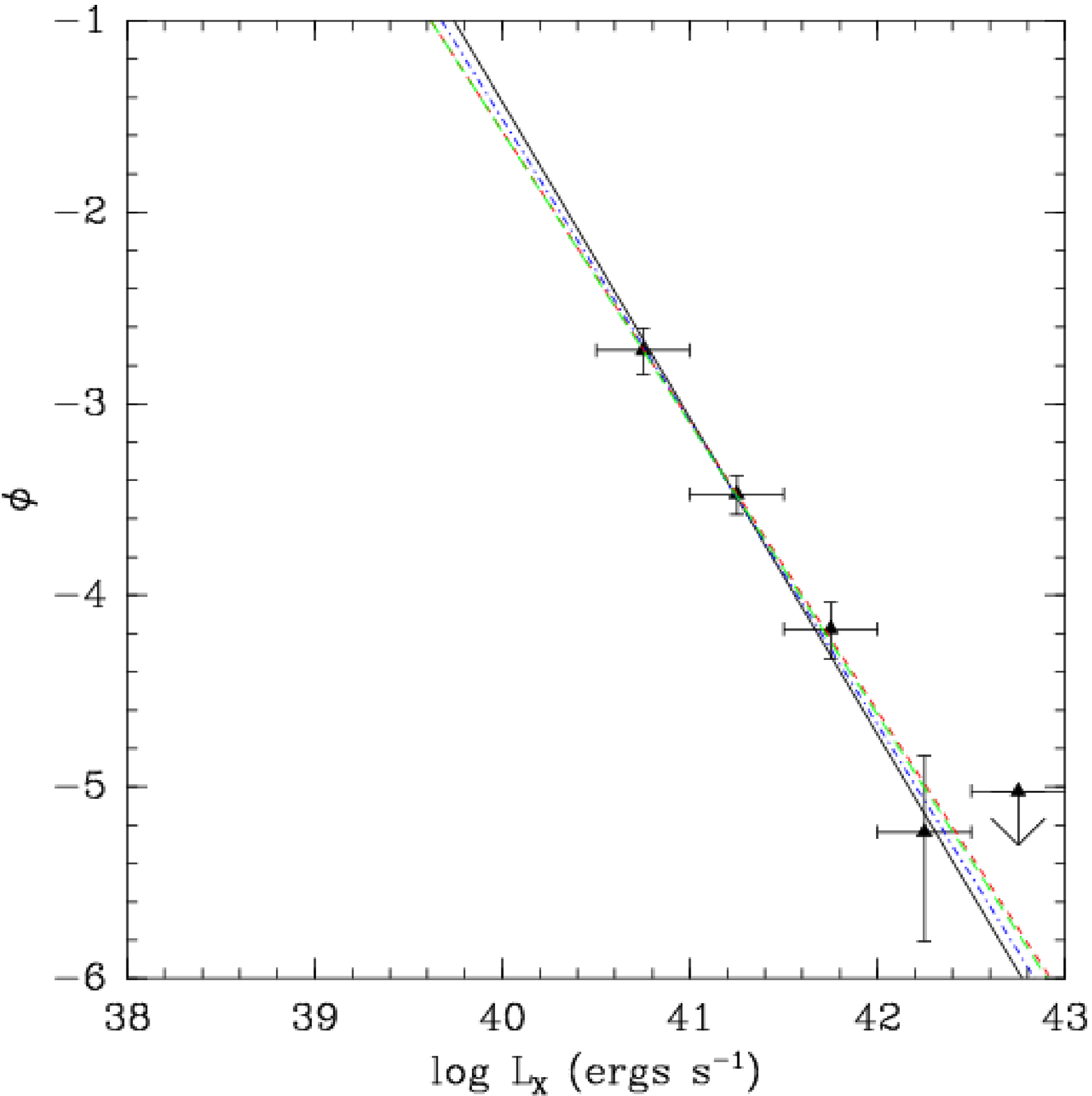}
\caption{Full-sample (early + late-type galaxy) XLFs from the GOODS
  fields, for the 
  redshift intervals z$<$0.5 (left) and 0.5 $<$ z $<$ 1.2 (right).
  The solid (black), dashed (red), dot-dashed (blue) and long-dashed
  lines (green) lines respectively show the best-fitting models from Method 1
  (survival analysis linear regression), Method 2 (least-squares
  excluding upper-limits), MCMC fitting with Method 1 results used to
  initialize the fit, and MCMC fitting with Method 2 results
  used to initialize the fit. \label{f_xlfs_gal_linear}}
\end{figure*}

\begin{figure*}
\plottwo{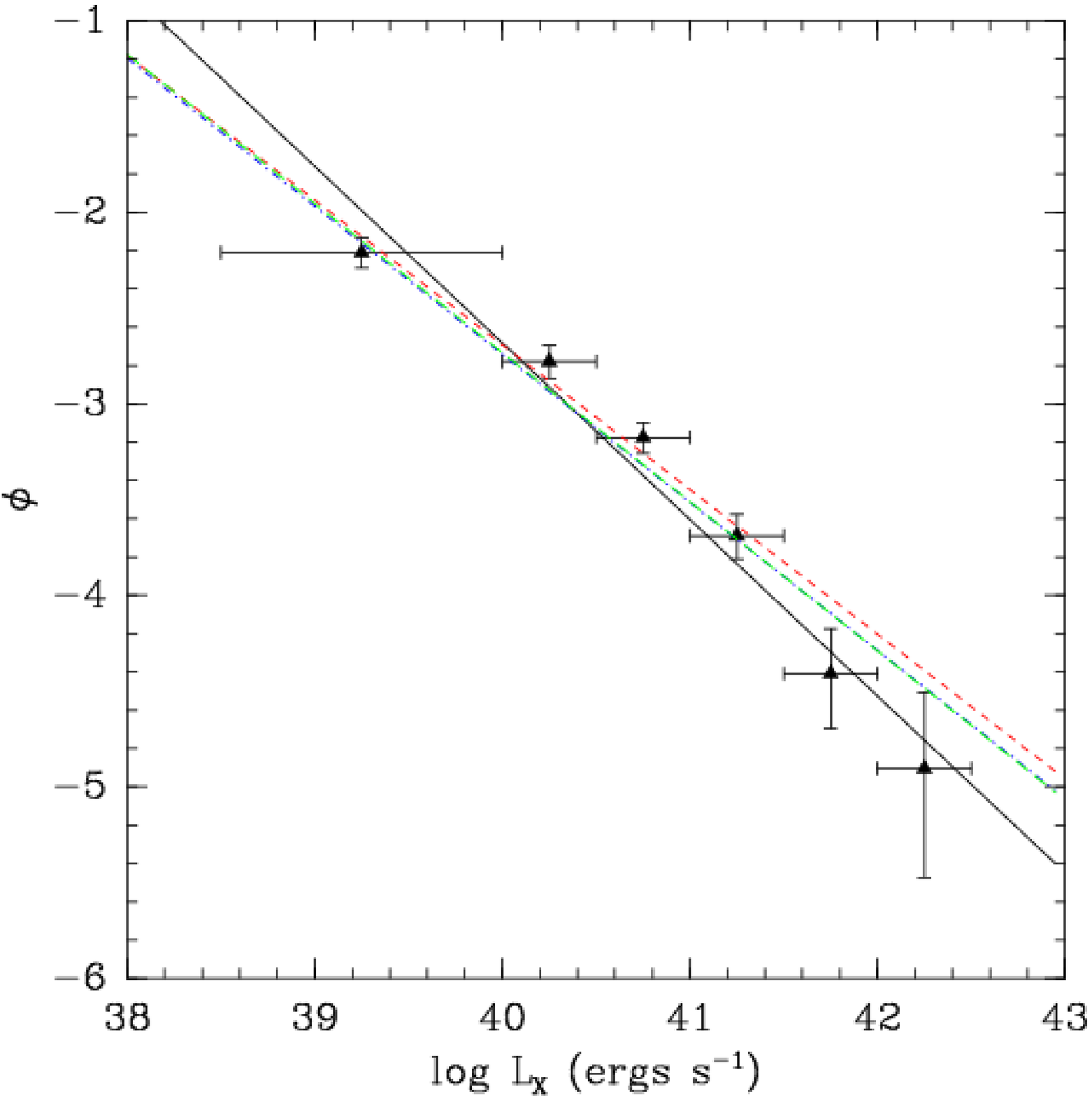}{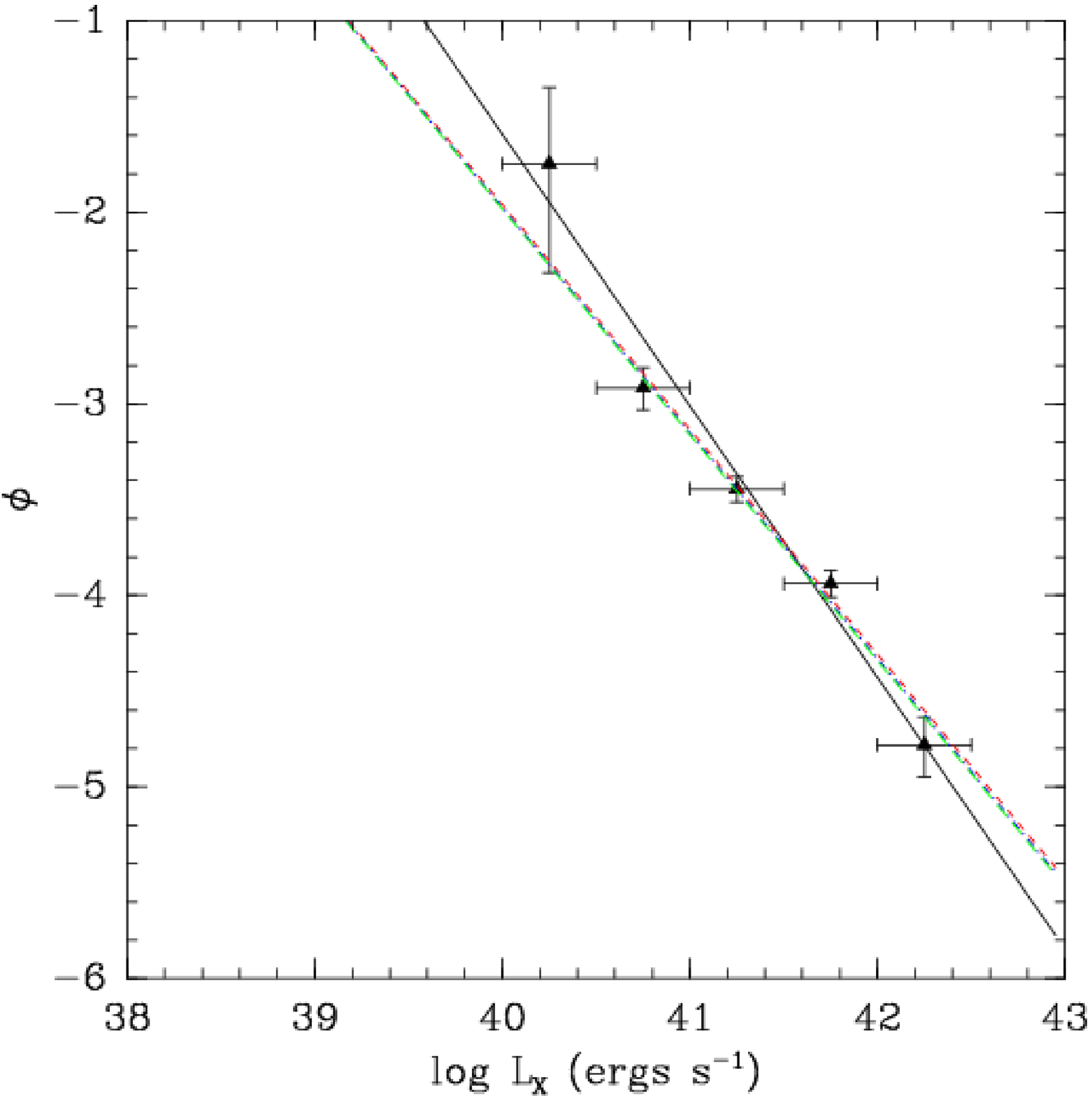}
\caption{The N04 XLFs for z$<$0.5 (left) and 0.5 $<$
  z $<$ 1.2 (right), with linear fits shown as in Figure
  \ref{f_xlfs_gal_linear}.
\label{f_xlfs_norman_linear}} 
\end{figure*}

\begin{figure*}
\plottwo{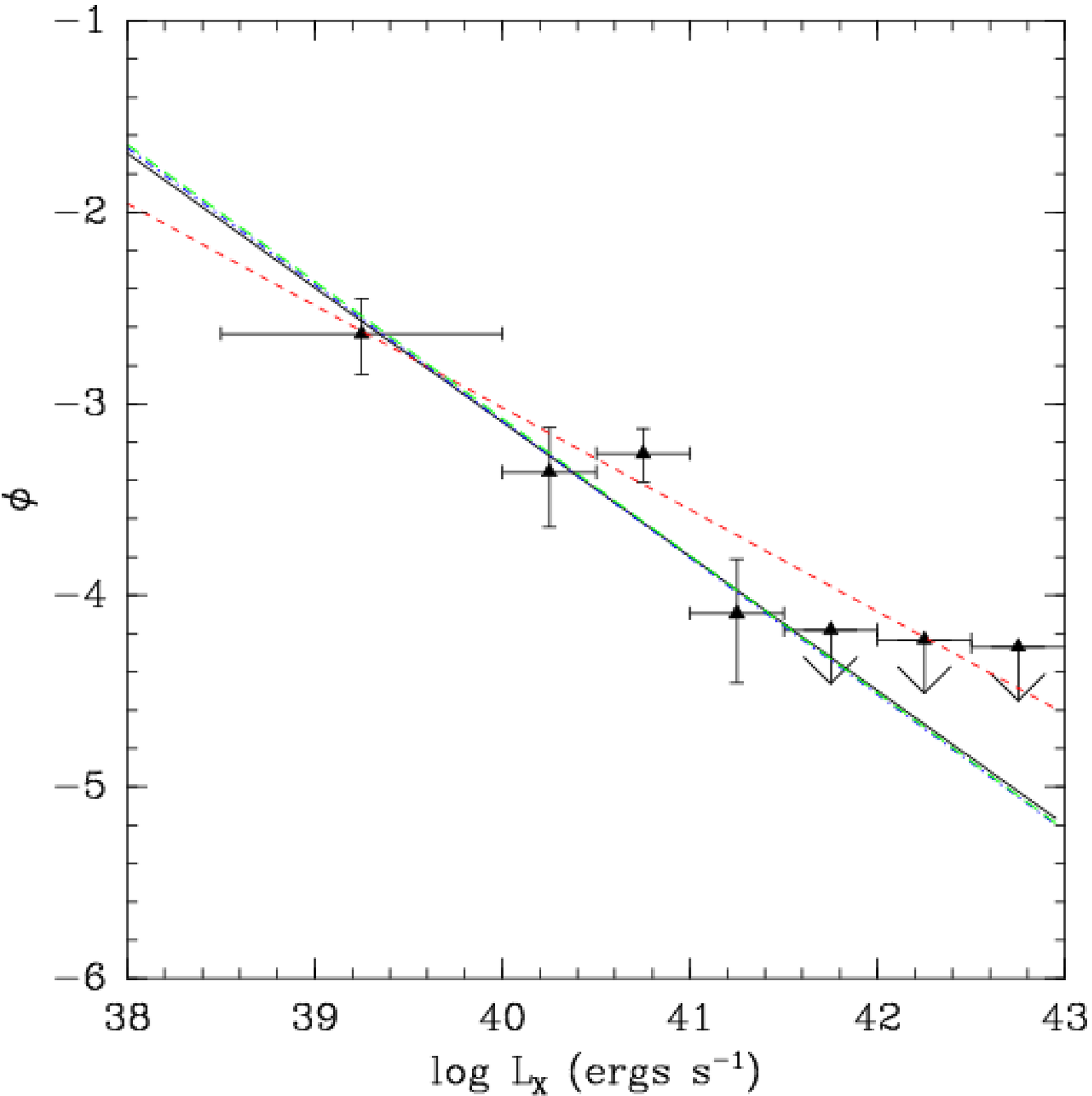}{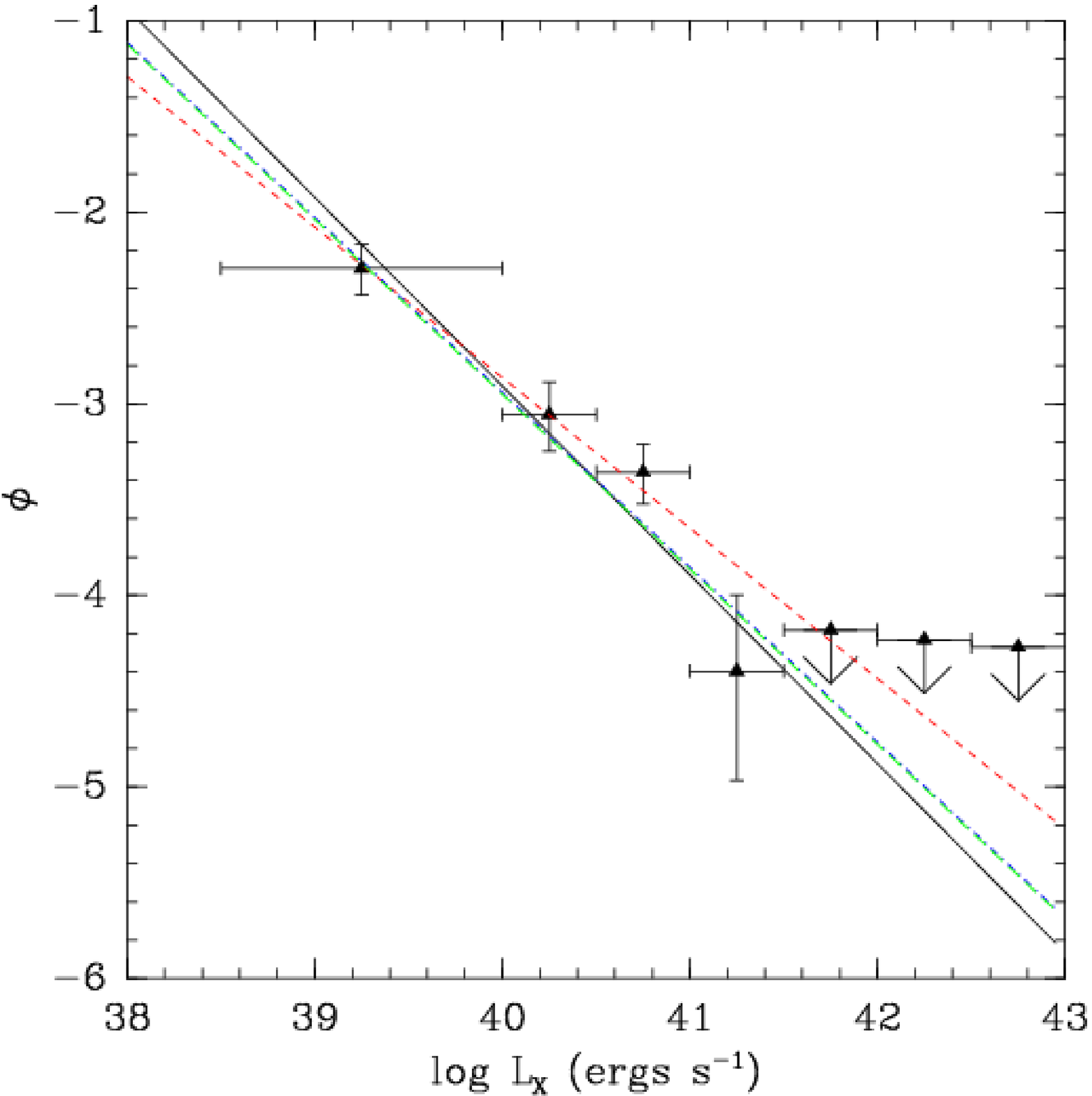}
\caption{$z < 0.5$ early-type (left) and late-type (right) galaxy XLFs
  from the GOODS fields.  Lines are as in Figure
  \ref{f_xlfs_gal_linear}. 
\label{f_xlfs_lowz_linear}} 
\end{figure*}

\begin{figure*}
\plottwo{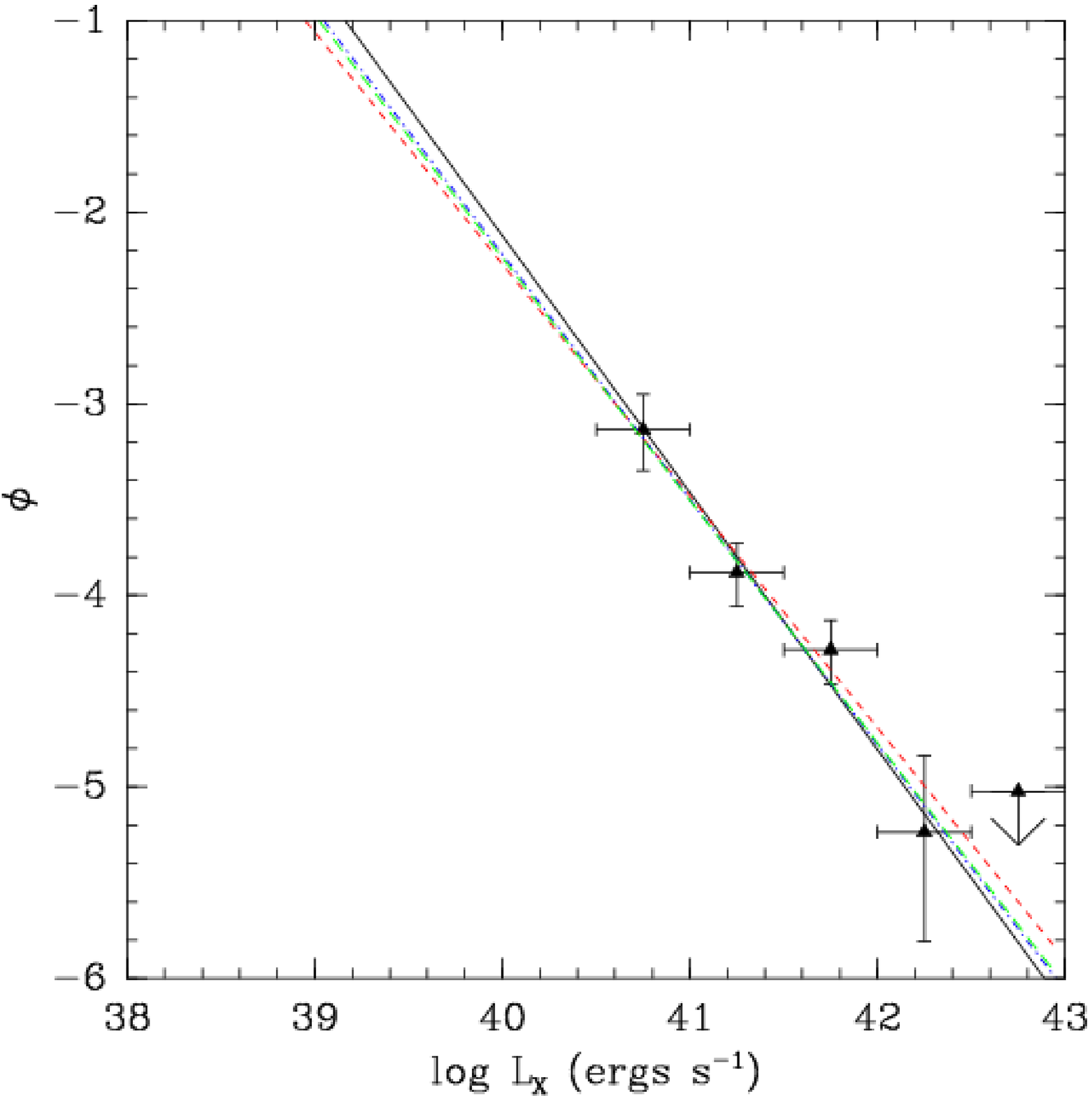}{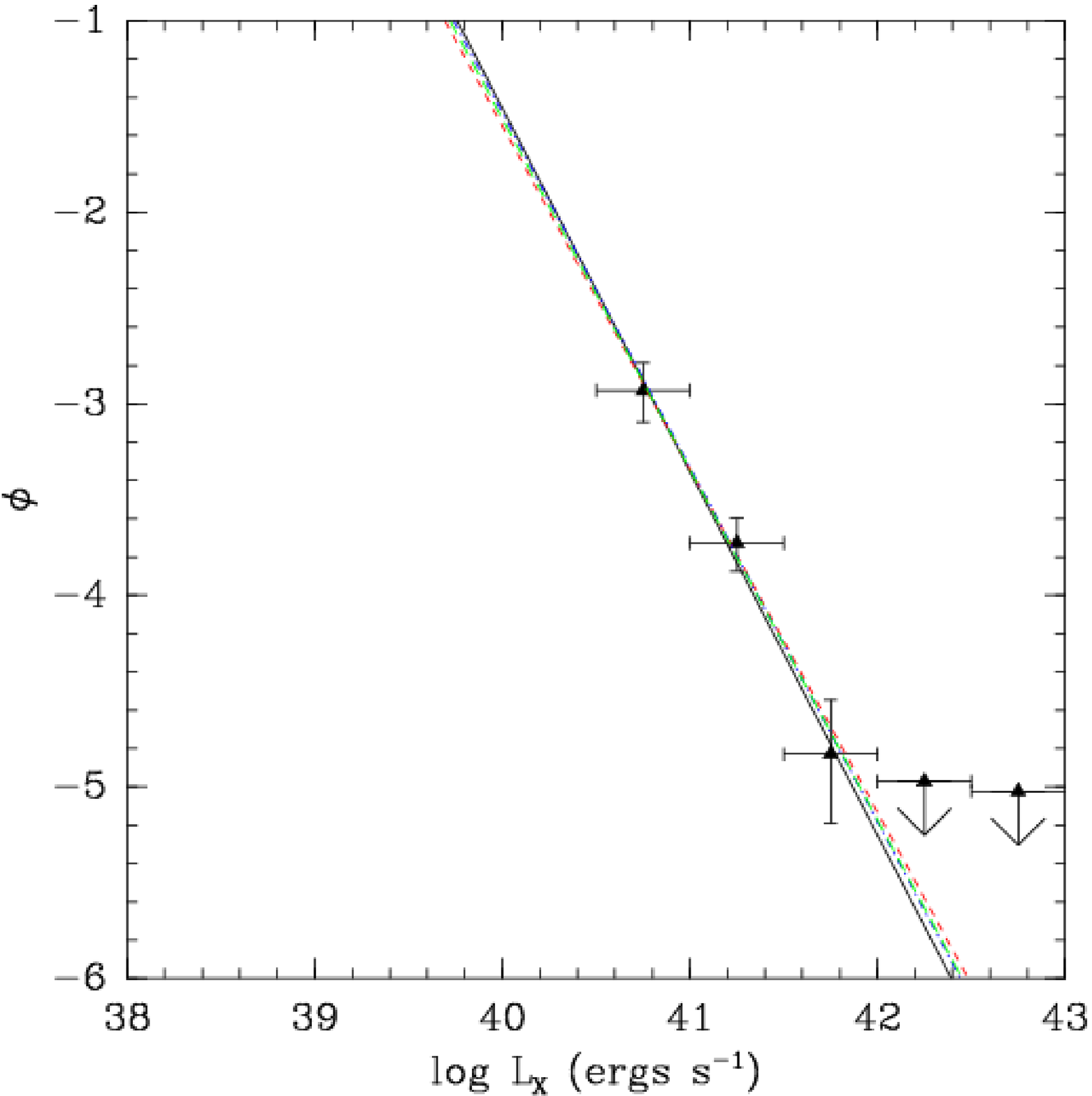}
\caption{$0.5 < z < 1.2$ early-type (left) and late-type (right) galaxy XLFs
  from the GOODS fields.  Lines are as in Figure
  \ref{f_xlfs_gal_linear}. 
\label{f_xlfs_hiz_linear}} 
\end{figure*}

In Figures \ref{f_xlfs_gal_linear}-\ref{f_xlfs_hiz_linear} we show the
XLFs derived 
from the GOODS fields based on the early-type, late-type and total
early + late-type galaxy samples along with the linear model fits.  We
also show the fits to the ``optimistic'' galaxy sample and the N04
XLFs for comparison.  The best-fit parameters and errors 
are given in Table \ref{t:slopes} along with the associated $\chi^2$
values and the probability with which the model can be rejected (again
with the caveat that this probability does not include upper limit data).

\begin{figure*}
\plottwo{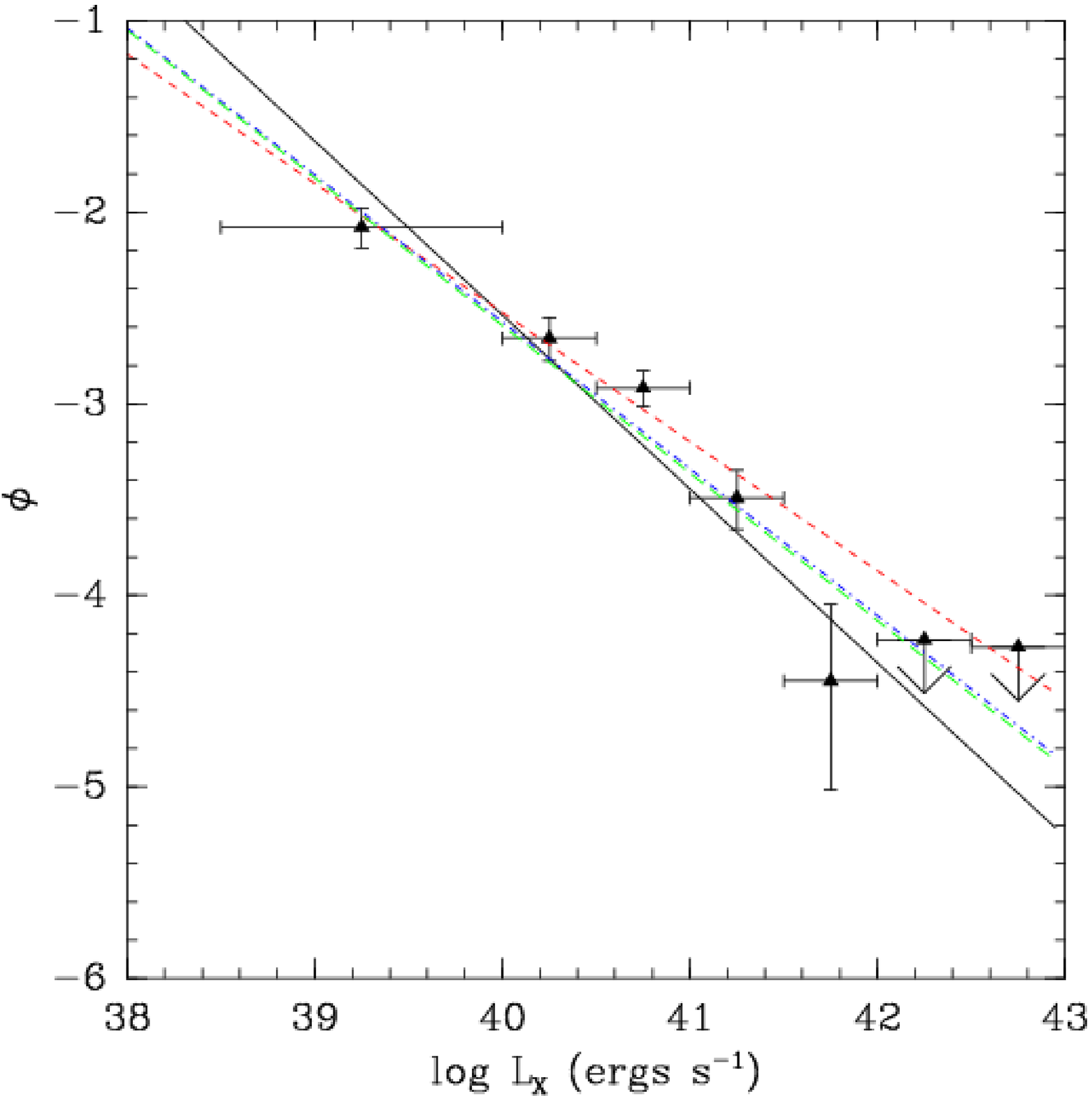}{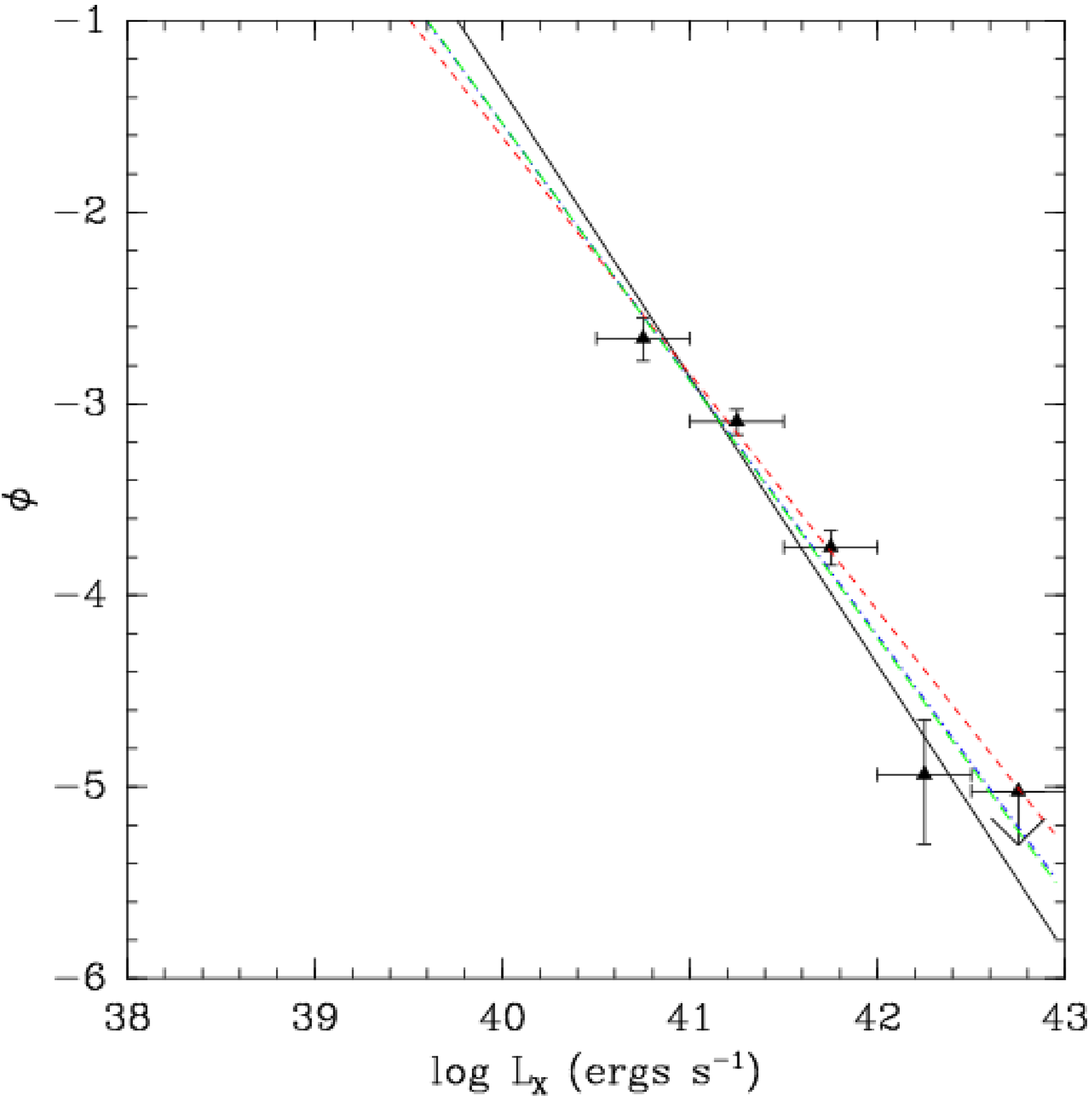}
\caption{``Optimistic'' sample (early + late-type galaxy) XLFs from the GOODS
  fields, for the 
  redshift intervals z$<$0.5 (left) and 0.5 $<$ z $<$ 1.2 (right).
  Lines are as in Figure \ref{f_xlfs_gal_linear}}
\end{figure*}


\begin{deluxetable*}{llllll}
\tablecaption{XLF Linear Fits \label{t:slopes}}
\tablehead{Sample & Method & $a$ & $b$ & $\chi^2/dof$ & $p_{\chi^2}$}
\startdata
Early-type galaxies, low-z & 1 & $-0.70 \pm 0.11$ & $25.0 \pm 4.5$ & 3.58/2& 83.2\%\\
 & 2 & $-0.53^{+0.13}_{-0.13}$ & $18.3^{+5.4}_{-5.4}$ & 6.34/2& 95.6\%\\
 & 3* &$-0.71^{+0.12}_{-0.09}$ & $25.4^{+5.2}_{-3.3}$ &3.66/2& 83.9\%\\
 & 3$^{\dagger}$ &$-0.72^{+0.11}_{-0.10}$ & $25.6^{+5.2}_{-3.3}$ &3.70/2& 84.2\%\\
Late-type galaxies, low-z & 1 & $-0.99 \pm 0.14$ & $36.5 \pm 5.5$ & 3.78/2& 84.8\%\\
 & 2 & $-0.79^{+0.11}_{-0.12}$ & $28.6^{+4.7}_{-4.6}$ & 5.55/2& 93.6\%\\
 & 3* &$-0.91^{+0.09}_{-0.11}$ & $33.6^{+4.1}_{-3.9}$ &2.95/2& 77.0\%\\
 & 3$^{\dagger}$ &$-0.91^{+0.12}_{-0.08}$ & $33.6^{+4.1}_{-4.0}$ &2.97/2& 77.2\%\\
All galaxies, low-z & 1 & $-0.90 \pm 0.11$ & $33.5 \pm 4.6$ & 8.95/2& 98.6\%\\
 & 2 & $-0.67^{+0.09}_{-0.08}$ & $24.1^{+3.4}_{-3.5}$ & 12.47/2& 99.5\%\\
 & 3* &$-0.81^{+0.06}_{-0.08}$ & $29.8^{+3.3}_{-2.3}$ &6.64/2& 96.2\%\\
 & 3$^{\dagger}$ &$-0.82^{+0.06}_{-0.08}$ & $30.1^{+3.1}_{-2.5}$ &6.64/2& 96.2\%\\
All galaxies (optimistic), low-z & 1 & $-0.91 \pm 0.11$ & $33.8 \pm 4.6$ & 15.98/3& 99.9\%\\
 & 2 & $-0.67^{+0.07}_{-0.07}$ & $24.4^{+2.9}_{-2.9}$ & 19.03/3& $>$99.9\%\\
 & 3* &$-0.77^{+0.06}_{-0.06}$ & $28.1^{+2.4}_{-2.4}$ &9.85/3& 98.0\%\\
 & 3$^{\dagger}$ &$-0.77^{+0.06}_{-0.06}$ & $28.2^{+2.2}_{-2.5}$ &9.70/3& 97.9\%\\
Norman et al. (2004), low-z & 1 & $-0.92 \pm 0.07$ & $34.2 \pm 2.8$ & 20.24/4& $>$99.9\%\\
 & 2 & $-0.76^{+0.05}_{-0.05}$ & $27.5^{+2.2}_{-2.1}$ & 13.85/4& 99.2\%\\
 & 3* &$-0.77^{+0.05}_{-0.04}$ & $28.2^{+2.0}_{-1.8}$ &10.56/4& 96.8\%\\
 & 3$^{\dagger}$ &$-0.78^{+0.04}_{-0.06}$ & $28.4^{+1.7}_{-2.1}$ &10.43/4& 96.6\%\\
Early-type galaxies, hi-z & 1 & $-1.34 \pm 0.10$ & $51.5 \pm 4.2$ & 1.17/2& 44.4\%\\
 & 2 & $-1.21^{+0.21}_{-0.21}$ & $46.2^{+8.7}_{-8.7}$ & 1.25/2& 46.3\%\\
 & 3* &$-1.28^{+0.20}_{-0.21}$ & $49.0^{+9.1}_{-8.0}$ &1.07/2& 41.4\%\\
 & 3$^{\dagger}$ &$-1.27^{+0.18}_{-0.23}$ & $48.4^{+9.5}_{-7.5}$ &1.07/2& 41.3\%\\
Late-type galaxies, hi-z & 1 & $-1.90 \pm 0.10$ & $74.5 \pm 4.2$ & 0.58/1& 53.9\%\\
 & 2 & $-1.79^{+0.29}_{-0.29}$ & $70.1^{+11.7}_{-11.7}$ & 0.52/1& 51.5\%\\
 & 3* &$-1.85^{+0.29}_{-0.23}$ & $72.6^{+12.4}_{-8.8}$ &0.54/1& 52.3\%\\
 & 3$^{\dagger}$ &$-1.83^{+0.19}_{-0.32}$ & $71.7^{+13.0}_{-8.1}$ &0.49/1& 50.5\%\\
All galaxies, hi-z & 1 & $-1.65 \pm 0.08$ & $64.6 \pm 3.3$ & 0.99/2& 38.9\%\\
 & 2 & $-1.51^{+0.16}_{-0.16}$ & $58.9^{+6.7}_{-6.7}$ & 0.62/2& 26.8\%\\
 & 3* &$-1.58^{+0.17}_{-0.14}$ & $61.7^{+5.7}_{-7.3}$ &0.64/2& 27.2\%\\
 & 3$^{\dagger}$ &$-1.52^{+0.13}_{-0.18}$ & $59.3^{+8.1}_{-4.9}$ &0.60/2& 25.9\%\\
All galaxies (optimistic), hi-z & 1 & $-1.50 \pm 0.17$ & $58.7 \pm 6.9$ & 11.75/2& 99.4\%\\
 & 2 & $-1.24^{+0.12}_{-0.12}$ & $47.8^{+5.1}_{-5.1}$ & 13.83/2& 99.6\%\\
 & 3* &$-1.33^{+0.11}_{-0.09}$ & $51.8^{+4.3}_{-4.1}$ &9.16/2& 98.7\%\\
 & 3$^{\dagger}$ &$-1.34^{+0.10}_{-0.10}$ & $52.2^{+3.9}_{-4.5}$ &9.12/2& 98.7\%\\
Norman et al. (2004), hi-z & 1 & $-1.42 \pm 0.10$ & $55.1 \pm 4.2$ & 15.25/3& 99.8\%\\
 & 2 & $-1.17^{+0.09}_{-0.09}$ & $45.0^{+3.9}_{-3.9}$ & 4.10/3& 74.9\%\\
 & 3* &$-1.18^{+0.10}_{-0.09}$ & $45.1^{+3.6}_{-4.1}$ &3.76/3& 71.1\%\\
 & 3$^{\dagger}$ &$-1.18^{+0.08}_{-0.10}$ & $45.2^{+3.6}_{-4.2}$ &3.76/3& 71.2\%\\

\enddata
\tablecomments{Fit parameters are given for $\log\phi = a\log L_X + b$. Method 1: linear regression, 2: fit to data excluding upper-limits, 3*: MCMC fit using Method 1 results to initialize the fit, 3$^{\dagger}$: MCMC fit using Method 2 results to initialize the fit (see text).  $p_{\chi^2}$ gives the $\chi^2$ probability at which the model fit can be rejected (note that $\chi^2$ is computed excluding upper limits).}
\end{deluxetable*}


As is evident from the fit results, the MCMC fitting was not very
sensitive as to whether the Method 1 or Method 2 fit values were used
as the prior means.  Interestingly, in every case the MCMC fits
resulted in curves that were intermediate to the Method 1 and Method 2
fits, and tended 
towards the Method 1 results when upper-limits were constraining
(e.g., the early-type and late-type galaxy XLFs) and towards Method 2
when the no upper limits were used or the upper limits were not constraining
(e.g., the Normal et al XLFs).  The marginalized posterior probability
distributions for $a$ and $b$ were close to Gaussian, resulting in
nearly symmetric errors. In general the linear models cannot be
rejected at high confidence (i.e., $> 3\sigma$).  Hereafter, the fit
results presented refer to MCMC fitting.

\subsection{Joint Linear Model Fits}
\begin{figure*}
\plottwo{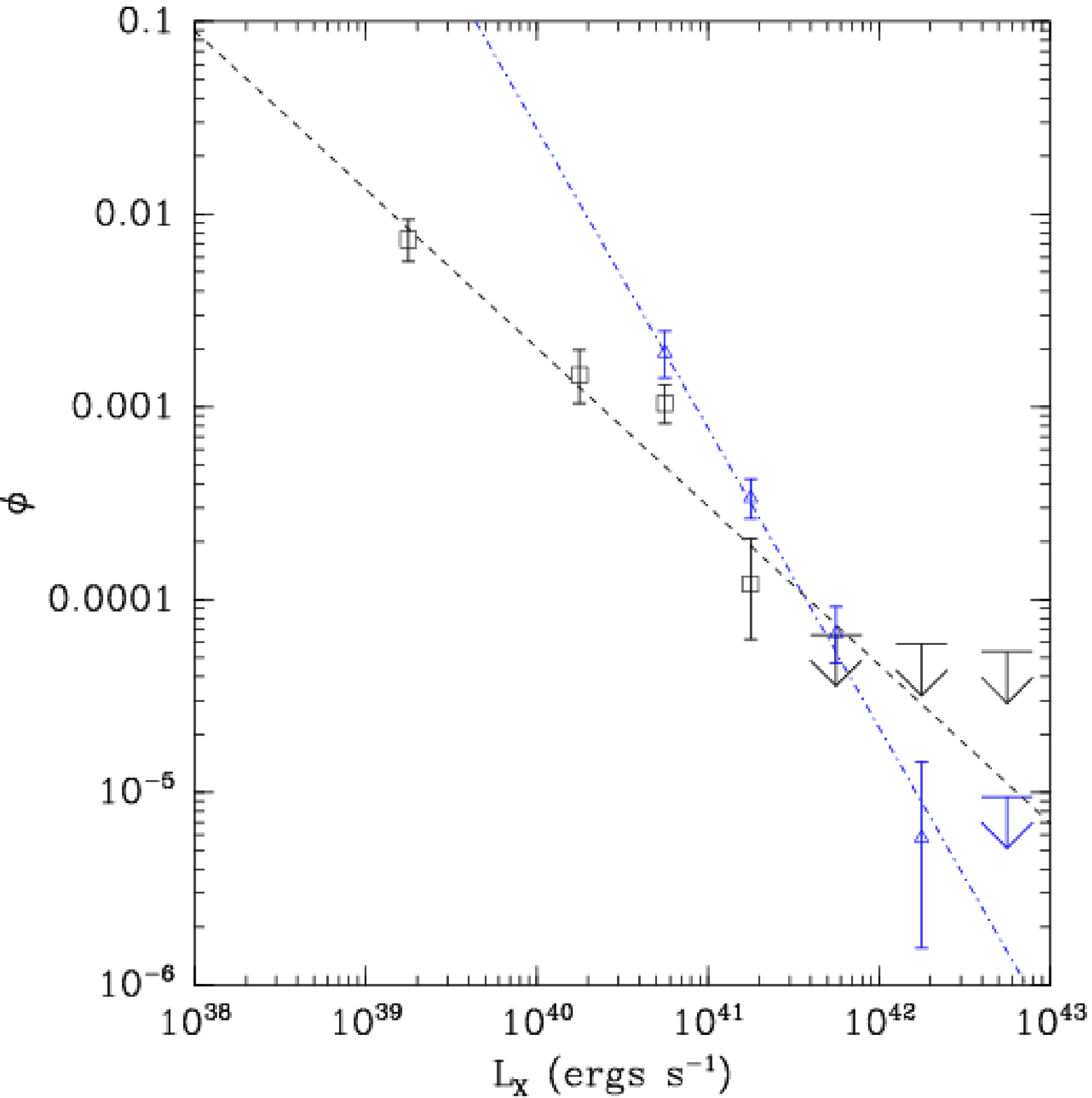}{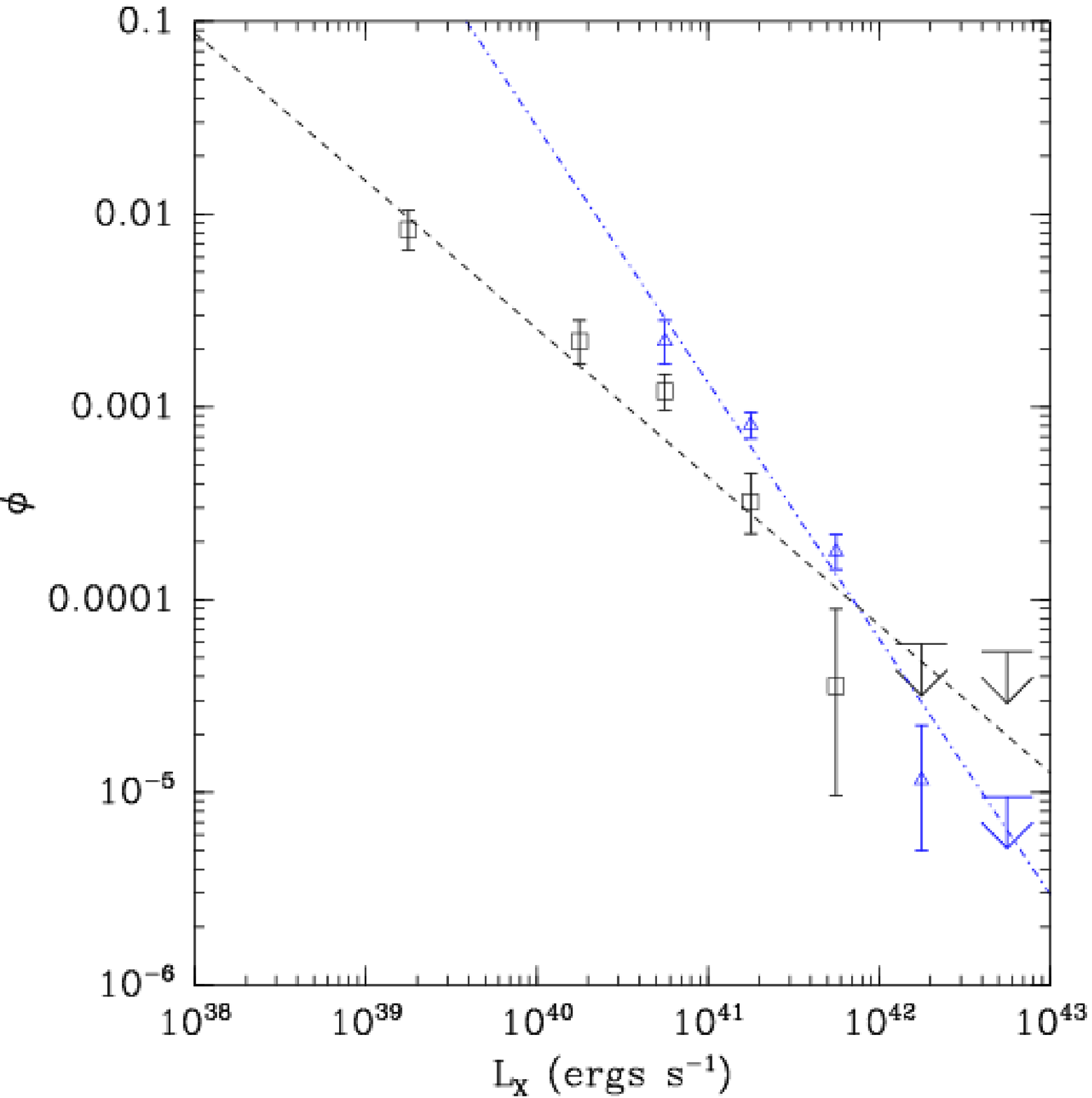}
\caption{Joint linear fits to the low-z and high-z XLFs for the full
  (early + late-type) galaxy sample (left) and the optimistic galaxy
  sample (right).  The low-z XLF points are marked with (black)
  squares and the high-z XLF points are marked with (blue) triangles.
  The dashed (black) lines and dot-dashed (blue) lines respectively show the fit to
  the low-z and high-z XLFs. \label{f_xlfs_gal_2linear}
}\end{figure*}

\begin{figure*}
\plottwo{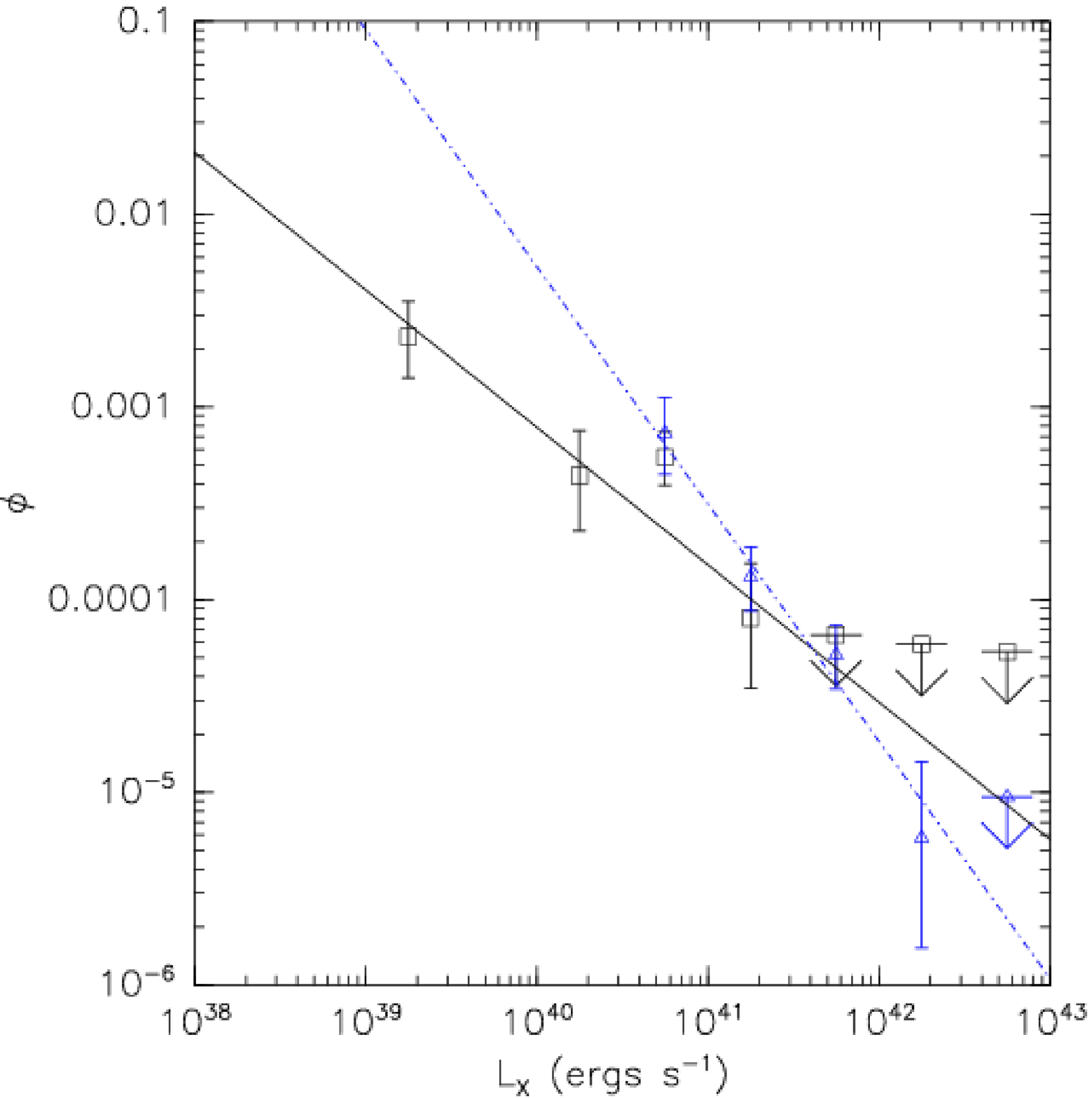}{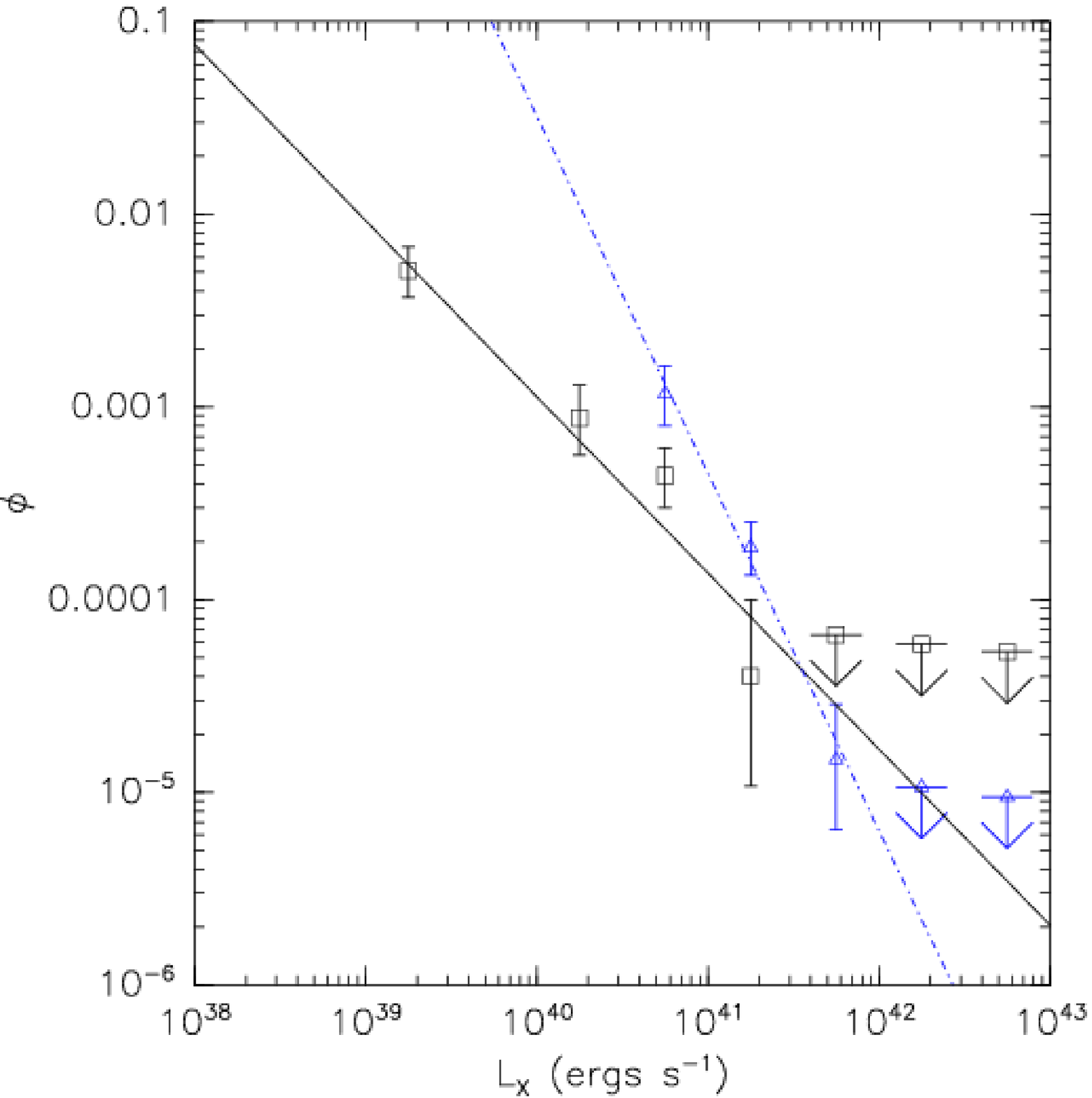}
\caption{As in Figure \ref{f_xlfs_gal_2linear}, with the left panel
  showing the early-type galaxy XLFs and the right panel showing the
  late-type galaxy XLFs. \label{f_xlfs_early_late_2linear}}
\end{figure*}

\begin{figure*}
\plottwo{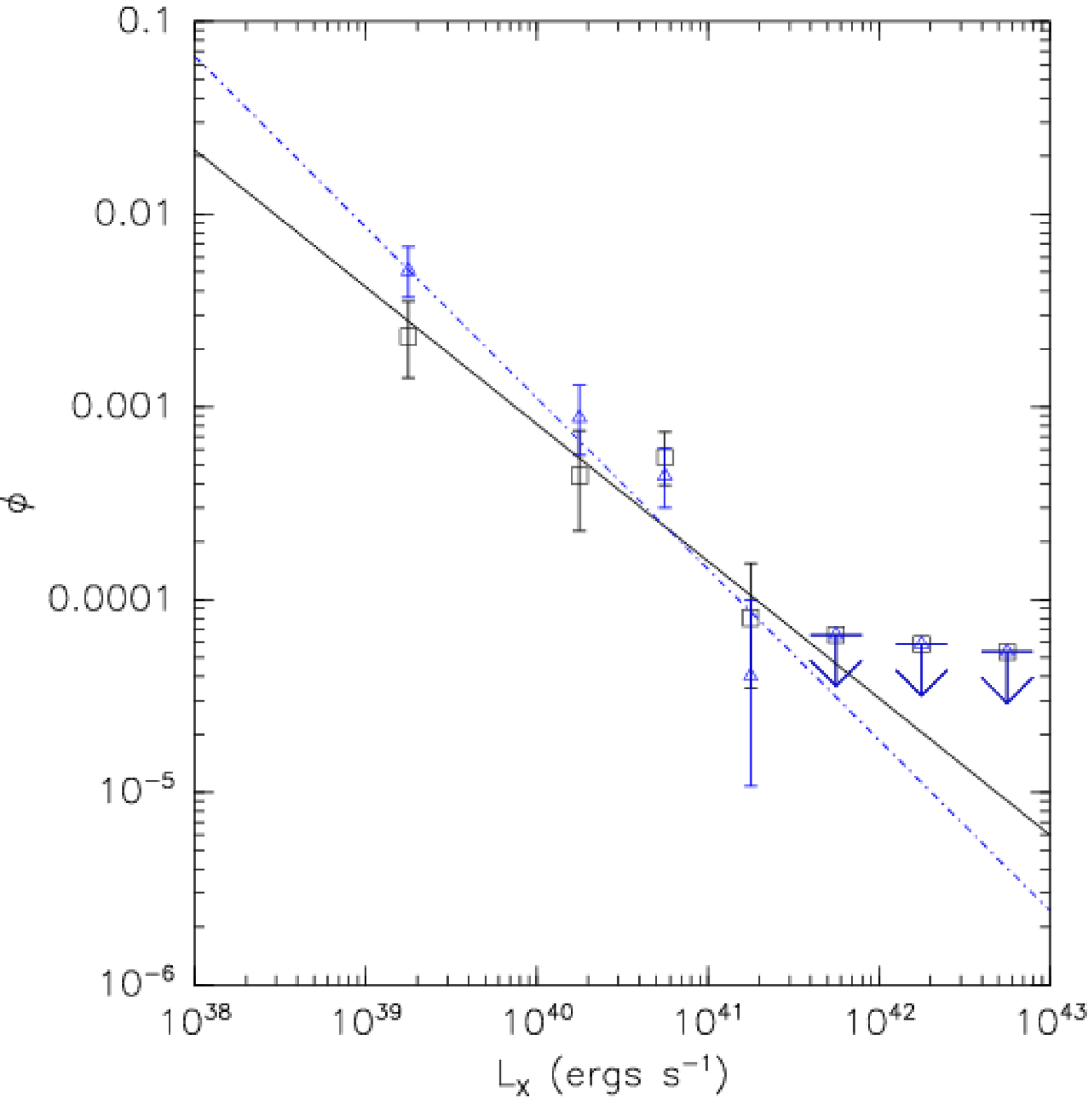}{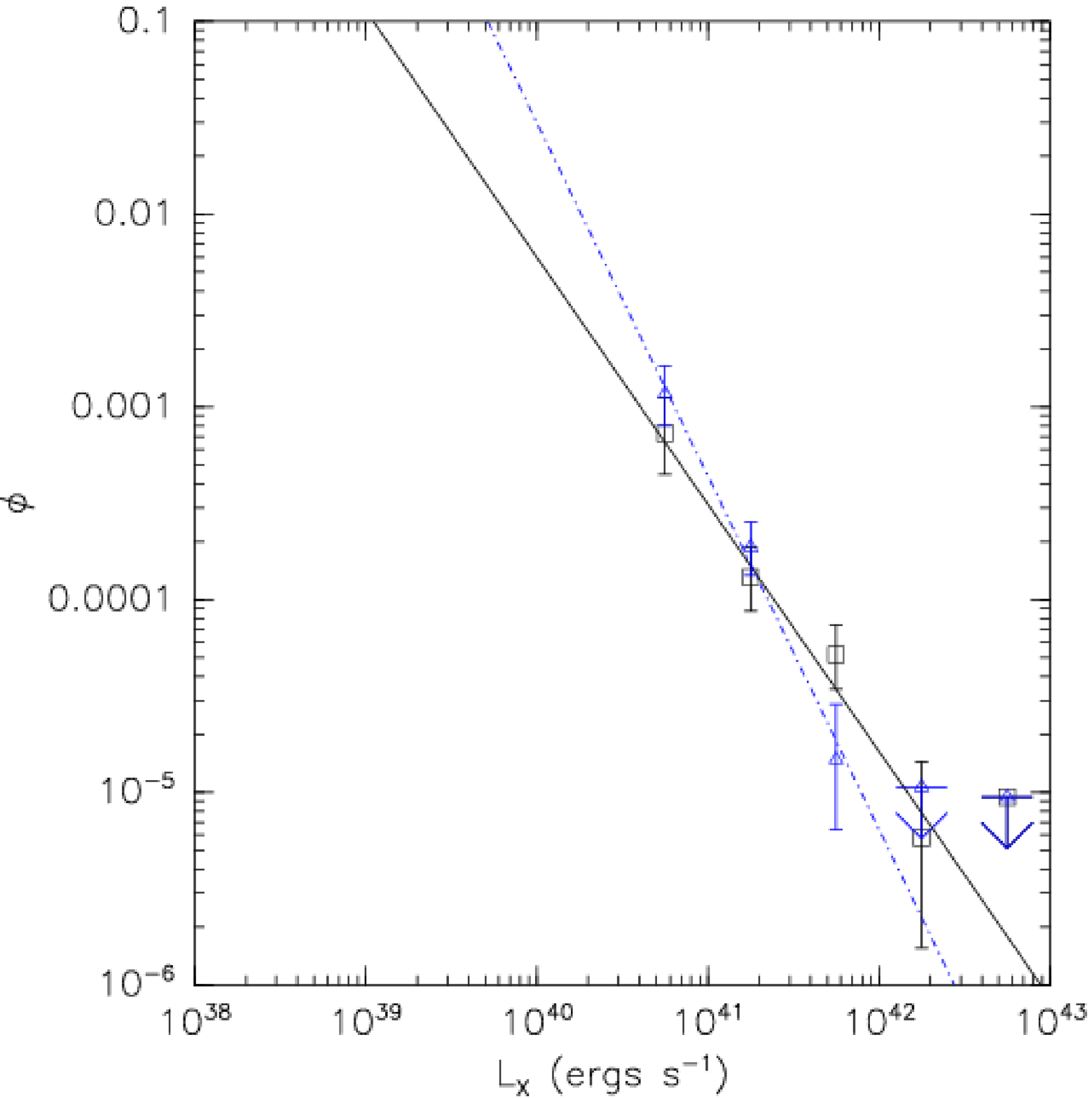}
\caption{Joint linear fits to the low-z early and late-type XLFs (left)
  high-z early and late-type XLFs (right).  The early-type XLF points
  are marked with (black)  squares and the late-type XLF points are
  marked with (blue) triangles. 
  The dashed (black) lines and dot-dashed (blue) lines show the fit to
  the early and late-type XLFs. \label{f_xlfs_gal_bytype}
}
\end{figure*}

In Figures \ref{f_xlfs_gal_2linear}-\ref{f_xlfs_early_late_2linear} we show
the results of fitting corresponding pairs of XLFs simultaneously
(i.e., the low and high redshift XLFs for a given sample, and the early
and late-type galaxy XLFs at low or high redshift), explicitly
fitting for the offsets in $a$ and $b$ between the XLFs. Figure 
\ref{f_xlfs_gal_bytype} shows the early and late-type XLFs similarly
being fitted jointly. 
Here also
the posterior probabilities were mostly Gaussian in shape, and the best-fit
parameter values and errors are given in Table \ref{t:mcmc_linear_corr_2epoch}.
These fits resulted in parameters for the given XLFs that were equivalent
to those obtained by fitting the XLFs seperately with the linear
model, however here we are deriving the probability distribution for
the offsets in slope and intercept.  As discussed above, this approach
avoids the need to propagate errors in determining the significance of
the change in the linear model parameters between XLFs.  However, in
these fits the final errors on $\Delta a$ and $\Delta b$ are very
similar to adding the errors obtained from the
individual fits in quadrature, which is perhaps not surprising given that the
posterior probabilities for these parameters were nearly Gaussian.
In every case, the best fit values of $\Delta a$ were $< 0$ and the
best fit values of $\Delta b$ were $> 0$.  In order to estimate the
significance of a change in the XLF, we determined the fraction of
simulations in which $\Delta a <  0$ and $\Delta b > 0$.  Since in
some cases this probability is very small, we performed 50 runs with a
very large chain length ($5 \times 10^6$) for these fits.  These
probabilities are also listed in Table \ref{t:mcmc_linear_corr_2epoch}.
\begin{figure}
\plotone{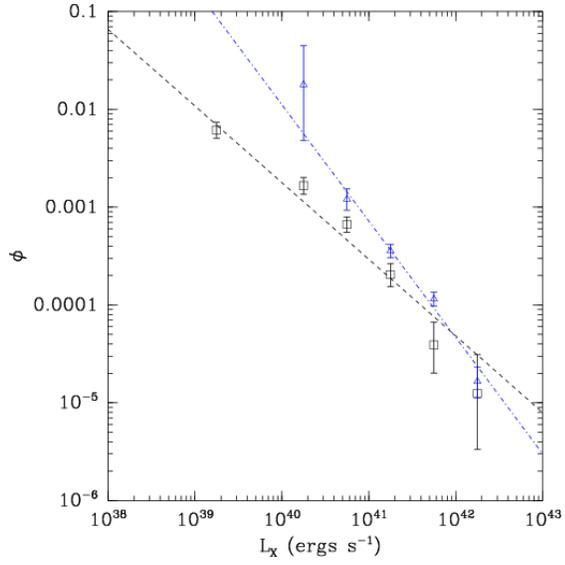}
\caption{As in Figure \ref{f_xlfs_gal_2linear} for the N04 XLFs.}
\end{figure}
\footnotesize
\begin{deluxetable*}{llllll}
\tablecaption{MCMC Joint Linear Fits \label{t:mcmc_linear_corr_2epoch}}
\tablehead{\colhead{Sample} & \colhead{$\Delta a$} &\colhead{$\Delta b$} &\colhead{$p_{\Delta a, \Delta b}$} &\colhead{$\chi^2/dof$} & \colhead{$p_{\chi^2}$}}
\startdata
All galaxies  
 & $-0.75^{+0.19}_{-0.15}$ & $30.4^{+6.9}_{-7.2}$ & $>$99.9\% &7.3/4 & 87.9\%\\
All galaxies (optimistic)  
 & $-0.56^{+0.10}_{-0.13}$ & $23.5^{+5.1}_{-4.5}$ & $>$99.9\% &18.9/5 & 99.8\%\\
Norman et al. (2004)  
 & $-0.38^{+0.09}_{-0.11}$ & $17.0^{+3.9}_{-4.7}$ & $>$99.9\% &14.0/7 & 94.9\%\\
Early-type galaxies  
 & $-0.58^{+0.25}_{-0.21}$ & $21.6^{+11.4}_{-7.7}$ & 99.4\% &4.7/4 & 68.4\%\\
Late-type galaxies  
 & $-1.03^{+0.33}_{-0.23}$ & $39.1^{+12.8}_{-9.9}$ & $>$99.9\% &3.5/3 & 67.4\%\\
Early/late-type galaxies, low-z  
 & $-0.19^{+0.15}_{-0.14}$ & $7.2^{+6.0}_{-5.6}$ & 89.8\% &6.6/4 & 84.4\%\\
Early/late-type galaxies, hi-z  
 & $-0.59^{+0.32}_{-0.34}$ & $22.8^{+15.4}_{-11.7}$ & 96.9\% &1.6/3 & 33.2\%\\
\enddata
\tablecomments{
Best-fitting change in slope ($\Delta a$) and intercept ($\Delta b$) from joint fits to the low and high-z samples.  The "Early/late-type" samples refer to comparing the early and late-type galaxy sample (i.e., see Figure \ref{f_xlfs_gal_bytype}).  $p_{\Delta a, \Delta b}$ refers to the probability that $\Delta a < 0$ and $\Delta b > 0$.
\\$p_{\chi^2}$ gives the $\chi^2$ probability at which the model fit can be rejected (note that $\chi^2$ is computed excluding upper limits).}
\end{deluxetable*}
\normalsize

\subsection{Log-Normal and Schechter Fits}
We fit the XLFs using the log-normal (Saunders et al. 1990) and 
Schechter  (Schechter et al. 1976) functions since these functional
forms fit the FIR (and hence star-forming galaxy) and optical
luminosity functions of galaxies well. 
Specifically the functional forms were:
\begin{equation}
\phi(L) =
\phi^*\left(\frac{L}{L^*}\right)^{1-\alpha}\exp\left[-\frac{1}{2\sigma^2}\log^2(1+\frac{L}{L^*})\right]
\end{equation}
\begin{equation}
\phi(L) = \ln(10)\phi^*\left(\frac{L}{L^*}\right)^{1+\alpha}\exp\left(-\frac{L}{L^*}\right)
\end{equation}
where in both cases the units of $\phi$ are galaxies Mpc$^{-3}$ log$L^{-1}$.
For both of these sets of fits
we placed very weak priors on $\log L^*$ and $\log \phi^*$.  We first
fitted the FIR LF published in Saunders et al. (1990) in order to
obtain prior information based on the local star-forming galaxy luminosity
function, and to validate our methodology.  We assumed that the errors
listed for each LF point were Gaussian and excluded upper limits.  Our results are shown in
Table \ref{t:FIR_LF_fits}, along with the original results of Saunders
et al. (1990) and the results of fitting the log-normal function to a
more recent FIR sample in Takeuchi et al. (2001).  Our results are
consistent with the fitting based on traditional methods within the
errors.  Since the Takeuchi et al. LF fits were based on more recent
data we set the log-normal fit prior means to their values, with $\log
L^*$ scaled to the X-ray band using $\log L_X = \log L_{60\ \rm \mu m}
- 3.65$ (Ranalli et al. 1990), and by $(1+z)^{3}$ where $z=0.25$ for the
$z<0.5$ XLFs and $z=0.75$ for the $0.5<z<1.2$ XLFs.  In the case of
the early-type or late-type galaxy XLFs we further reduced the prior
mean for $\phi^*$ by a factor of 2 since these samples were $\sim 50\%$
of the total galaxy sample.  We conservatively
assumed the MCMC errors in Table \ref{t:FIR_LF_fits} which were
similar to, but larger than, the Takeuchi et al. errors. We set the
prior standard deviations to 5 times these errors for $\alpha$ and
$\sigma$ and 50 times these errors for $\log L^*$ and $\log \phi^*$
(effectively weak).  In practice, however, different values of
$\alpha$ were preferred by the data, and so we refit the data after
setting the prior mean (and initial starting values) to the best-fitting
$\alpha$ values from the original fits.  The fits were not stable if
we allowed the slope prior widths to be broader, and in the case of
the early-type galaxy XLFs, the prior for $\alpha$ was required to be
50\% smaller in order to result in stable fits.

\begin{deluxetable}{llll}
\tablecaption{Log-Normal Fits to the 60 $\mu$m Luminosity Function \label{t:FIR_LF_fits}}
\tablehead{\colhead{Parameter} & \colhead{Saunders et al. (1990)} &
  \colhead{Takeuchi et al. (2001)} & \colhead{MCMC (This Work)}}
\startdata
$\alpha$ & $1.09 \pm 0.12$ & $1.23 \pm 0.04$ &  $1.04^{+0.07}_{-0.08}$\\
$\sigma$ & $0.724 \pm 0.031$ & $0.724 \pm 0.010$ & $0.751^{+0.014}_{-0.015}$ \\
$\phi \ h^3\rm\ Mpc^{-3}$ & $0.026 \pm 0.008$ & $0.026 \pm 0.003$ & $0.026^{+0.003}_{-0.003}$\\
$L^*\ h^{-2}\rm\ L_{\odot}$ & $10^{8.47 \pm 0.23}$ & $(4.34 \pm 0.86) \times 10^{8}$  & $10^{8.39^{+0.12}_{-0.15}}$\\
\enddata
\tablecomments{The data used in our MCMC fitting of the 60 $\mu$m LF
  was taken from Saunders et al. (1990), with several upper-limit data
  points excluded.}
\end{deluxetable}

\begin{figure*}
\plottwo{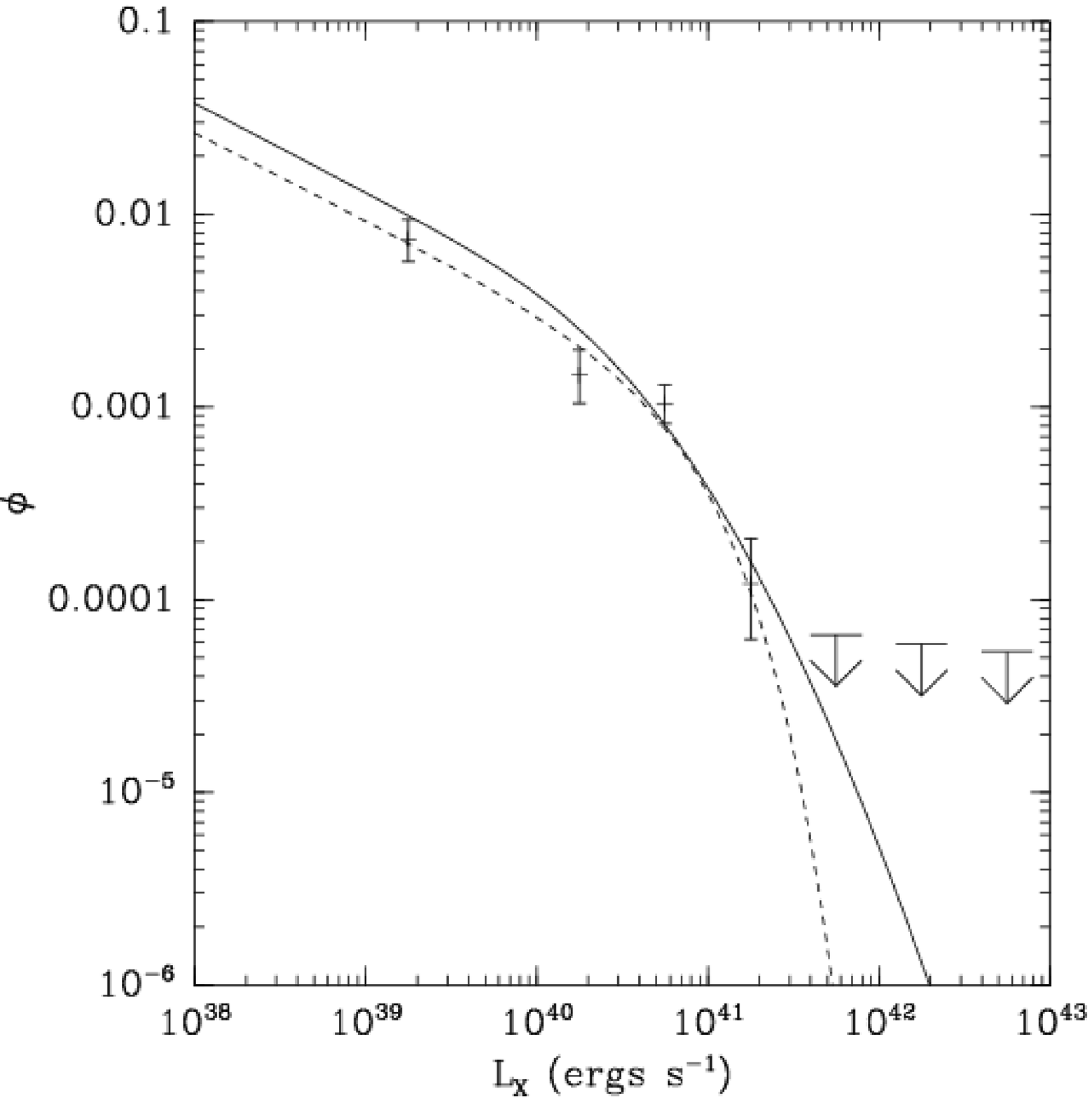}{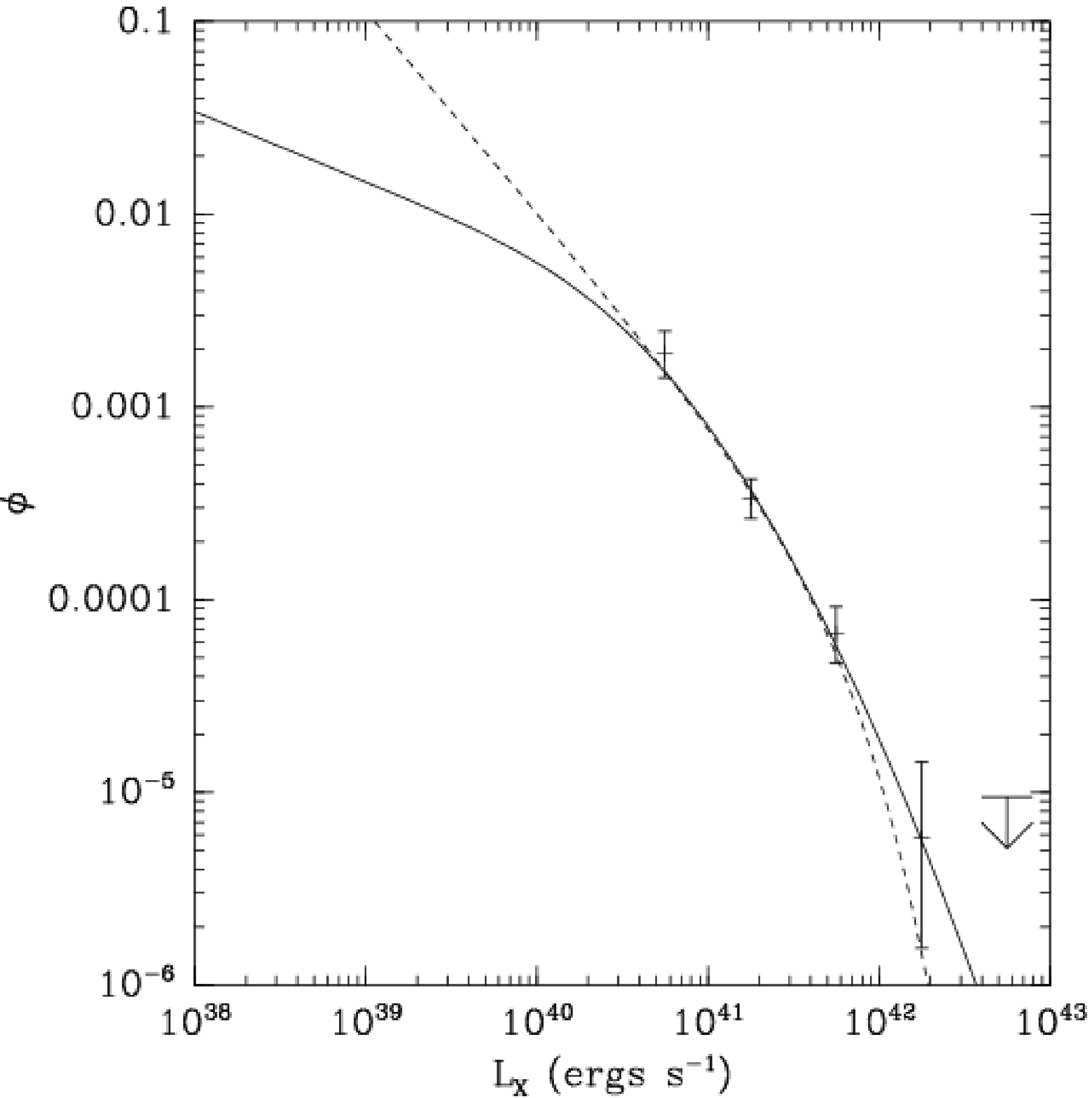}
\caption{Log-normal (solid curves) and Schechter function (dashed
  curves) fits to the full (early +  late-type) galaxy sample.  The
  left panel shows the $z<0.5$ sample and the right panel shows the
  $0.5<z<1.2$ sample.\label{f_all_gal_sch_logn}}
\end{figure*}

\begin{figure*}
\plottwo{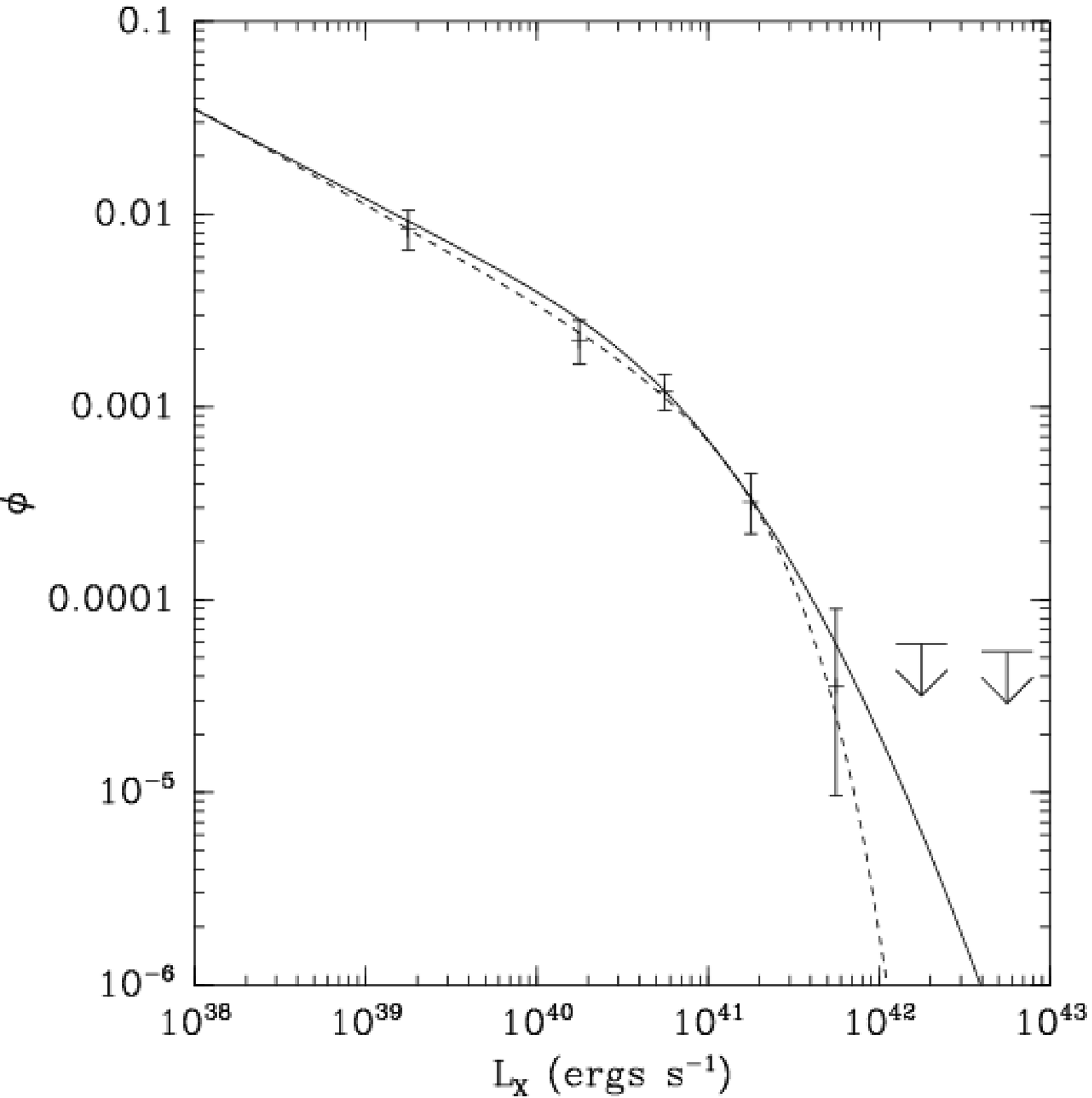}{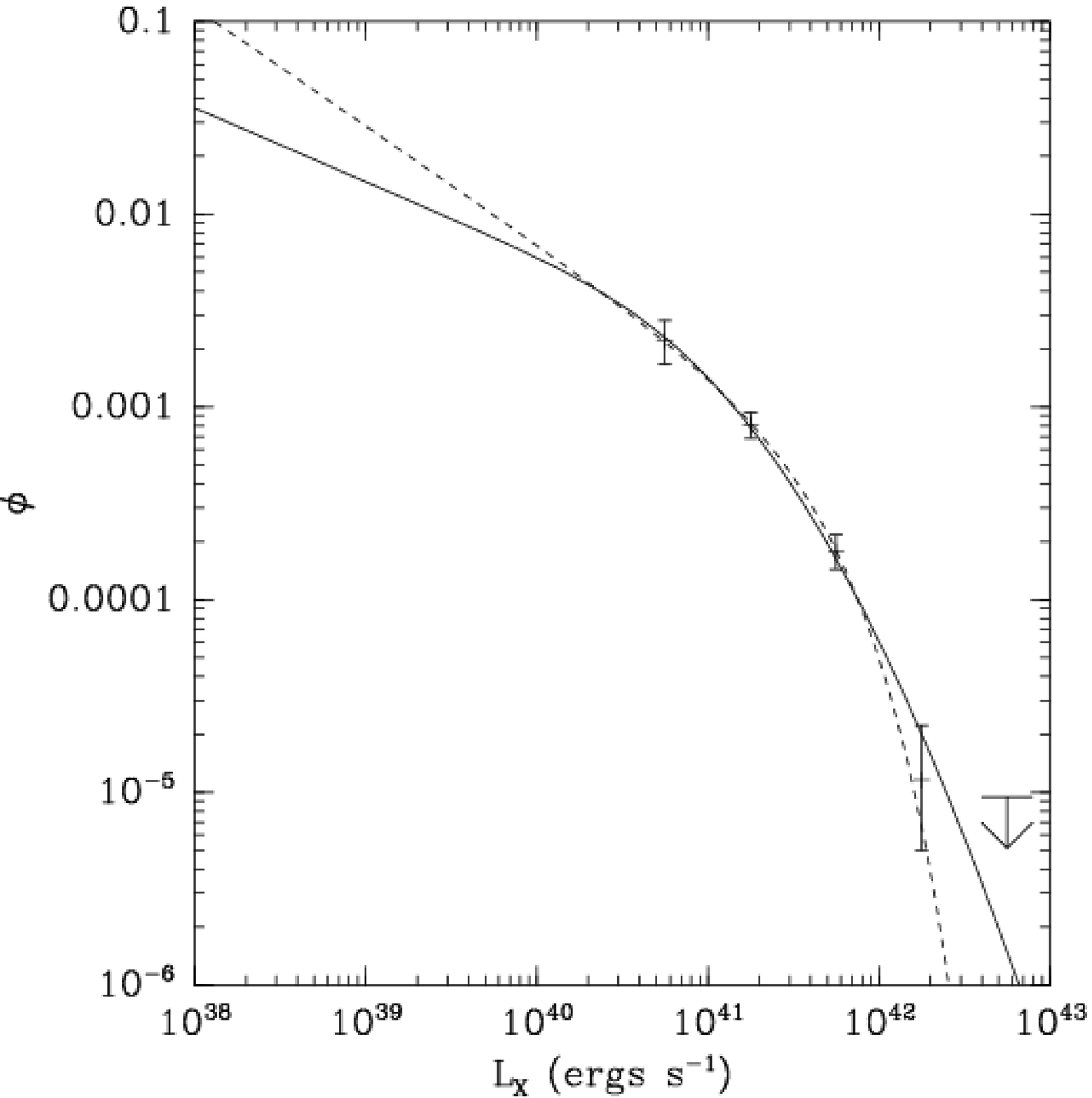}
\caption{Log-normal (solid curves) and Schechter function (dashed
  curves) fits to the optimistic galaxy sample.  The
  left panel shows the $z<0.5$ sample and the right panel shows the
  $0.5<z<1.2$ sample.\label{f_all_xgal_sch_logn}}
\end{figure*}

\begin{figure*}
\plottwo{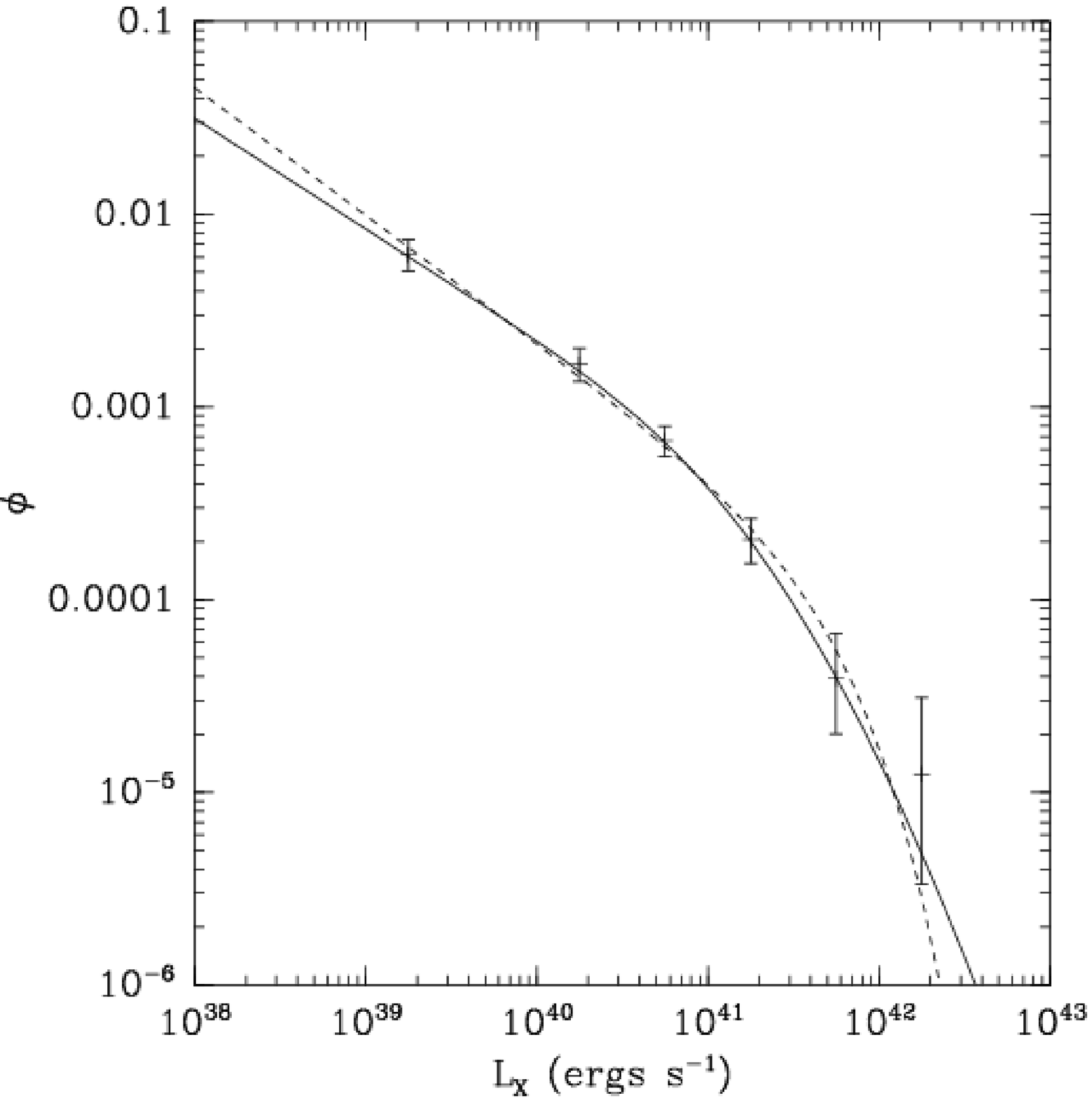}{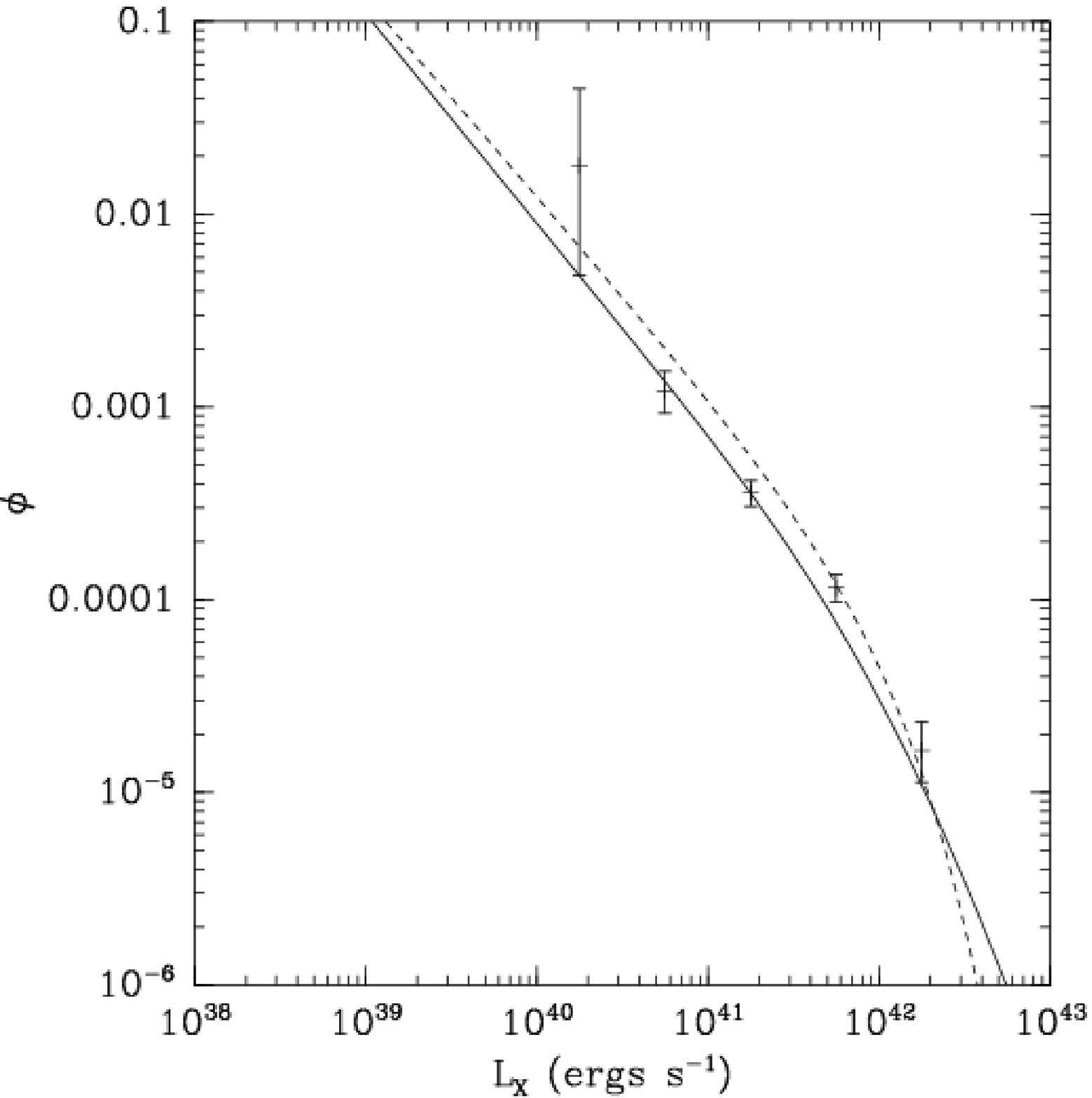}
\caption{Log-normal (solid curves) and Schechter function (dashed
  curves) fits to the N04 galaxy sample.  The
  left panel shows the $z<0.5$ sample and the right panel shows the
  $0.5<z<1.2$ sample.\label{f_norman_sch_logn}}
\end{figure*}

\begin{figure*}
\plottwo{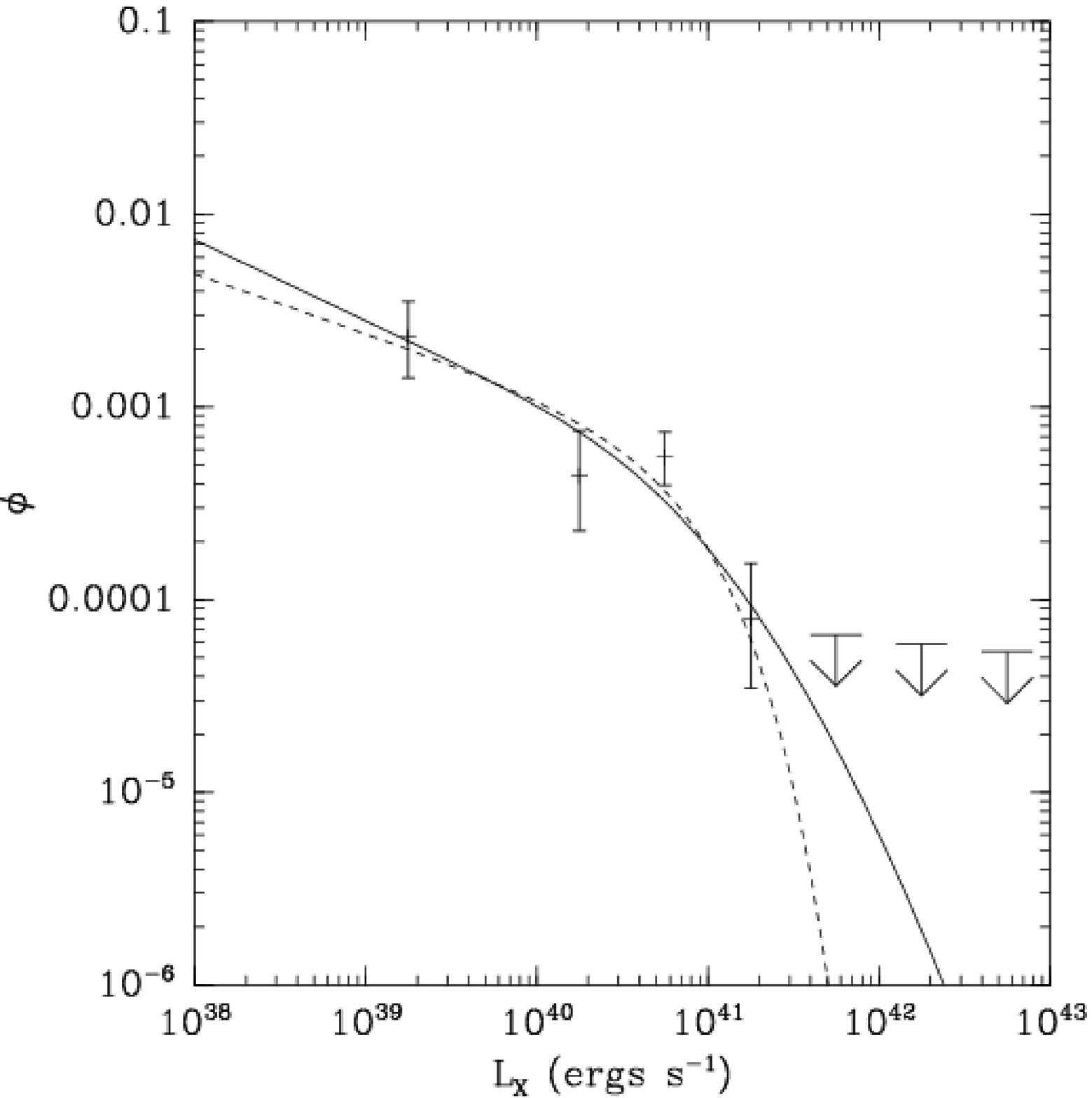}{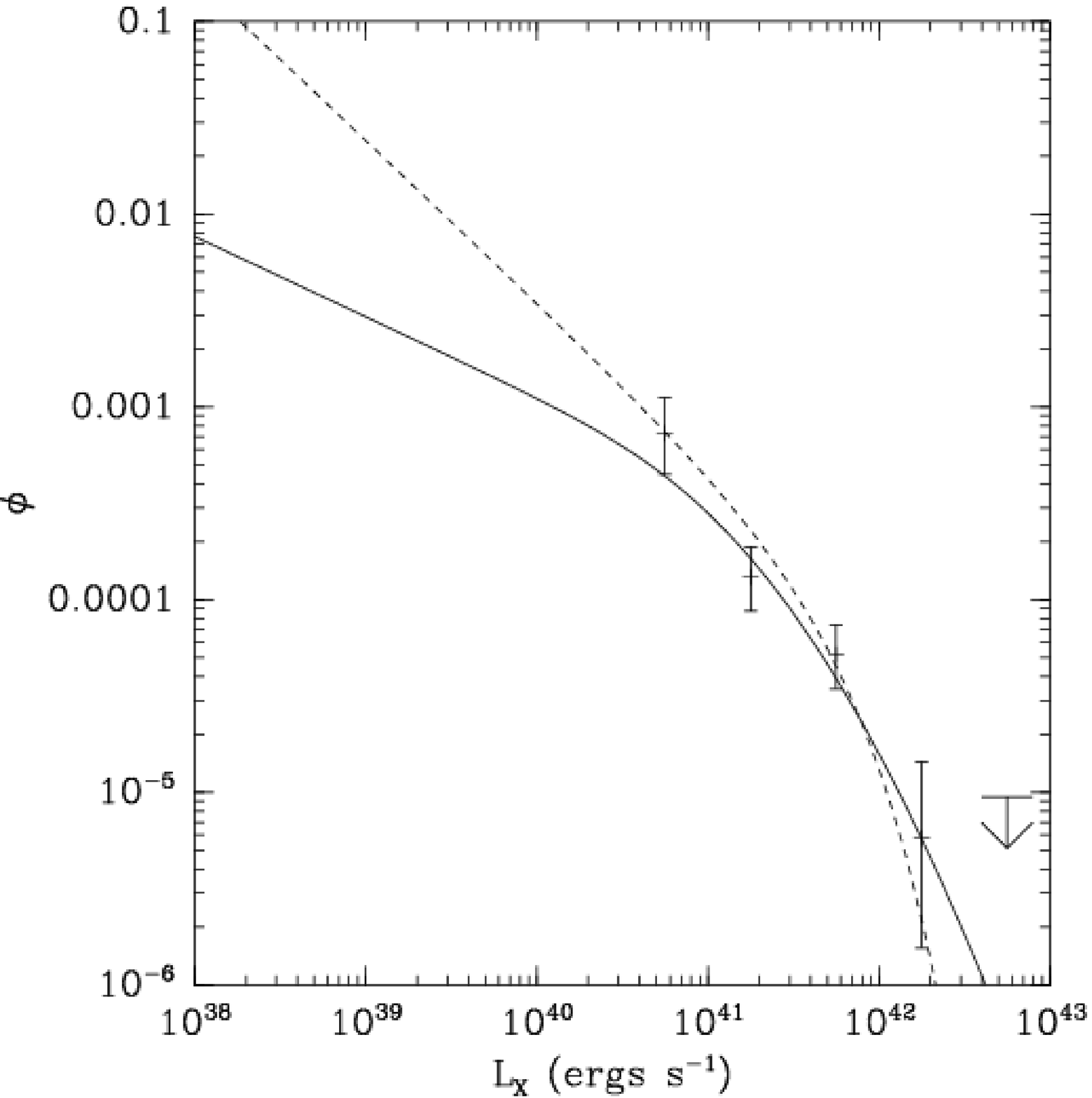}
\caption{Log-normal (solid curves) and Schechter function (dashed
  curves) fits to the early-type galaxy sample.  The
  left panel shows the $z<0.5$ sample and the right panel shows the
  $0.5<z<1.2$ sample.\label{f_early_sch_logn}}
\end{figure*}

\begin{figure*}
\plottwo{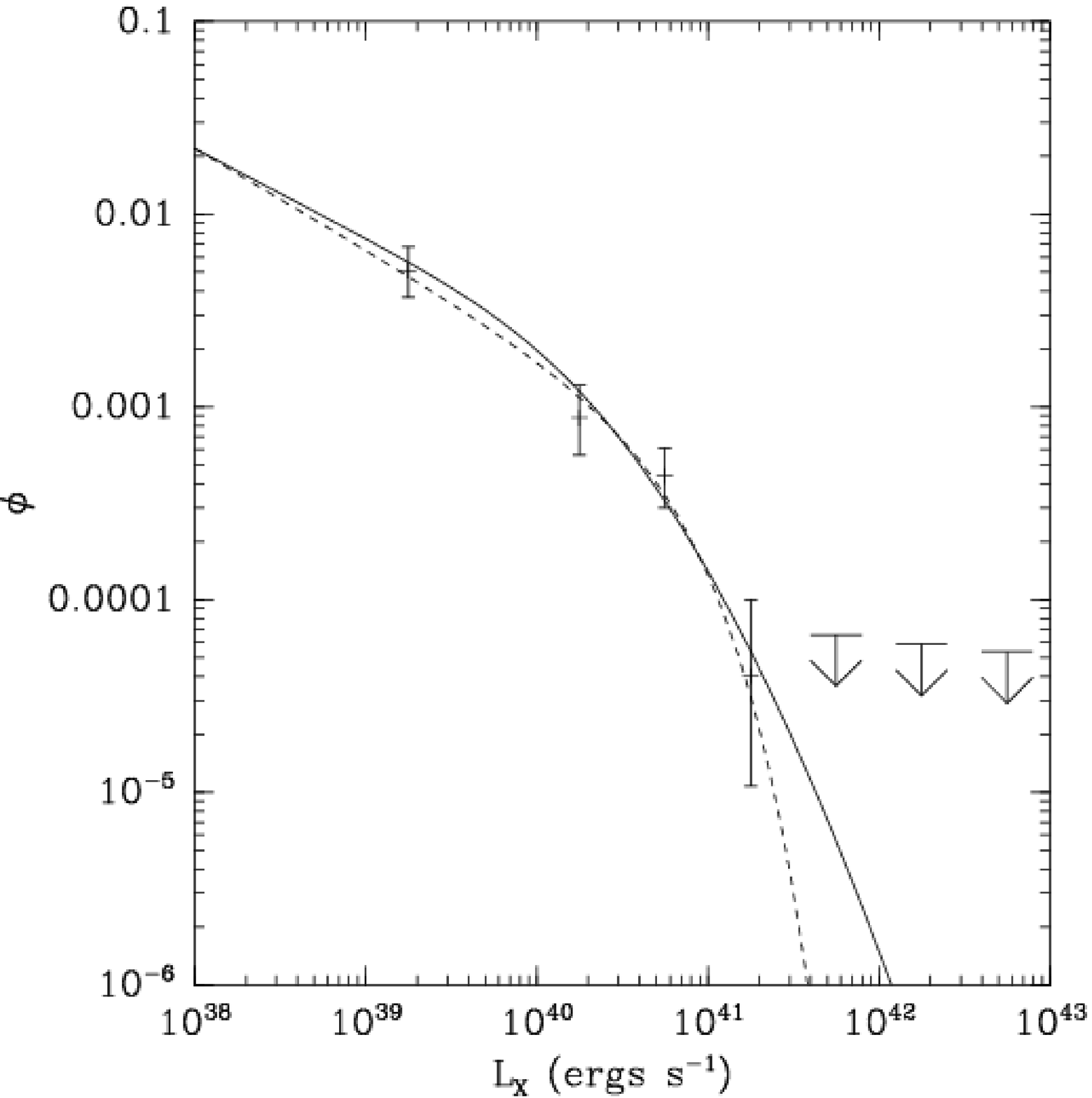}{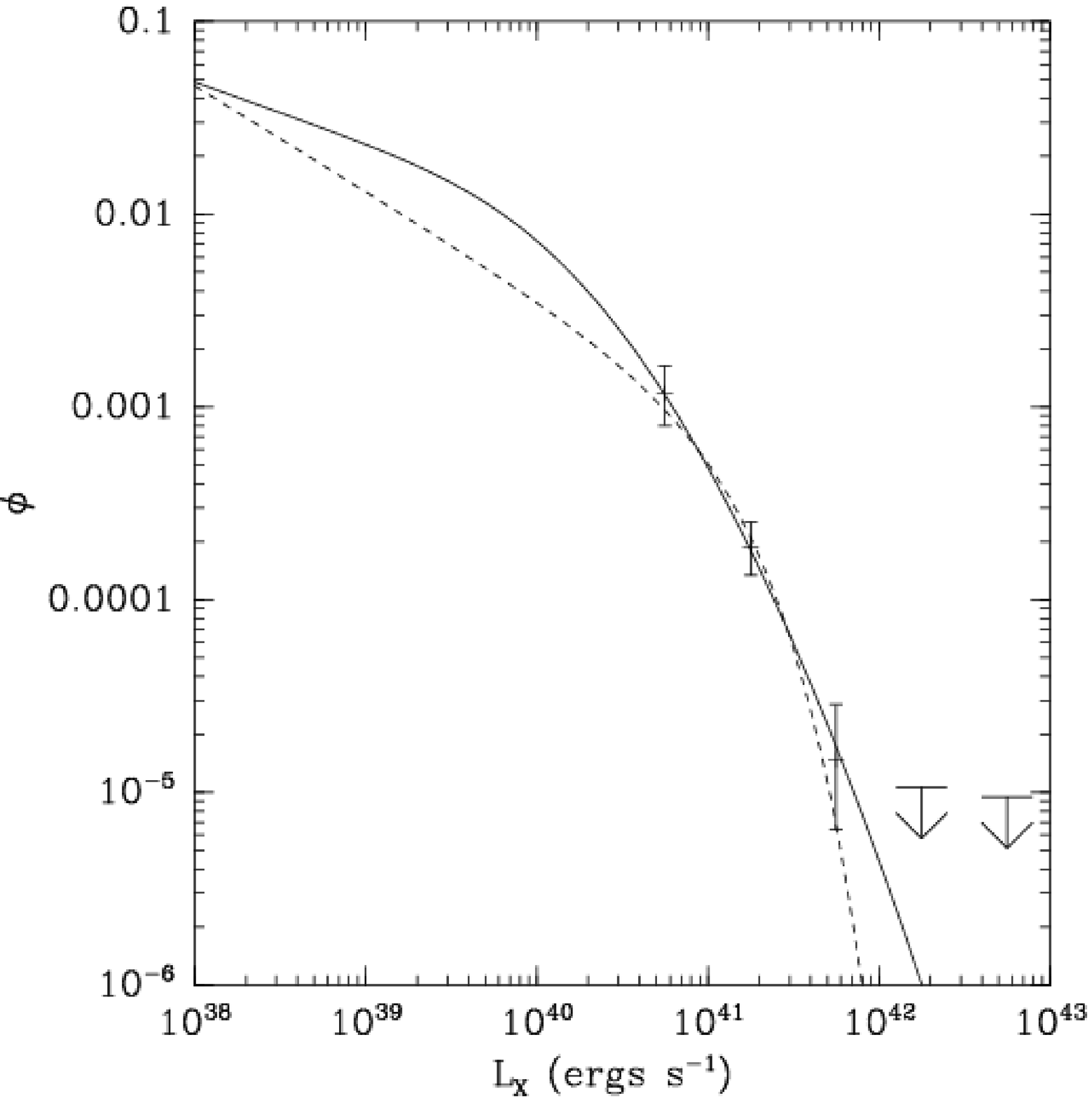}
\caption{Log-normal (solid curves) and Schechter function (dashed
  curves) fits to the late-type galaxy sample.  The
  left panel shows the $z<0.5$ sample and the right panel shows the
  $0.5<z<1.2$ sample.\label{f_late_sch_logn}}
\end{figure*}

\begin{figure*}
\plotone{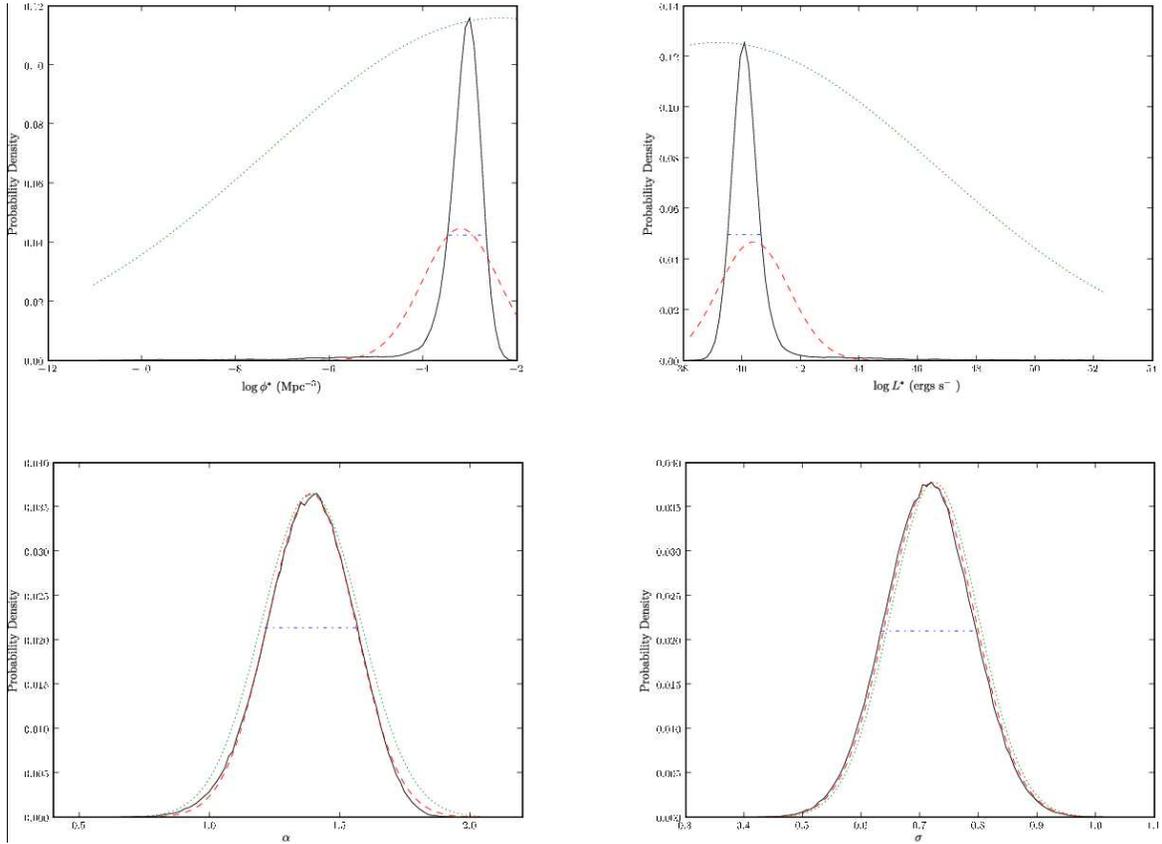}
\caption{Marginalized posterior probabilities for the fit parameters
  $\log \phi^*$, $\log L^*$, $\alpha$ and $\sigma$ for the log-normal
  fit to the $z<0.5$ early-type galaxy XLF.  The solid (black) lines
  show the posterior probability, the dotted (green) line show the
  prior, the dashed (red) lines shows a Gaussian distribution with the
  same mean and standard deviation as the 
  posterior, and the dot-dashed line shows the 68\% confidence
  interval. \label{f_early_lowz_lognormal_hists}}
\end{figure*}

\begin{figure*}[htbp]
\plottwo{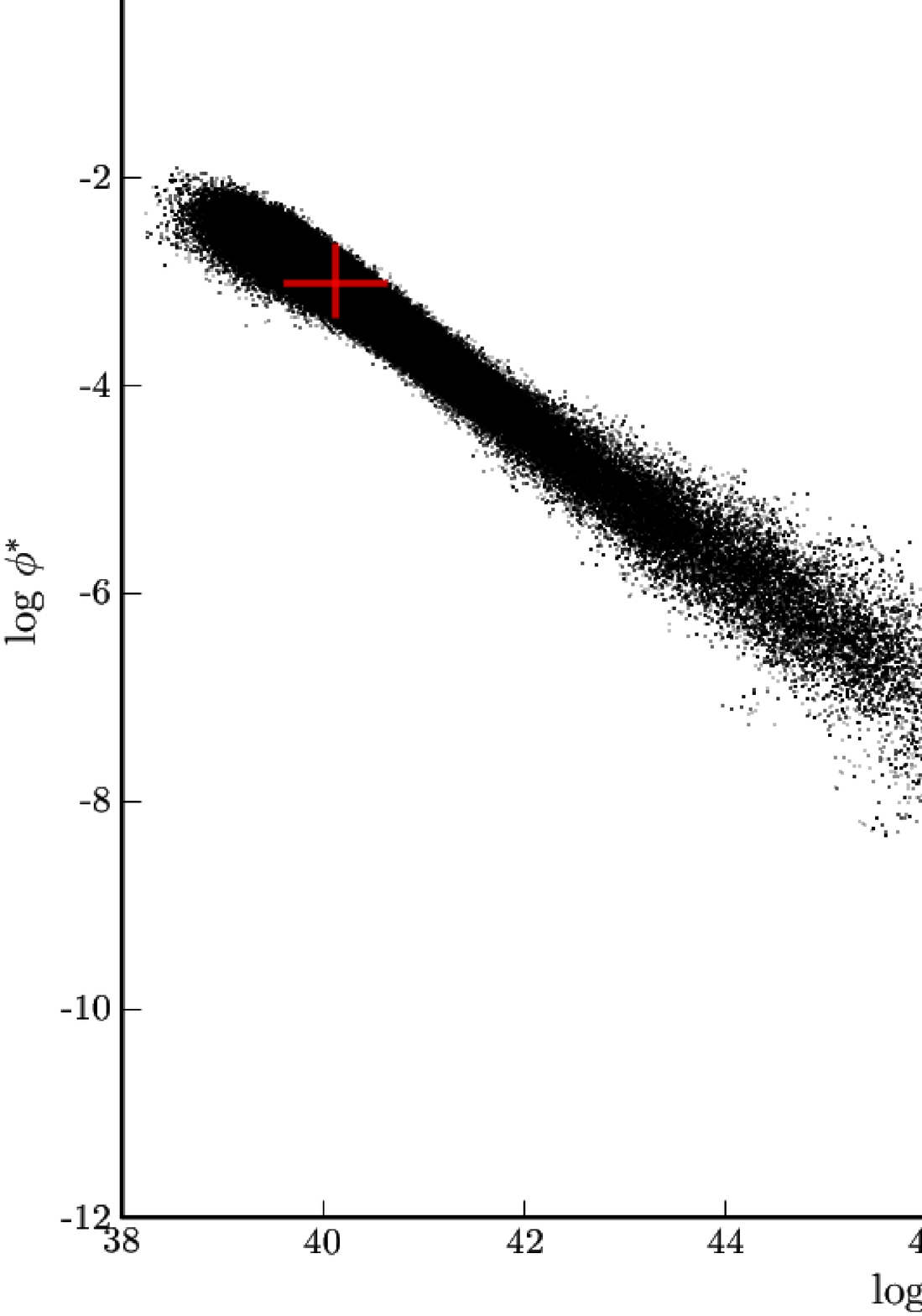}{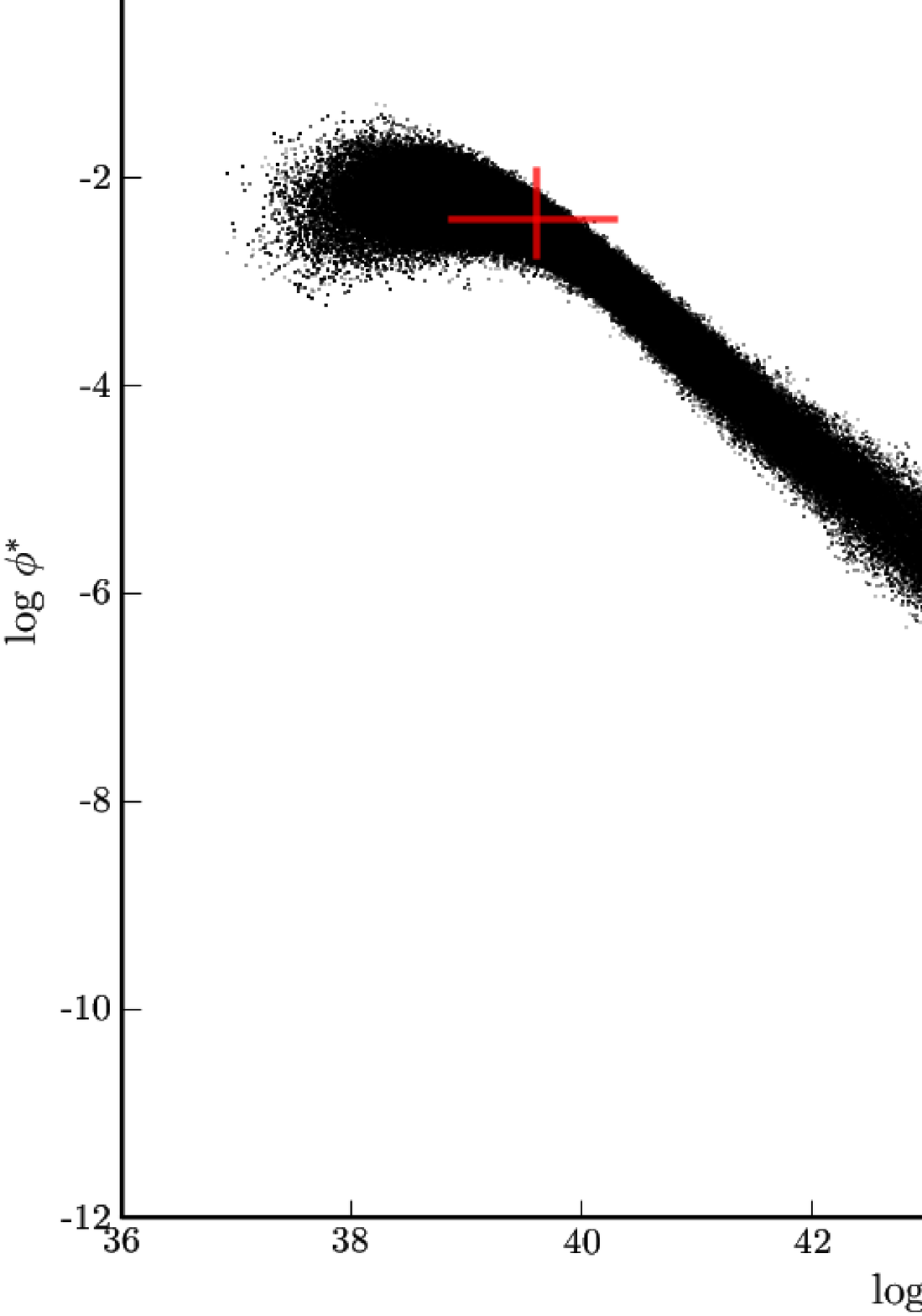}
\caption{The $z<0.5$ early-type (left) and late-type (right) galaxy
  MCMC draws for $\log \phi^*$ and $\log L^*$ in the log-normal fit.
  The solid lines show the 68\% confidence intervals determined from
  the marginalized (and hence one-dimensional) posterior probabilities
  for the parameters.\label{f_early_gal_logphi_loglstar}}
\end{figure*}

The log-normal XLF fits to the XLFs are shown in Figures
\ref{f_all_gal_sch_logn}-\ref{f_late_sch_logn}, with the
best-fit parameters and errors given in Table \ref{t:mcmc_lognormal}.
The posterior probability distributions for the fit parameters for the 
$z < 0.5$ early-type 
sample XLF are shown in Figure \ref{f_early_lowz_lognormal_hists},
and the probability distributions for the fit parameters from the
other log-normal fits have similar shapes.
Note that using a
tighter prior for $\alpha$ for the early-type galaxy fits did not
change the best-fit parameter values significantly but resulted in
$\sim 25\%$ smaller errors for $\log L^*$ and $\log \phi^*$. The
posterior probability distributions for $\sigma$ were obviously 
completely dominated by the prior distribution.   This is,
nevertheless, an improvement over simply ``fixing'' fit parameters,
which is the common procedure when parameters are not sufficiently
constrained by the data since the width of the priors are scaled from the
``physical prior'' of the 60 $\mu$m LF fit results.  A broad
tail in the probability densities toward low values of $\log
\phi^*$ and high values of $\log L^*$ is due to degeneracy between
these parameters\footnote{Note that the prior is not biasing this
  result since in both cases the prior peaks on the {\it opposite}
  side of the probability density to the broad tail.}.  This also can
be seen in Figure \ref{f_early_gal_logphi_loglstar} where the MCMC
draws are plotted.

\footnotesize
\begin{deluxetable*}{llllllll}
\tablecaption{MCMC Log-Normal Fits \label{t:mcmc_lognormal}}
\tablehead{\colhead{Sample} & \colhead{$\log \phi^*$} &\colhead{$\log L^*$} &\colhead{$\alpha$} &\colhead{$\sigma$} &\colhead{$\log \rho$} &\colhead{$\chi^2/dof$} & \colhead{$p_{\chi^2}$}}
\startdata
All galaxies, low-z  
 & $-2.28^{+0.18}_{-0.39}$ & $39.84^{+0.57}_{-0.43}$ & $1.46^{+0.20}_{-0.26}$$^{\ddagger}$ & $0.71^{+0.08}_{-0.07}$$^{\dagger}$ & $37.94^{+0.07}_{-0.09}$ & 8.1 & N/A \\
All galaxies (optimistic), low-z  
 & $-2.46^{+0.31}_{-0.27}$ & $40.17^{+0.44}_{-0.49}$ & $1.46^{+0.17}_{-0.19}$$^{\ddagger}$ & $0.73^{+0.08}_{-0.07}$$^{\dagger}$ & $38.11^{+0.07}_{-0.08}$ & 1.8/1 & 79.9\%\\
Norman et al. (2004), low-z  
 & $-2.88^{+0.34}_{-0.38}$ & $40.40^{+0.51}_{-0.40}$ & $1.57^{+0.15}_{-0.10}$ & $0.74^{+0.08}_{-0.07}$$^{\dagger}$ & $37.96^{+0.08}_{-0.07}$ & 0.5/2 & 21.7\%\\
Early-type galaxies, low-z  
 & $-3.03^{+0.36}_{-0.30}$ & $40.13^{+0.48}_{-0.47}$ & $1.42^{+0.14}_{-0.19}$$^{\ddagger}$ & $0.74^{+0.06}_{-0.09}$$^{\dagger}$ & $37.62^{+0.13}_{-0.16}$ & 3.0 & N/A \\
Late-type galaxies, low-z  
 & $-2.41^{+0.34}_{-0.47}$ & $39.61^{+0.74}_{-0.67}$ & $1.46^{+0.35}_{-0.30}$$^{\ddagger}$ & $0.73^{+0.07}_{-0.08}$$^{\dagger}$ & $37.63^{+0.12}_{-0.10}$ & 1.6 & N/A \\
All galaxies, hi-z  
 & $-2.14^{+0.48}_{-0.48}$ & $39.85^{+0.68}_{-0.60}$ & $1.36^{+0.45}_{-0.36}$$^{\dagger}$ & $0.75^{+0.07}_{-0.08}$$^{\dagger}$ & $38.27^{+0.24}_{-0.17}$ & 0.9 & N/A \\
All galaxies (optimistic), hi-z  
 & $-2.34^{+0.27}_{-0.31}$ & $40.32^{+0.52}_{-0.48}$ & $1.38^{+0.37}_{-0.34}$$^{\dagger}$ & $0.71^{+0.08}_{-0.07}$$^{\dagger}$ & $38.44^{+0.14}_{-0.10}$ & 1.2 & N/A \\
Norman et al. (2004), hi-z  
 & $-3.4^{+1.0}_{-2.2}$ & $41.2^{+2.1}_{-1.3}$ & $2.09^{+0.16}_{-0.27}$$^{\ddagger}$ & $0.73^{+0.08}_{-0.07}$$^{\dagger}$ & $38.22^{+0.56}_{-0.09}$ & 6.0/1 & 94.1\%\\
Early-type galaxies, hi-z  
 & $-3.13^{+0.54}_{-0.40}$ & $40.43^{+0.42}_{-0.60}$ & $1.42^{+0.20}_{-0.20}$$^{\dagger}$ & $0.73^{+0.08}_{-0.07}$$^{\dagger}$ & $37.83^{+0.16}_{-0.13}$ & 1.6 & N/A \\
Late-type galaxies, hi-z  
 & $-1.76^{+0.81}_{-0.66}$ & $39.43^{+0.57}_{-0.79}$ & $1.32^{+0.40}_{-0.37}$$^{\dagger}$ & $0.72^{+0.08}_{-0.06}$$^{\dagger}$ & $38.24^{+0.34}_{-0.35}$ & 0.1 & N/A \\
\enddata
\tablecomments{
Best-fitting parameters from fitting a Log-Normal function  to the XLFs.\\
$^{\dagger}$ Parameter is tightly constrained by prior\\
$^{\ddagger}$ Parameter is moderately constrained by prior\\Luminosities are in ergs s$^{-1}$ in the 0.5-2.0 keV bandpass.
\\$\rho$ is in ergs s$^{-1}$ Mpc$^{-3}$ in the 0.5-2.0 keV bandpass.
\\$p_{\chi^2}$ gives the $\chi^2$ probability at which the model fit can be rejected (note that $\chi^2$ is computed excluding upper limits).}
\end{deluxetable*}
\normalsize

\begin{figure*}[htbp]
\plotone{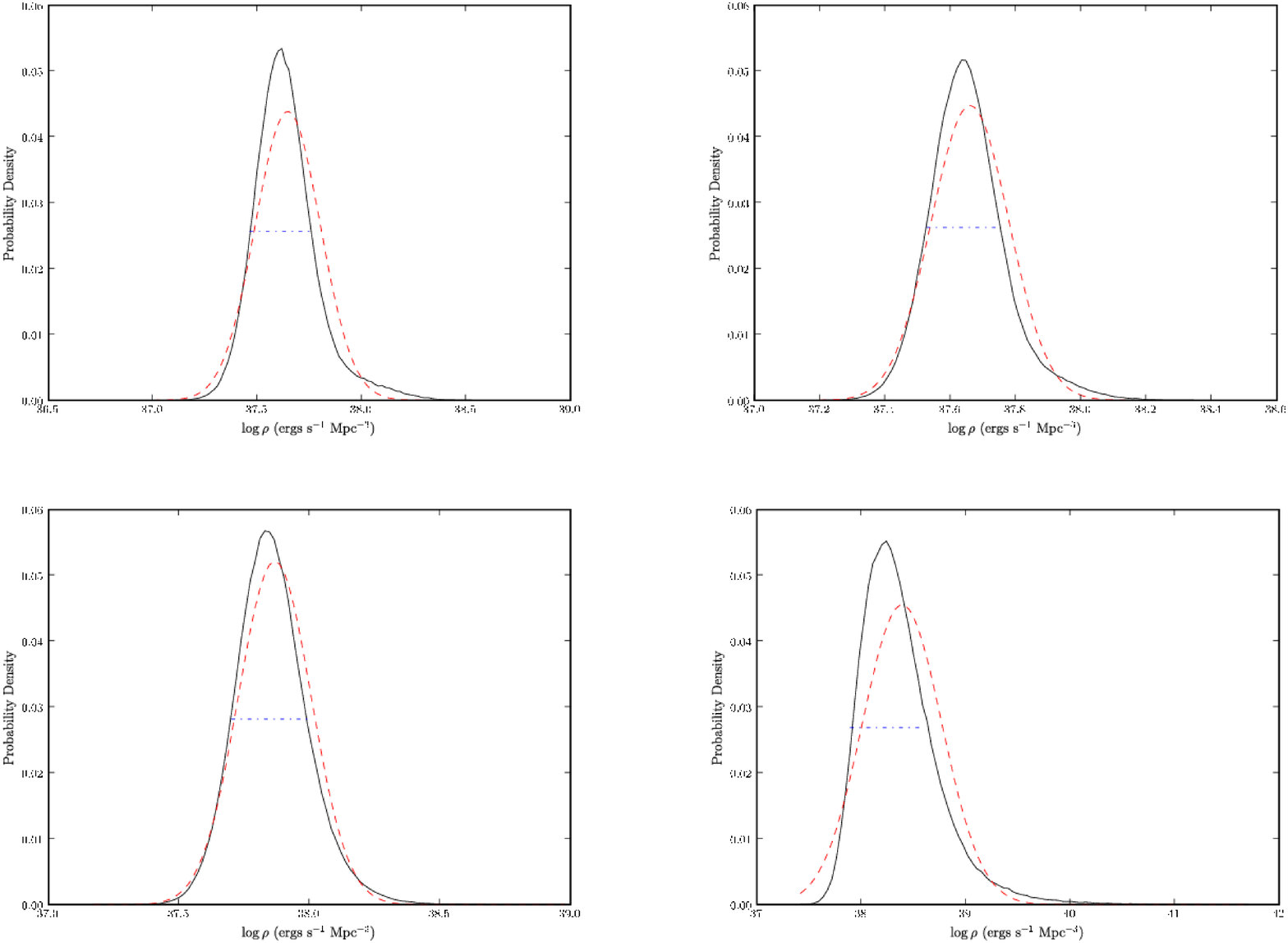}
\caption{The posterior probability distributions for the luminosity
  density ($\rho$) derived from log-normal fits to the early-type (left 
  panels) and late-type (right panels) galaxy samples.  The 
  $z<0.5$ results are shown in the top row and the $0.5<z<1.2$ results
  are shown in the bottom row.  The lines are as shown in Figure
  \ref{f_early_lowz_lognormal_hists} (there is no prior for $\rho$ since it
  is derived from other fit parameters).\label{f_lognormal_lumdens_hists}}
\end{figure*}

We computed the luminosity density, $\rho = \int \phi(L)L
d\log L$, for each MCMC draw by
numerically integrating the log-normal function over the range
$10^{37} < L_X < 10^{43}$ \ergs.  The posterior probabilities for
$\rho$ are shown in Figure \ref{f_lognormal_lumdens_hists}.  Note that
this results in a statistically-correct estimation of $\rho$ and its
error since no (usually questionable) propagation of errors is
required.

\begin{figure*}
\plotone{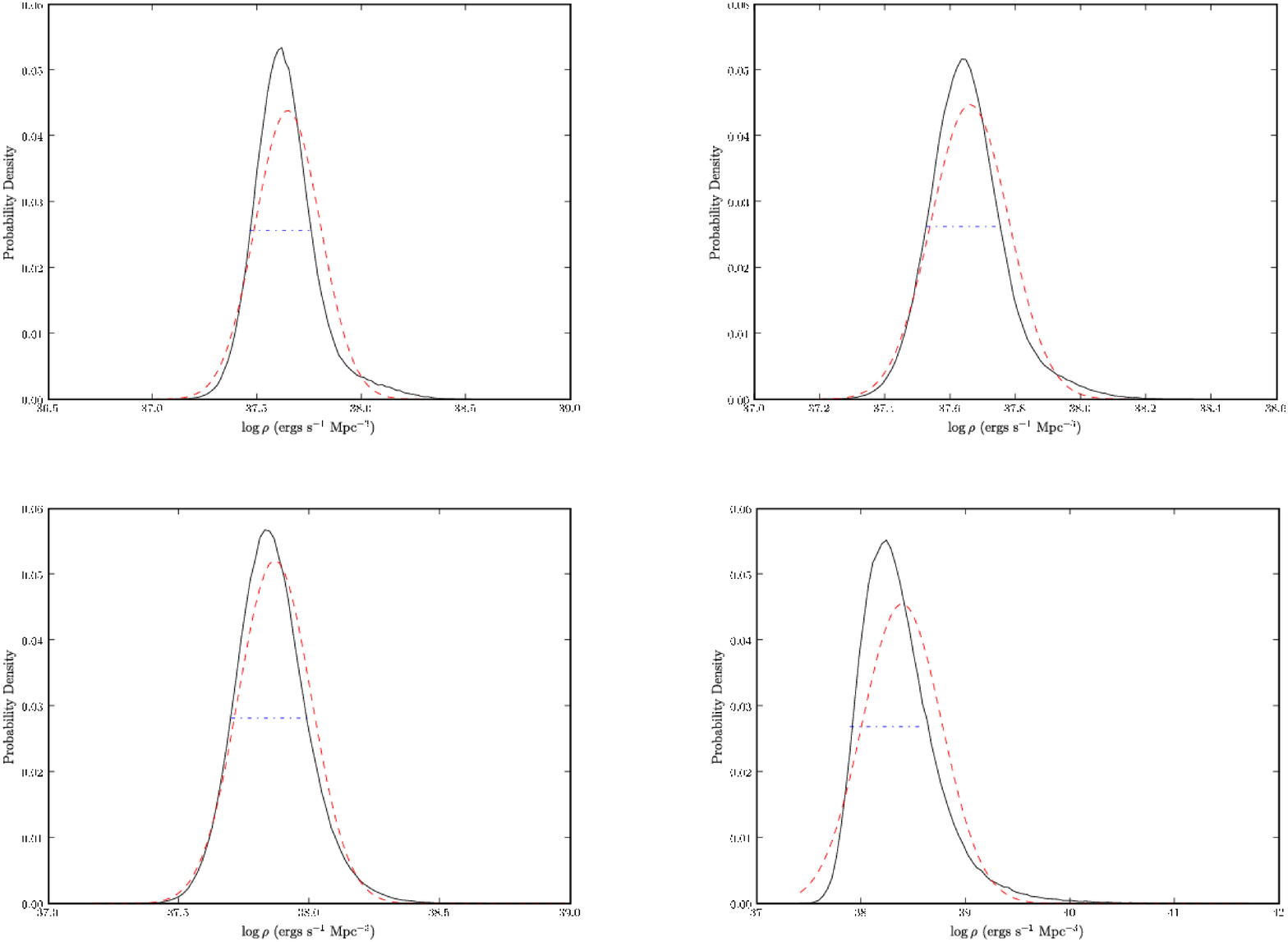}
\caption{Marginalized posterior probabilities for the fit parameters
  $\log \phi^*$, $\log L^*$, and $\alpha$ for the Schechter
  function fit to the $z<0.5$ early-type galaxy XLF. Also shown is the probability distribution
  for the luminosity density $\rho$ derived from the fit parameters.
  Lines are as shown in Figure \ref{f_early_lowz_lognormal_hists}.  \label{f_early_lowz_schechter_hists}
}
\end{figure*}

In the case of the Schechter function
fits we assumed $\alpha=1.0$ with a prior width of $50$\% (i.e.,
prior $\mu_{\sigma}=-1.0, \sigma_{\alpha} = 0.5$) since the GOODS-S
J-band $0.1<z<0.5$ luminosity function $\alpha$ values where in the
range $-1.4 <\alpha < -0.5$ (Dahlen et al. 2005).  We also used the
early-type galaxy $0.1<z<0.5$ J-band LF fit parameters from Dahlen et
al. (2005) for $M_J^*$ (-22.97) and $\phi^*$ ($8.6 \times 10^{-4} \rm
\ Mpc^{-3}\ mag^{-1}$) to estimate the initial XLF fit parameters (and
prior mean values) for $\log L^{*}$ and $\phi^*$. We converted $M_J$
to the X-ray band by computing the mean k-corrected value of $\log F_X  +0.4J +
5.1$ (see also Appendix C) to be -3.3 for normal galaxies.  $\phi^*$
was rescaled by 2.5 to be in the units of Mpc$^{-3}$ dex$^{-1}$.  As
with the log-normal fits the prior mean for $\phi^*$ was reduced by a
factor of 2 for the early-type and late-type galaxy sample XLFs.
The best-fitting Schechter models are also shown in Figures
\ref{f_all_gal_sch_logn}-\ref{f_late_sch_logn}, with the 
best-fitting parameter values and errors given in Table \ref{t:mcmc_schechter}.
The posterior probability densities for $\log L^*$, $\log \phi^*$,
$\alpha$ and $\rho$ are shown in Figure
\ref{f_early_lowz_schechter_hists} 
for the $z<0.5$ early-type galaxy sample, and again other Schechter
fits resulted in posterior probabilities with roughly similar shapes.

\footnotesize
\begin{deluxetable*}{lllllll}
\tablecaption{MCMC Schechter Fits \label{t:mcmc_schechter}}
\tablehead{\colhead{Sample} & \colhead{$\log \phi^*$} &\colhead{$\log L^*$} &\colhead{$\alpha$} &\colhead{$\log \rho$} &\colhead{$\chi^2/dof$} & \colhead{$p_{\chi^2}$}}
\startdata
All galaxies, low-z  
 & $-3.27^{+0.28}_{-0.26}$ & $40.93^{+0.24}_{-0.15}$ & $-1.45^{+0.16}_{-0.12}$ & $37.90^{+0.08}_{-0.08}$ & 2.9/1 & 87.9\%\\
All galaxies (optimistic), low-z  
 & $-3.45^{+0.24}_{-0.27}$ & $41.28^{+0.23}_{-0.18}$ & $-1.50^{+0.11}_{-0.11}$ & $38.08^{+0.08}_{-0.07}$ & 0.4/2 & 16.9\%\\
Norman et al. (2004), low-z  
 & $-4.18^{+0.27}_{-0.41}$ & $41.74^{+0.38}_{-0.22}$ & $-1.66^{+0.06}_{-0.09}$ & $37.99^{+0.09}_{-0.08}$ & 2.1/3 & 45.0\%\\
Early-type galaxies, low-z  
 & $-3.57^{+0.35}_{-0.36}$ & $40.94^{+0.36}_{-0.22}$ & $-1.31^{+0.24}_{-0.22}$ & $37.55^{+0.13}_{-0.14}$ & 3.3/1 & 89.7\%\\
Late-type galaxies, low-z  
 & $-3.51^{+0.41}_{-0.57}$ & $40.83^{+0.43}_{-0.26}$ & $-1.52^{+0.18}_{-0.25}$ & $37.63^{+0.11}_{-0.12}$ & 0.9/1 & 62.9\%\\
All galaxies, hi-z  
 & $-4.14^{+0.59}_{-0.85}$ & $41.71^{+0.54}_{-0.25}$ & $-2.04^{+0.26}_{-0.35}$$^{\ddagger}$ & $38.27^{+0.87}_{-0.21}$ & 1.5/1 & 75.7\%\\
All galaxies (optimistic), hi-z  
 & $-3.54^{+0.25}_{-0.31}$ & $41.67^{+0.20}_{-0.14}$ & $-1.61^{+0.23}_{-0.25}$ & $38.42^{+0.20}_{-0.13}$ & 0.3/1 & 43.5\%\\
Norman et al. (2004), hi-z  
 & $-4.38^{+0.47}_{-0.94}$ & $42.05^{+0.65}_{-0.26}$ & $-2.03^{+0.28}_{-0.12}$ & $38.27^{+0.41}_{-0.14}$ & 19.8/2 & 99.5\%\\
Early-type galaxies, hi-z  
 & $-4.31^{+0.53}_{-1.06}$ & $41.77^{+0.74}_{-0.32}$ & $-1.84^{+0.41}_{-0.32}$$^{\ddagger}$ & $37.77^{+0.42}_{-0.20}$ & 4.2/1 & 92.2\%\\
Late-type galaxies, hi-z  
 & $-3.41^{+0.29}_{-0.37}$ & $41.14^{+0.22}_{-0.17}$ & $-1.55^{+0.50}_{-0.38}$$^{\ddagger}$ & $37.83^{+0.40}_{-0.20}$ & 0.9 & N/A \\
\enddata
\tablecomments{
Best-fitting parameters from fitting a Schechter function  to the XLFs.\\
$^{\ddagger}$ Parameter is moderately constrained by prior\\Luminosities are in ergs s$^{-1}$ in the 0.5-2.0 keV bandpass.
\\$\rho$ is in ergs s$^{-1}$ Mpc$^{-3}$ in the 0.5-2.0 keV bandpass.
\\$p_{\chi^2}$ gives the $\chi^2$ probability at which the model fit can be rejected (note that $\chi^2$ is computed excluding upper limits).}
\end{deluxetable*}
\normalsize

\begin{figure*}
\plottwo{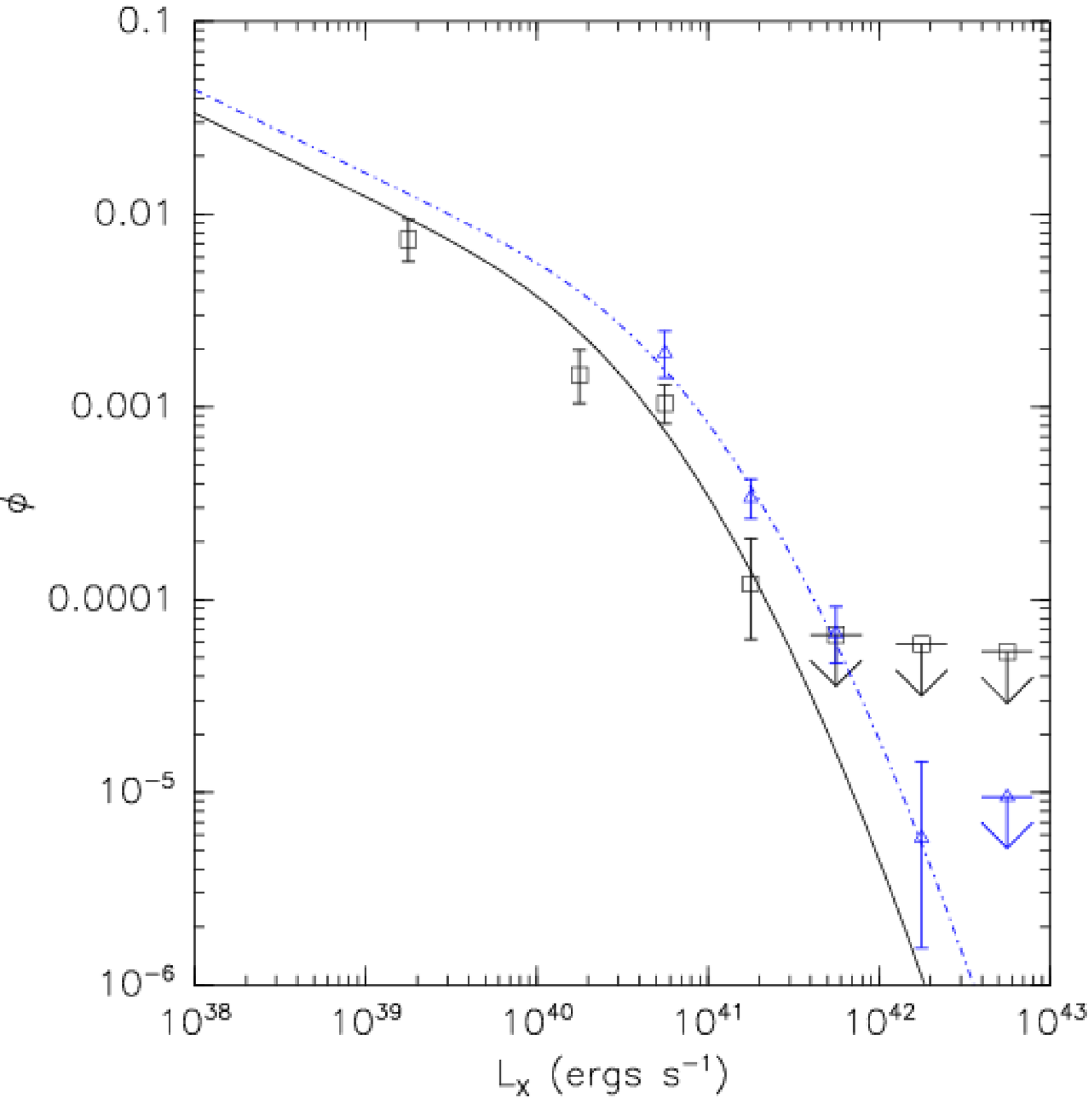}{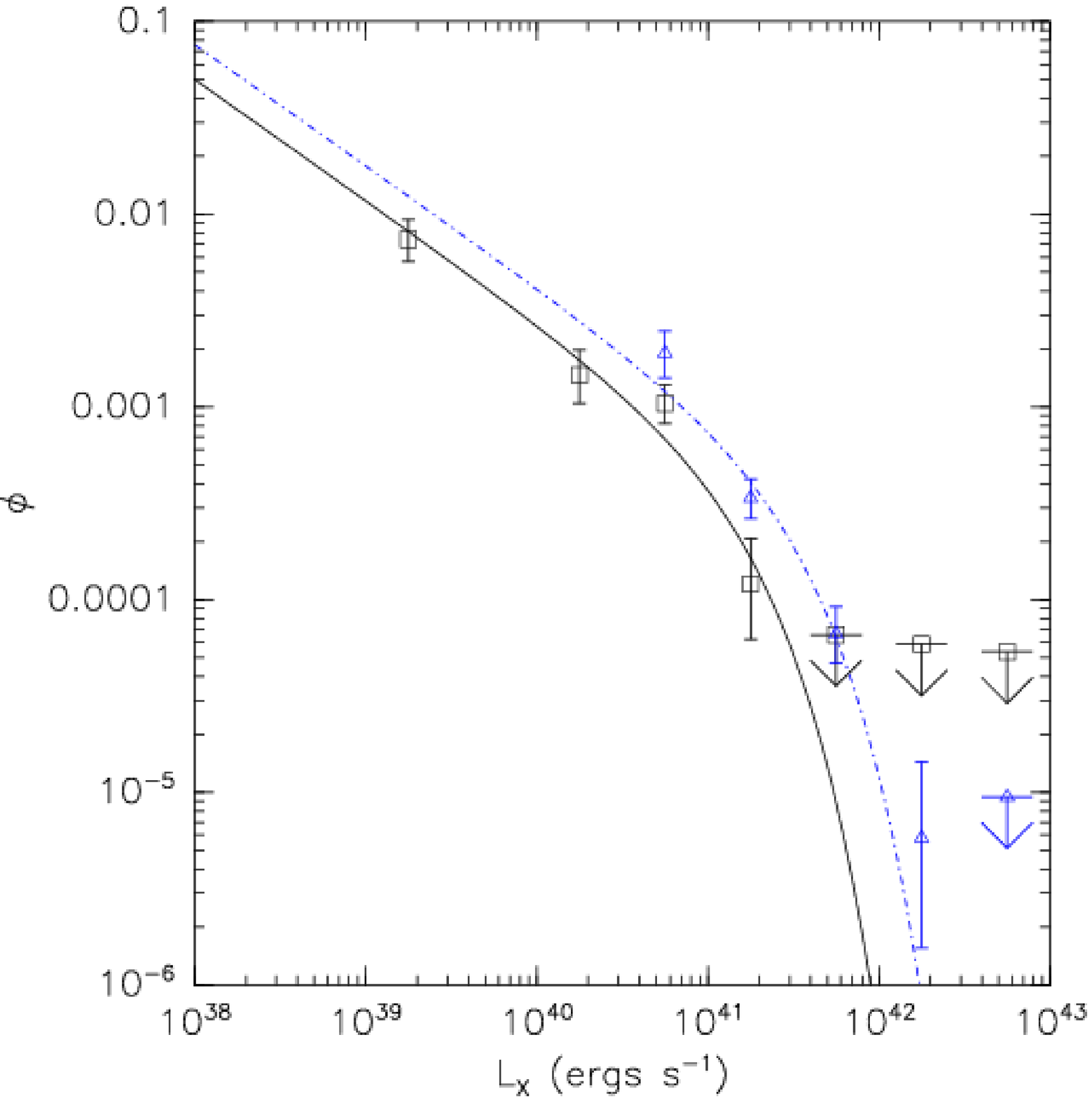}
\caption{Joint fits to the  low-z and high-z XLFs for the full
  (early + late-type) galaxy sample with the log-normal (left) and
  Schechter functions (right).  In these fits the function parameters
  are tied between the two XLFs, and the offset in $\log L^*$ is
  introduced as an additional fit parameter (i.e., pure luminosity
  evolution is assumed).   The low-z XLF points are marked with (black)
  squares and the high-z XLF points are marked with (blue) triangles.
  The dashed (black) lines and dot-dashed (blue) lines show the fit to
  the low-z and high-z XLFs. \label{f_all_gal_sch_logn_2ep}
}
\end{figure*}
\begin{figure*}
\plottwo{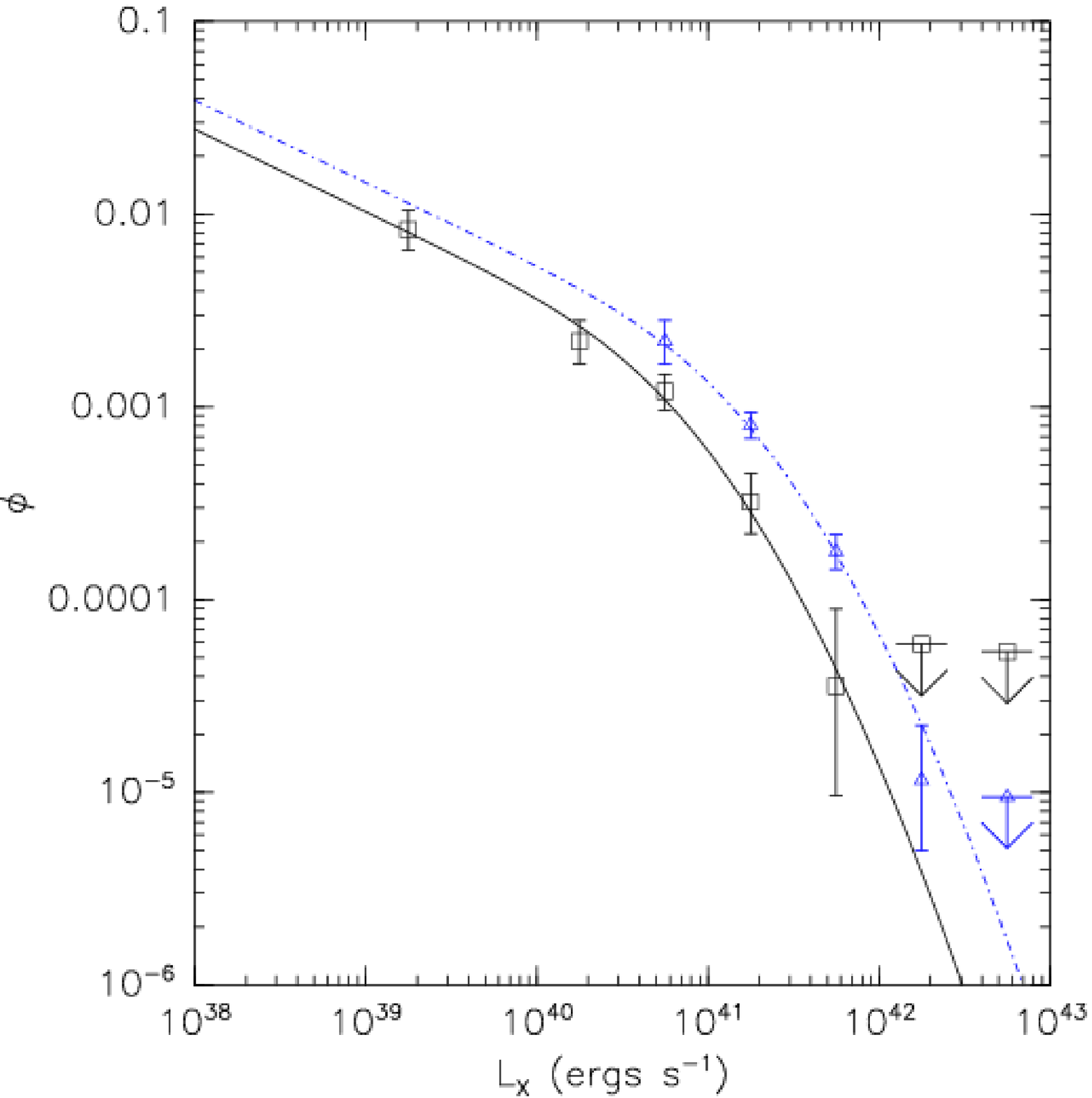}{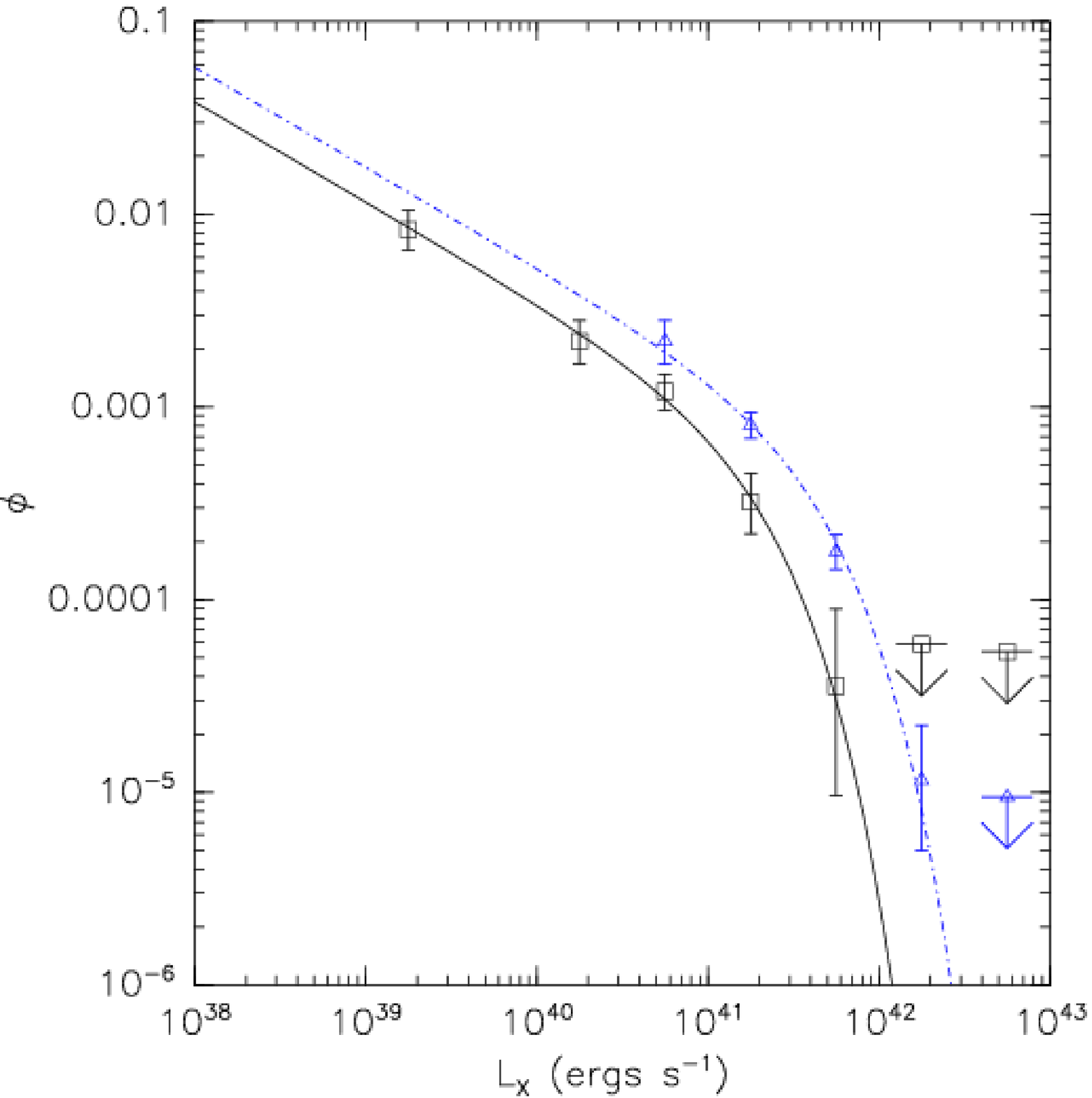}
\caption{As in Figure \ref{f_all_gal_sch_logn_2ep} for the optimistic
  galaxy sample. \label{f_all_xgal_sch_logn_2ep}}
\end{figure*}
\begin{figure*}
\plottwo{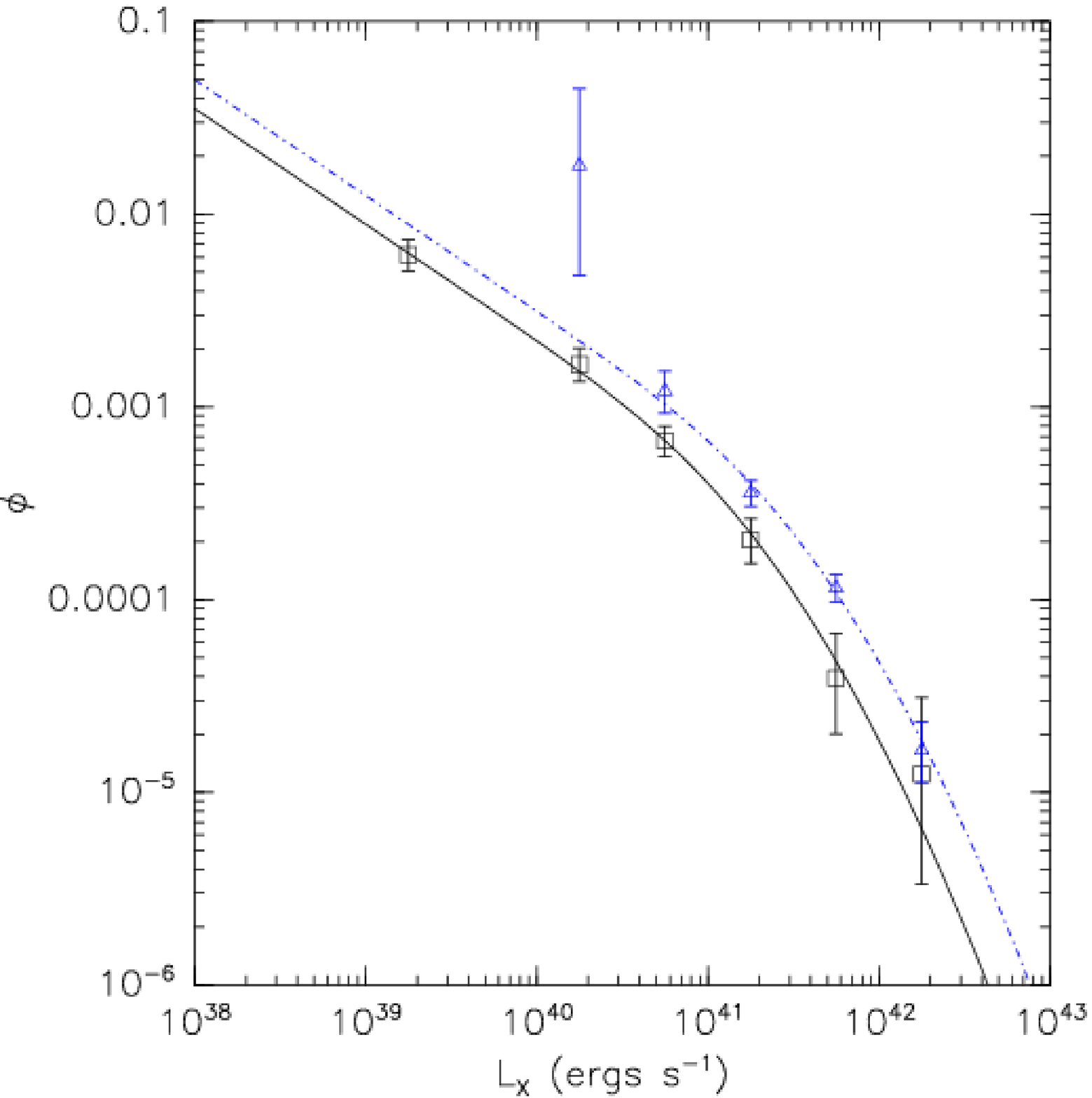}{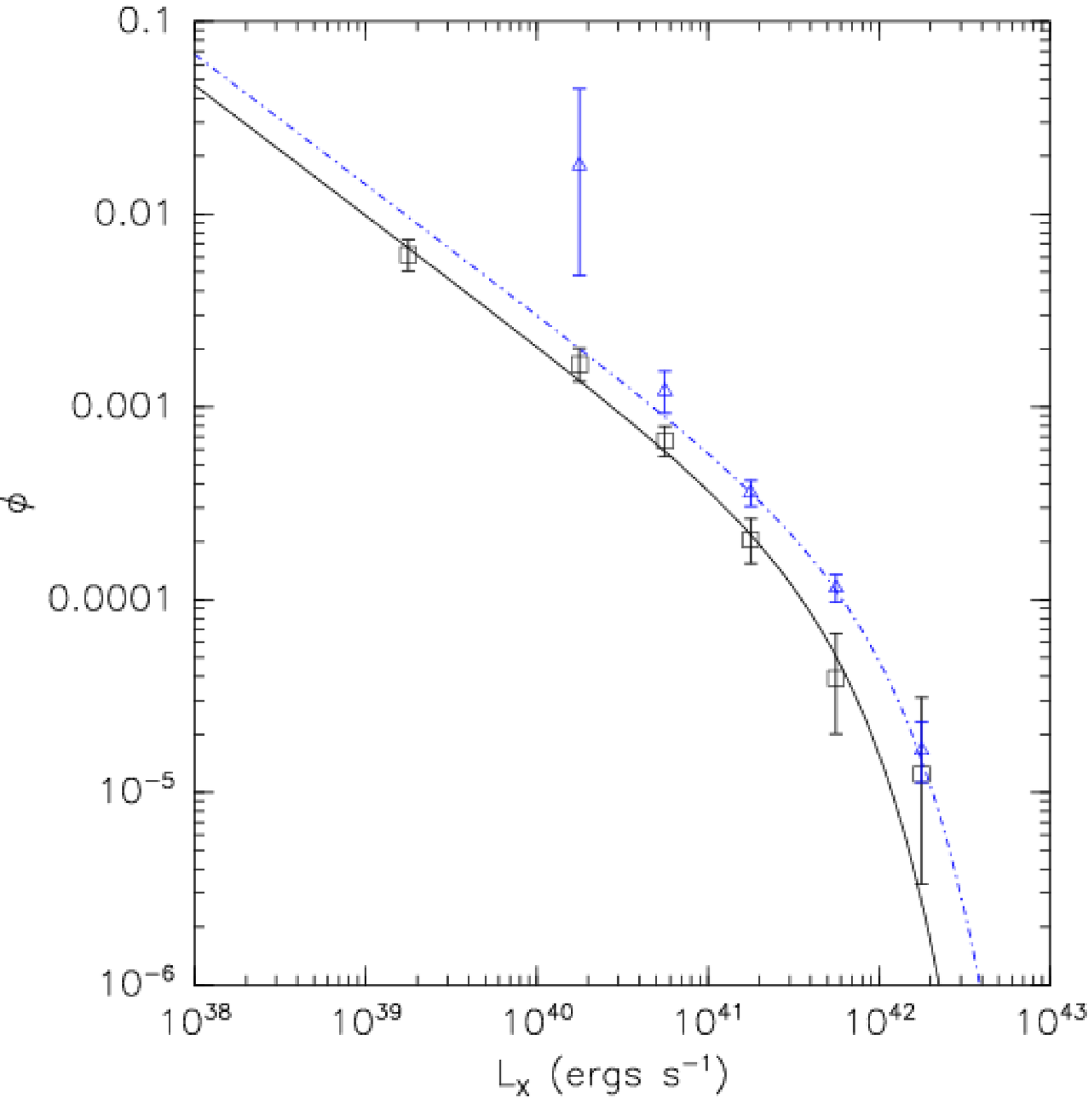}
\caption{As in Figure \ref{f_all_gal_sch_logn_2ep} for the N04
  sample. \label{f_all_norm_sch_logn_2ep}} 
\end{figure*}
\begin{figure*}
\plottwo{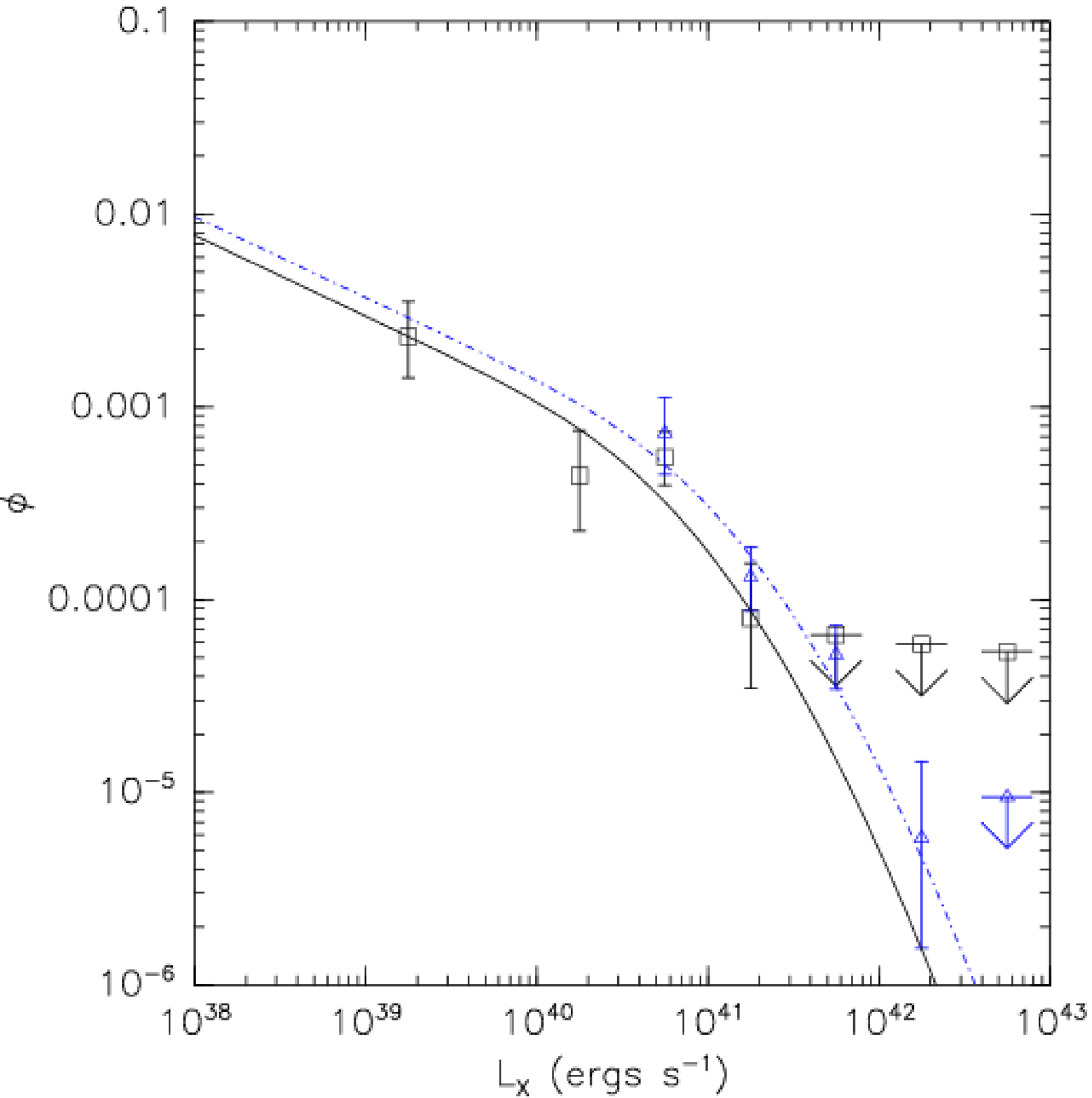}{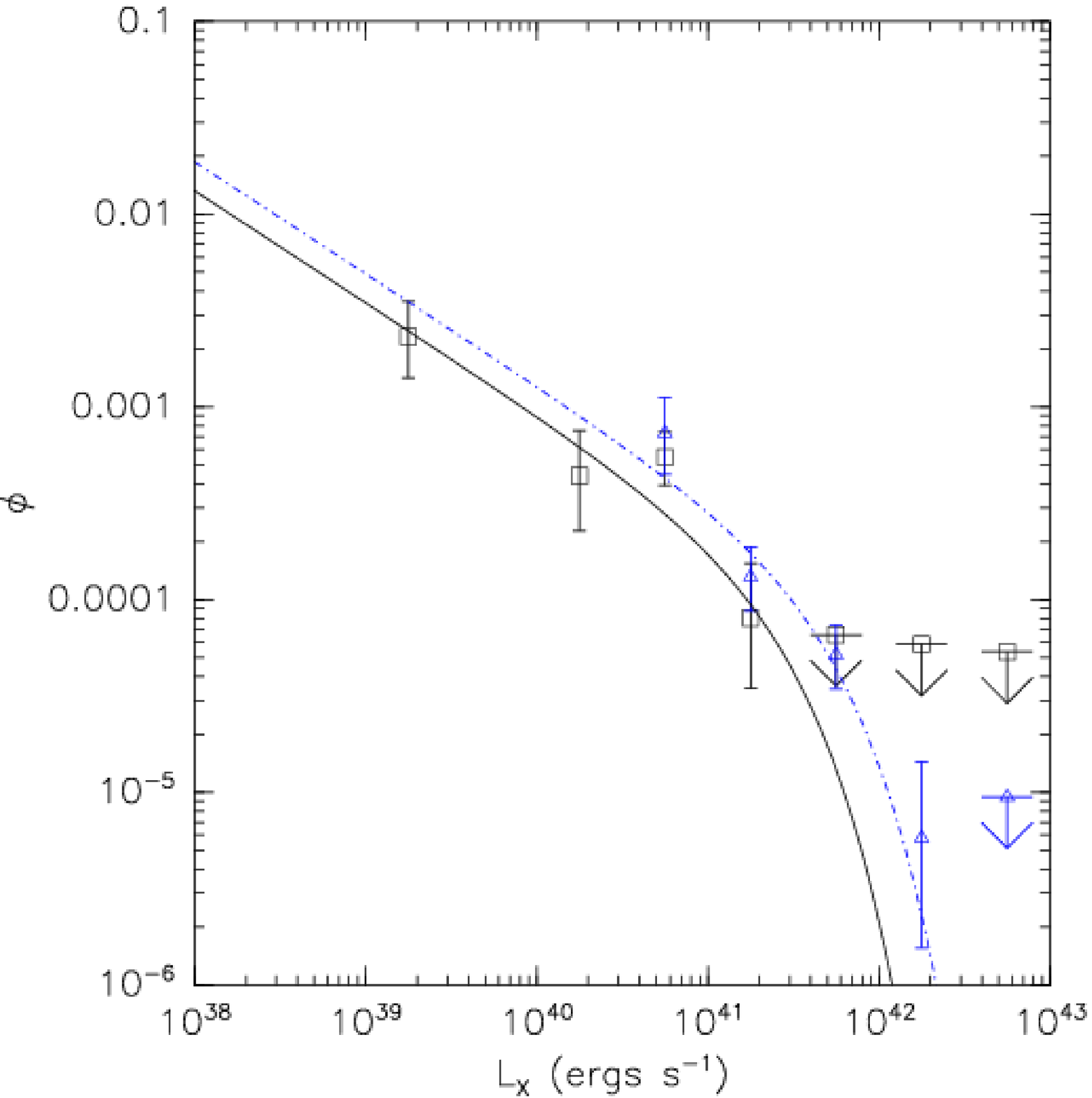}
\caption{As in Figure \ref{f_all_gal_sch_logn_2ep} for the early-type
  galaxy sample. \label{f_early_sch_logn_2ep}}
\end{figure*}
\begin{figure*}
\plottwo{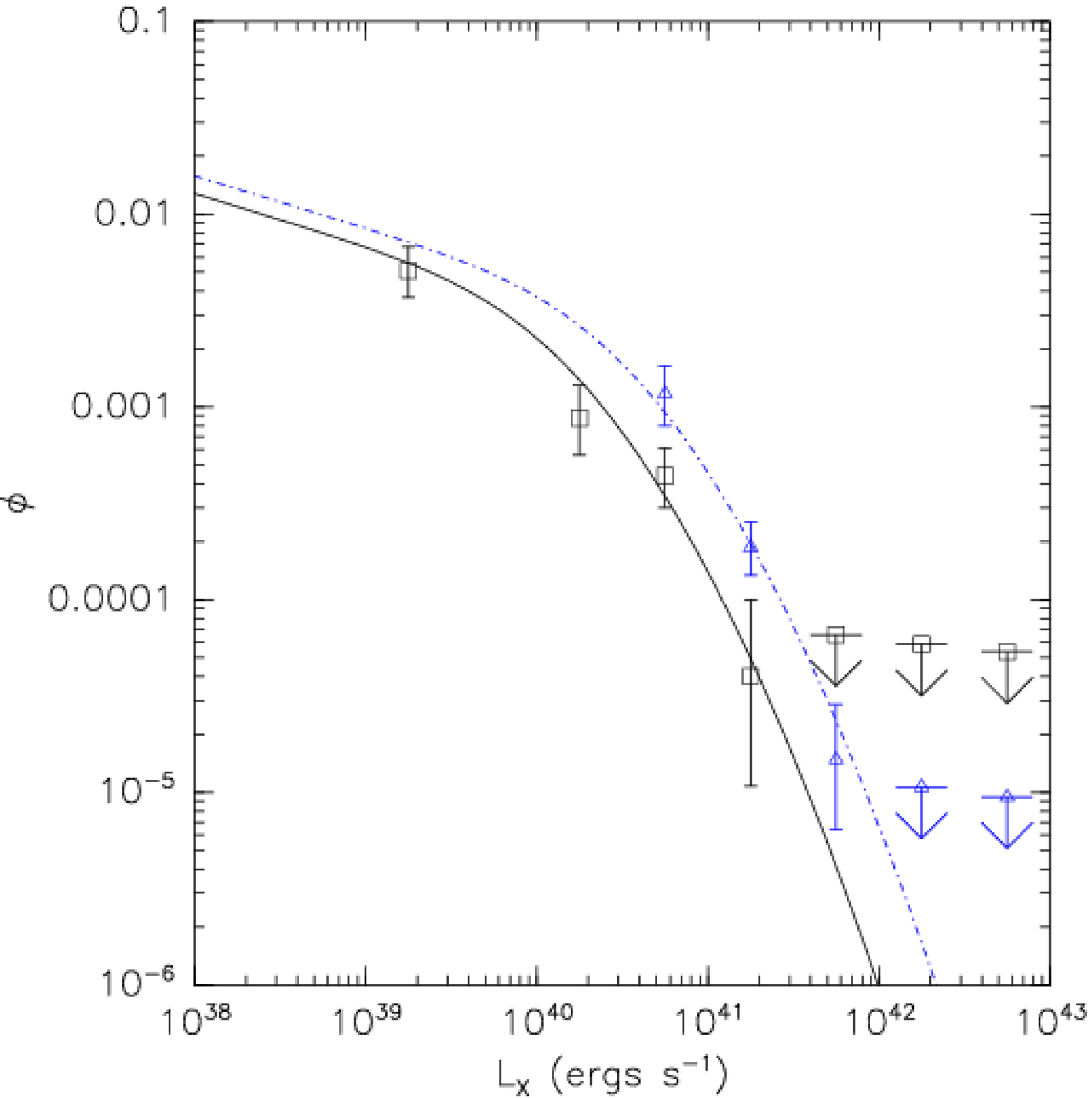}{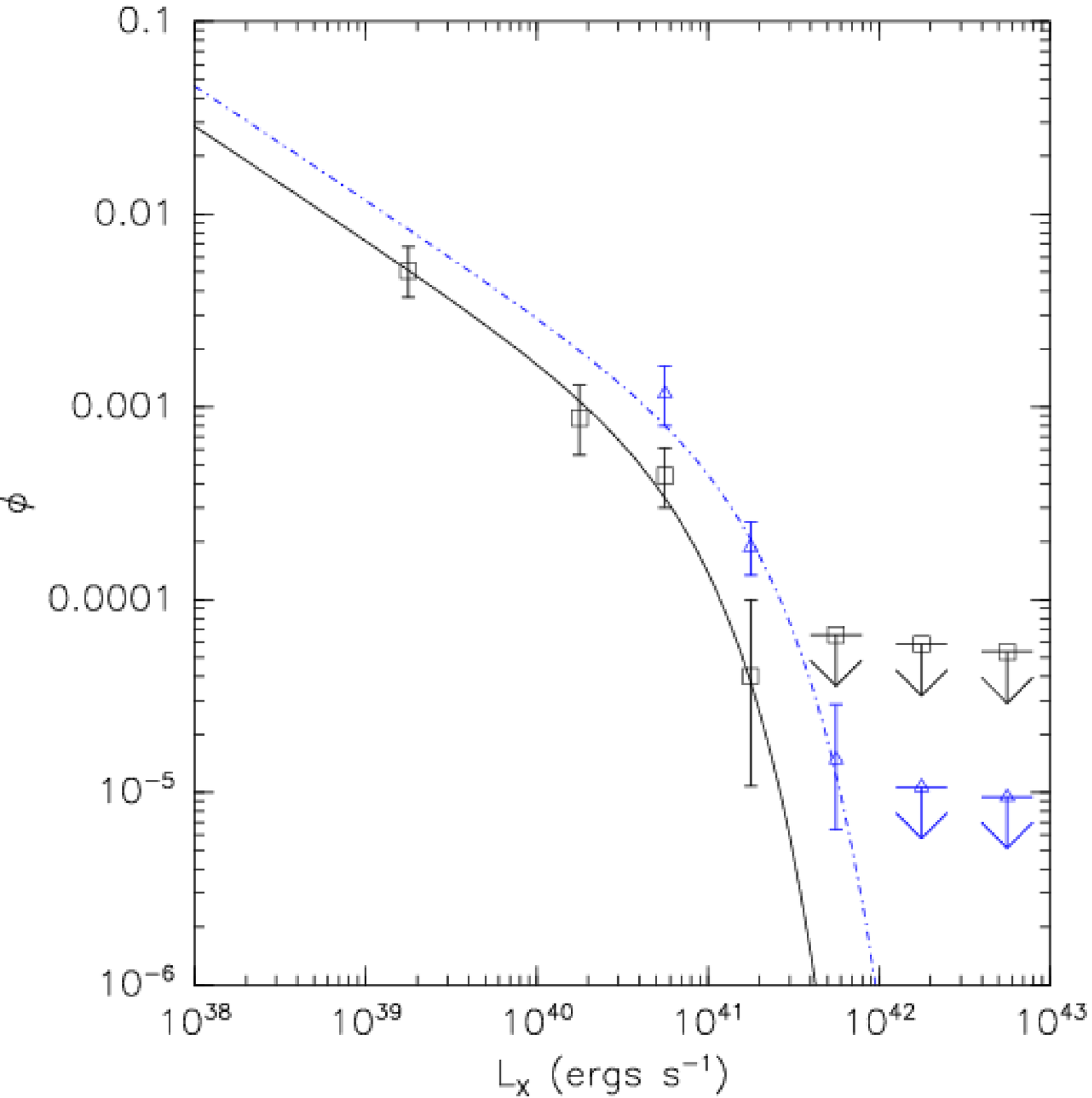}
\caption{As in Figure \ref{f_all_gal_sch_logn_2ep} for the late-type
  galaxy sample. \label{f_late_sch_logn_2ep}}
\end{figure*}
\begin{figure*}
\plottwo{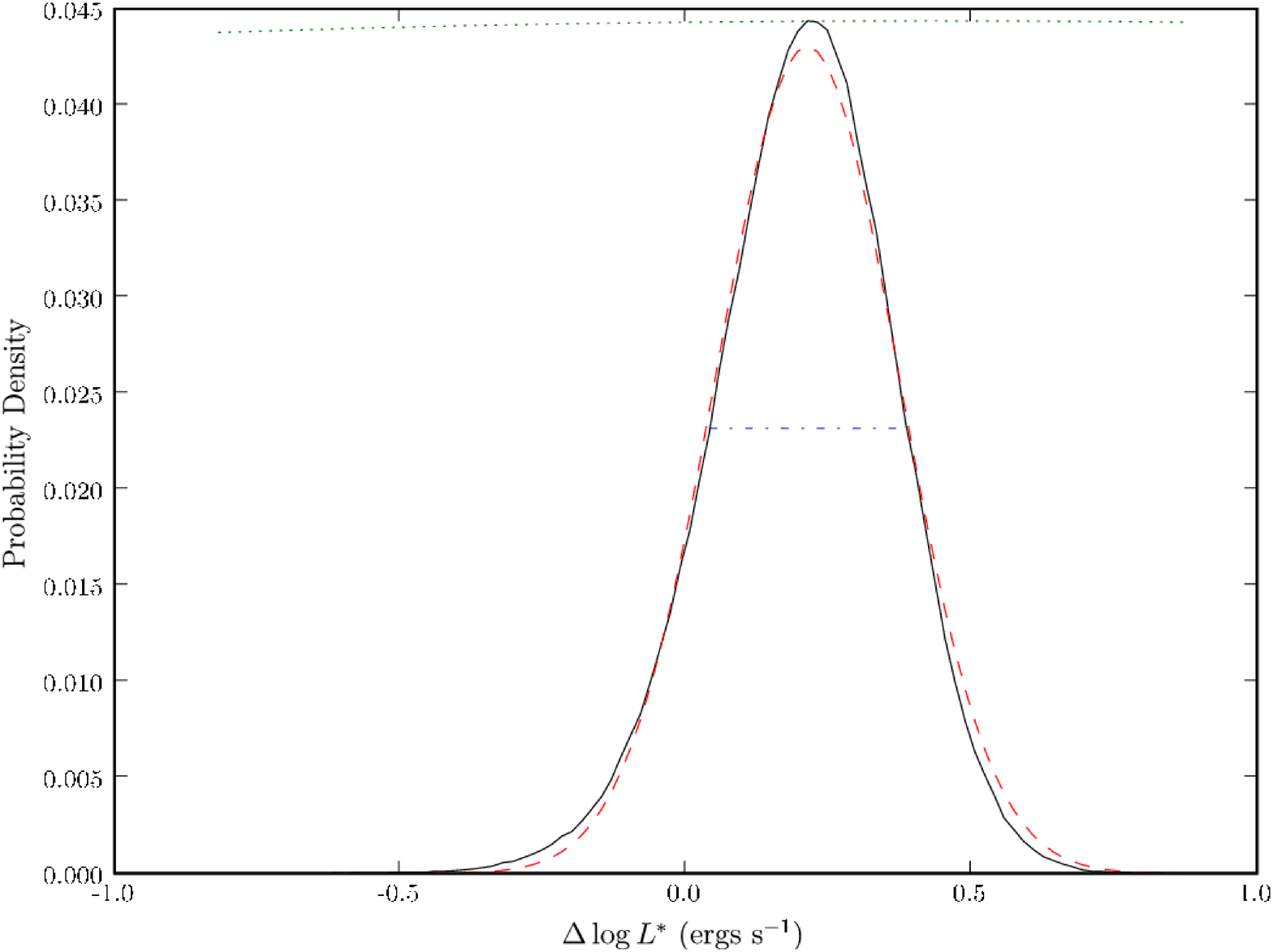}{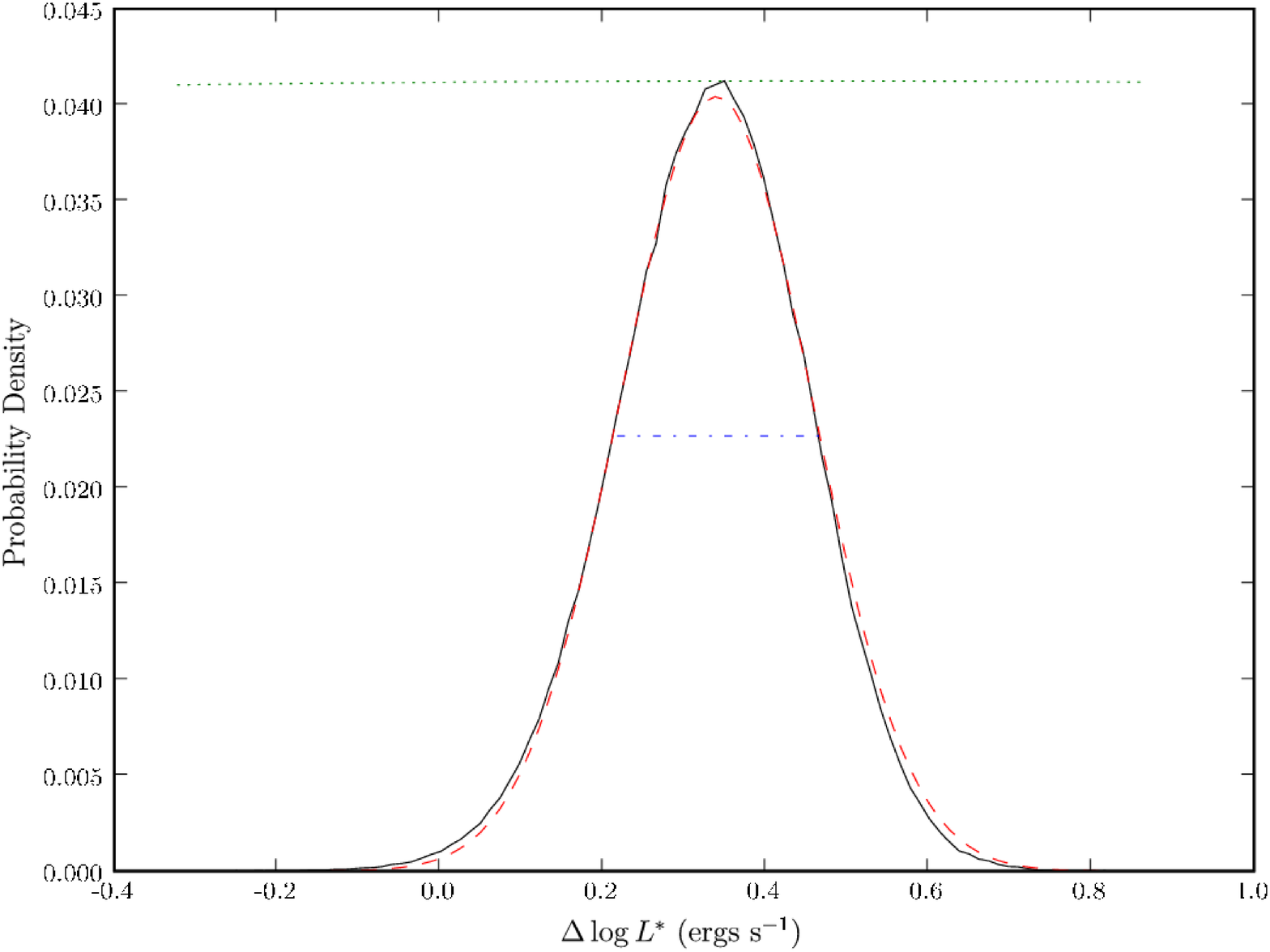}
\caption{The posterior probability distributions for $\Delta \log L^*$
  for the early-type (left) and late-type (right) galaxy samples based
  on log-normal fits (very similar results were obtained from the
  Schechter function fits).  The solid solid (black) lines
  show the posterior probability, the dashed (red) lines shows a
  Gaussian distribution with the same mean and standard deviation as the 
  posterior, the dotted (green) line show the
  prior, and the dot-dashed line shows the 68\% confidence
  interval.\label{f_deltalogL_post}}
\end{figure*}

We also fitted the low and high-z XLfs simultaneously, in this case
only allowing the  $\log L^*$ to vary between the low and high-z
models.  This is, by definition, the case of pure luminosity evolution
(PLE). The results of these fits are shown in Figures
\ref{f_all_gal_sch_logn_2ep}-\ref{f_late_sch_logn_2ep} and the fit
parameters are listed in Tables \ref{t:mcmc_lognormal_2epoch} and
\ref{t:mcmc_schechter_2epoch}. In all cases the posterior probability
for $\Delta \log L^*$ was nearly Gaussian, and we show the cases of
the log-normal fits to the early-type and late-type galaxy XLFs in
Figure \ref{f_deltalogL_post}.

\clearpage
\footnotesize
\begin{deluxetable*}{lllllllll}
\tablecaption{MCMC Joint Log-Normal Fits \label{t:mcmc_lognormal_2epoch}}
\tablehead{\colhead{Sample} & \colhead{$\log \phi^*$} &\colhead{$\log L^*$} &\colhead{$\alpha$} &\colhead{$\sigma$} &\colhead{$\Delta \log L^*$} &\colhead{$\Delta \log \rho$} &\colhead{$\chi^2/dof$} & \colhead{$p_{\chi^2}$}}
\startdata
All galaxies  
 & $-2.23^{+0.16}_{-0.29}$ & $39.74^{+0.48}_{-0.42}$ & $1.43^{+0.16}_{-0.30}$$^{\ddagger}$ & $0.72^{+0.07}_{-0.07}$$^{\ddagger}$ & $0.28^{+0.09}_{-0.09}$ & $0.29^{+0.09}_{-0.11}$ & 8.2/3 & 95.7\%\\
All galaxies (optimistic)  
 & $-2.47^{+0.29}_{-0.13}$ & $40.14^{+0.30}_{-0.42}$ & $1.43^{+0.13}_{-0.19}$$^{\ddagger}$ & $0.69^{+0.08}_{-0.06}$$^{\dagger}$ & $0.35^{+0.08}_{-0.08}$ & $0.35^{+0.08}_{-0.08}$ & 2.6/4 & 38.0\%\\
Norman et al. (2004)  
 & $-2.96^{+0.29}_{-0.24}$ & $40.52^{+0.32}_{-0.35}$ & $1.60^{+0.10}_{-0.10}$ & $0.74^{+0.08}_{-0.07}$$^{\dagger}$ & $0.25^{+0.08}_{-0.06}$ & $0.27^{+0.07}_{-0.09}$ & 2.4/6 & 12.2\%\\
Early-type galaxies  
 & $-2.98^{+0.31}_{-0.28}$ & $40.08^{+0.53}_{-0.52}$ & $1.42^{+0.21}_{-0.27}$$^{\ddagger}$ & $0.73^{+0.07}_{-0.07}$$^{\dagger}$ & $0.23^{+0.15}_{-0.16}$ & $0.23^{+0.15}_{-0.18}$ & 5.0/3 & 82.7\%\\
Late-type galaxies  
 & $-2.28^{+0.18}_{-0.29}$ & $39.43^{+0.50}_{-0.47}$ & $1.27^{+0.34}_{-0.25}$$^{\ddagger}$ & $0.69^{+0.08}_{-0.06}$$^{\dagger}$ & $0.34^{+0.12}_{-0.11}$ & $0.33^{+0.14}_{-0.11}$ & 3.4/2 & 81.4\%\\
\enddata
\tablecomments{
Best-fitting parameters from fitting a Log-Normal function  jointly to the low and high-z XLFs, allowing only $\log L^*$ to vary (i.e., assuming PLE).
\\
$^{\dagger}$ Parameter is tightly constrained by prior\\
$^{\ddagger}$ Parameter is moderately constrained by prior\\Luminosities are in ergs s$^{-1}$ in the 0.5-2.0 keV bandpass.
\\$\rho$ is in ergs s$^{-1}$ Mpc$^{-3}$ in the 0.5-2.0 keV bandpass.
\\$p_{\chi^2}$ gives the $\chi^2$ probability at which the model fit can be rejected (note that $\chi^2$ is computed excluding upper limits).}
\end{deluxetable*}
\normalsize

\footnotesize
\begin{deluxetable*}{llllllll}
\tablecaption{MCMC Joint Schechter Fits \label{t:mcmc_schechter_2epoch}}
\tablehead{\colhead{Sample} & \colhead{$\log \phi^*$} &\colhead{$\log L^*$} &\colhead{$\alpha$} &\colhead{$\Delta \log L^*$} &\colhead{$\Delta \log \rho$} &\colhead{$\chi^2/dof$} & \colhead{$p_{\chi^2}$}}
\startdata
All galaxies  
 & $-3.70^{+0.19}_{-0.22}$ & $41.24^{+0.19}_{-0.14}$ & $-1.63^{+0.10}_{-0.09}$ & $0.29^{+0.11}_{-0.09}$ & $0.30^{+0.10}_{-0.11}$ & 6.3/4 & 82.0\%\\
All galaxies (optimistic)  
 & $-3.50^{+0.12}_{-0.17}$ & $41.32^{+0.14}_{-0.10}$ & $-1.52^{+0.07}_{-0.09}$ & $0.35^{+0.07}_{-0.09}$ & $0.34^{+0.08}_{-0.09}$ & 1.0/5 & 4.1\%\\
Norman et al. (2004)  
 & $-4.23^{+0.17}_{-0.15}$ & $41.75^{+0.16}_{-0.11}$ & $-1.68^{+0.06}_{-0.06}$ & $0.24^{+0.07}_{-0.08}$ & $0.23^{+0.08}_{-0.08}$ & 4.1/7 & 23.4\%\\
Early-type galaxies  
 & $-4.26^{+0.30}_{-0.31}$ & $41.47^{+0.34}_{-0.23}$ & $-1.58^{+0.14}_{-0.13}$ & $0.26^{+0.17}_{-0.18}$ & $0.24^{+0.19}_{-0.18}$ & 5.0/4 & 70.8\%\\
Late-type galaxies  
 & $-3.62^{+0.28}_{-0.24}$ & $40.91^{+0.20}_{-0.17}$ & $-1.59^{+0.16}_{-0.13}$ & $0.35^{+0.11}_{-0.12}$ & $0.35^{+0.11}_{-0.14}$ & 1.6/3 & 35.0\%\\
\enddata
\tablecomments{
Best-fitting parameters from fitting a Schechter function  jointly to the low and high-z XLFs, allowing only $\log L^*$ to vary (i.e., assuming PLE).
\\Luminosities are in ergs s$^{-1}$ in the 0.5-2.0 keV bandpass.
\\$\rho$ is in ergs s$^{-1}$ Mpc$^{-3}$ in the 0.5-2.0 keV bandpass.
\\$p_{\chi^2}$ gives the $\chi^2$ probability at which the model fit can be rejected (note that $\chi^2$ is computed excluding upper limits).}
\end{deluxetable*}
\normalsize

\section{Discussion \label{s:disc}}
We used the GOODS survey to 
derive X-ray
luminosity functions for sources segregated by
optical/NIR spectral type.  
 We split the sources into low ($z < 0.5$) and high ($0.5  < z < 1.2$)
 redshift samples in order to 
investigate evolution. We also explored an ``optimistic'' sample (when
our galaxy selection criterion is relaxed) and the N04 XLFs for
comparison.  We implemented MCMC techniques for 
linear, log-normal and Schechter function fits to the binned XLFs.  This has
given us a reliable statistical assessment of the XLFs and any
evolution.  While in general either a log-normal or Schechter function
could fit a given XLF well, this is due in part to the relatively
sparse sampling in the XLFs presented here.  Better data would be
required to constrain the shapes of the XLFs.  A consequence of this
is that the faint-end slopes of the XLFs are somewhat uncertain, with
the log-normal and Schechter function fits giving divergent
predictions for the numbers of galaxies expected in deeper exposures.

\subsection{Comparison with Local XLFs for Different Spectral Types}
\begin{figure}
\plotone{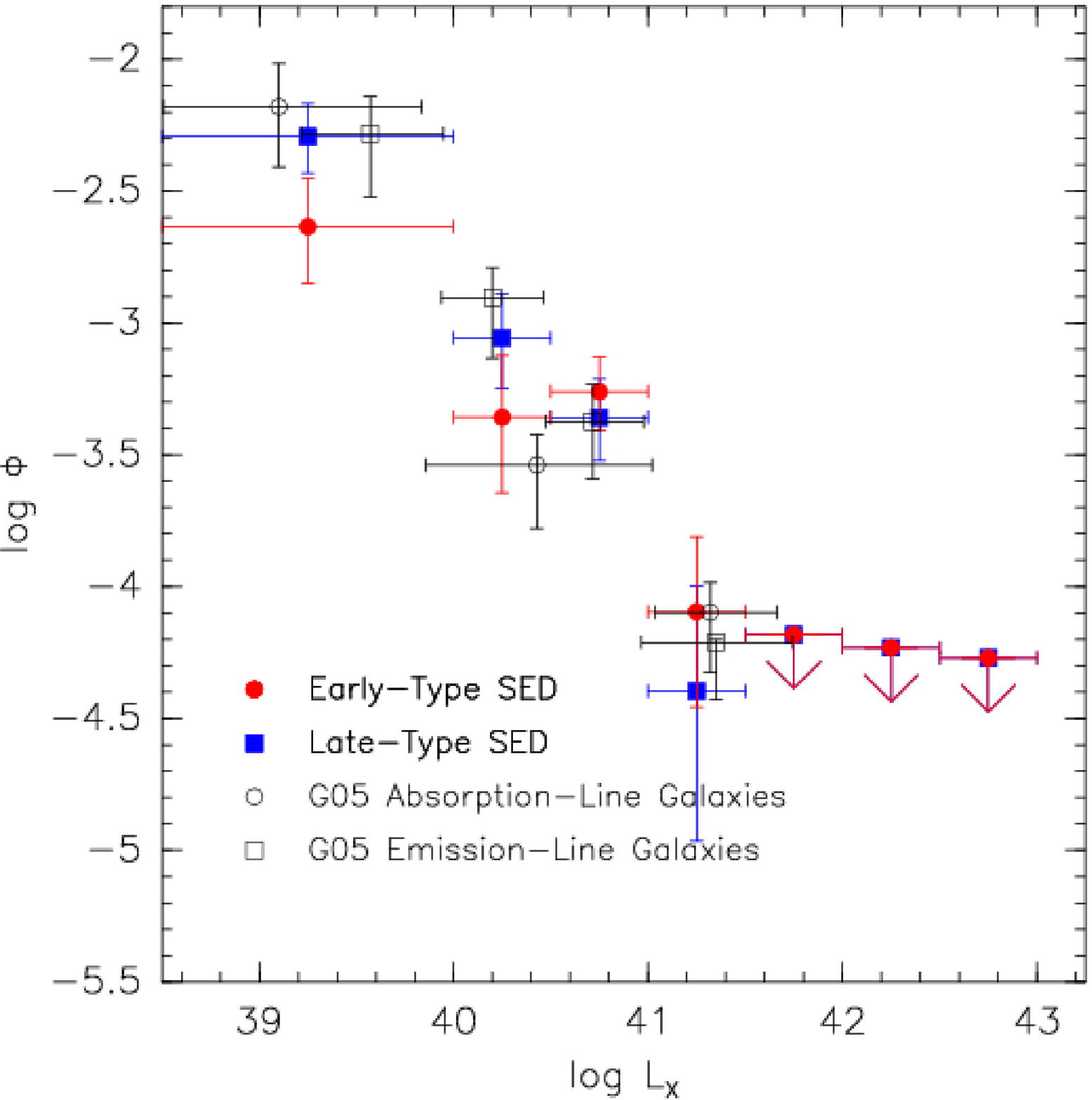}
\caption{The low-z spectral type and galaxy selected XLFs from this work shown
  along with the XLFs from G05. \label{f:xlf_g05}}
\end{figure}

We showed that the difference in the early and 
late-type galaxy XLFs was only significant at the $\sim 90\%$ and
$\sim 97\%$ level for the z$\sim 0.25$ and $z \sim 0.75$
XLFs. This is suggestive that there is a difference between the
spectral type-selected XLFs at each redshift but clearly better data
will be required 
to strengthen this result. Early and late-type XLFs were also derived in
\citet[][hereafter G05]{Georg05} for a sample of
galaxies at z$<0.2$.  In Figure \ref{f:xlf_g05} we plot our low-z
early and late-type normal/starburst sample XLFs with the G05 XLFs also
shown.  There is good overall agreement between the two sets of XLFs,
with the exception of the G05 early-type point at $L_X \sim 10^{39} \
\rm ergs \ s^{-1}$ being marginally higher than our corresponding
point.  The mean redshift of our low-z galaxy XLFs is $\sim 0.3$ for
both the early-type and late-type samples, tentatively implying that
there has been little or no evolution between z $\sim 0.1$ and z $\sim
0.3$.  However, $\sim 40\%$ of the G05 galaxy sample was comprised of
CDF sources, most of which probably are also in our
normal/starburst galaxy sample.  Having overlapping sources between
our GOODS and the G05 samples would obviously dilute any difference
between the XLFs.

\subsection{Evolution}
In all of the samples discussed here the low-z and high-z XLFs
differ at a confidence of $>99\%$, showing there is 
statistically-significant evolution with redshift. When log-normal and
Schechter functions are fit to the individual XLFs, not surprisingly
the resulting $\phi^*$  
and $L^*$ values differ for a given sample, however the implied
luminosity densities $\rho$ do not in general differ significantly.
In other words, our computed values of $\rho$ are insensitive to
whether the log-normal or Schechter functions are used, giving
additional confidence in the resulting values.  We also fit the low
and  high redshift XLFs simultaneously with the log-normal and
Schechter functions, with the main function parameters (i.e., $\log
\phi^*$, $\log L^*$ and the slopes) tied between the two XLFs but
assuming pure luminosity evolution by fitting also for $\Delta \log
L^*$.  The results for $\Delta \log L^*$ also do not depend strongly
on whether the log-normal or Schechter functions are used.  The
corresponding $\Delta \log \rho$ values were nearly identical to the
$\Delta \log L^*$ values, which might be expected since $\log \rho$
depends linearly on $\log L^*$ when all other parameters are fixed
(and the differences observed between $\Delta \log L^*$ and $\Delta \log
\rho$ show the impact of jointly varying $\log \phi^*$ and the
slopes).  

The early and
late-type galaxy $\Delta \log L^*$ values were 0.23 and
0.34 (from the log-normal fits), showing stronger evolution for
late-type galaxies. This corresponds to early and late-type galaxies
being factors of $\sim 1.7$ and $\sim 2.2$ brighter, respectively,
between $z\sim 0.25$ and $z\sim 0.75$.  It is not clear whether the
evolution observed in early-type galaxies is due to  
passive evolution of the low-mass X-ray binary population
\citep{Ghosh01, Ptak2001} or is due to enhanced star formation as expected in
the case of late-type galaxies.  However we note that our galaxy type
selection is based on galaxy SED type rather than morphological
type, and therefore these galaxies have red colors.  This suggests
that they are not actively star forming.  The lack
of any redshift dependence in the $F_X$/NIR flux ratio (see Appendix
C) may be constraining to LMXRB evolution models since LMXRB do not
``turn on'' instantenously as
discussed in \citet{Ghosh01}, however the scatter here is large.

Pure luminosity evolution is often expressed as $L^*(z) =
(1+z)^pL^*(z=0)$.  With that parametrization and $\Delta \log L^*$
being measured between two redshifts $z_1$ and $z_2$, $p =
\frac{\Delta \log L^*}{\log(1+z_2)-\log(1+z_1)}$ or $p \sim 6.8\Delta
  \log L^*$ for $z_1 \sim 0.25$ and $z_2 \sim 0.75$. 
The $\Delta \log L^*$ values then
correspond to $p_{\rm early}=1.57^{+1.09}_{-1.03}$ and $p_{\rm
late}=2.33^{+0.75}_{-0.82}$.  The late-type galaxy evolution is
consistent with the FIR evolution of $p \sim 3$. Also note that
\citet{Georgakakis07} similarly found $p \sim 2.4$ for star-forming
galaxies from the GOODS-N, using methods somewhat independent of those
discussed here (although in both studies low $L_X$, X-ray hardness, and
X-ray/optical flux ratio were among the selection criteria). The
full sample $\Delta L^*$ was 0.29, not 
surprisingly intermediate to the early-type and late-type galaxy XLF
values.  The optimistic sample resulted in a $\Delta \log L^*$
value of 0.35, basically the same as the late-type galaxy value.
However this may be somewhat coincidental since the optimistic sample
is most likely also introducing low-luminosity AGN. However, any AGN
activity is probably not dominating the near-IR-optical and the X-ray
bandpass since otherwise the X-ray/optical ratios and/or the X-ray
hardnesses would have resulted in an AGN classification (see
Appendix A).  Nevertheless the results of our analysis of the optimistic
sample probably represents a reasonable limit to the maximum amount of
evolution expected for the soft X-ray emission from normal/starburst
galaxies between $z \sim 0.25$ and $z \sim 0.75$.  We also note that
the luminosity densities inferred for the full sample XLFs are very
similar to the luminosity densities inferred for the N04 XLFs, while
the optimistic sample XLFs resulting in luminosity densities $\sim
0.2$ dex higher. Better data (either from deeper Chandra exposures or
future X-ray missions) would of course result in smaller errors on the
X-ray properties of the sources which in turn improve the
classification probabilities, as well as increasing the number of
sources populating the low-luminosity end of the XLFs. 

\section{Summary}
We have computed XLFs for normal/starburst galaxies in the GOODS for 
sources with X-ray counterparts, and fit the XLFs with linear models
using ``traditional'' techniques as well as Markov-Chain Monte Carlo
techniques. From the photometric redshift fitting procedure we
classified 40 galaxies as early-type and 46 galaxies as late-type
based on their SEDs.  The early-type galaxy XLFs tend to be slightly flatter
than those for late-type galaxies, although from the MCMC analysis the
significance of this result is only at the $1-2\sigma$ level. The
early and late-type galaxy sample XLFs at z$\sim 0.25$ are consistent
with the low-redshift early and late-type XLFs of Georgantopoulos et
al. (2005). We used the MCMC approach to also fit the XLFs with
log-normal and Schechter functions.  The XLFs discussed here all show
significant evolution between $z \sim 0.25$ and $z \sim 0.75$.  We
jointly fit the low and high-redshift XLFs assuming pure luminosity
evolution, allowing only $L^*$ to vary between the XLFs, which
resulted in evolution of $(1+z)^{1.6}$ and $(1+z)^{2.3}$ for the
early-type and late-type galaxy samples.  The late-type galaxy
evolution derived here is consistent with the star-forming galaxy
X-ray evolution given in \citet{Georgakakis07}. 
Including sources with ambiguous classification results in an
``optimistic'' galaxy sample with a total
galaxy X-ray evolution of $(1+z)^{2.4}$, essentially the same value as
derived for late-type galaxies.  The optimistic sample XLF evolution
suggests that the maximum amount of evolution in the X-ray emission of
normal/starburst galaxies at these redshifts is $(1+z)^{2.4 \pm 0.5}$.

The Bayesian fitting approach here could be expanded to include
additional uncertainties that might impact this analysis, such as the
redshift errors \citep[see, e.g.,][]{Dahlen05}, the uncertainties
in the completeness correction and the uncertainties the X-ray and
optical fluxes of the sources.  A larger impact on our results would likely result
from including radio and FIR data from the GOODS fields, which we will
explore in future work.  This will improve both the SED fitting (and
hence the galaxy type determination) and give an independent
star-formation rate estimation \citep[see also][]{Georgakakis07}.  It
may also be possible to simultaneously fit for the multi-variate
luminosity functions and the individual galaxy types, activity types and
redshifts (with any spectroscopic redshifts used as tight priors), at
least in an iterative fashion (i.e., where the current luminosity
function estimates guide the galaxy type and photometric redshift
probabilities). Finally, advanced Bayesian model selection
techiniques, as discussed in \citet{trotta07} and references therein,
can be applied here to guide the parameters of future observations by
predicting the ability of future data to prefer a given model and
arrive at a given set of constraints.  

\acknowledgements
We thank the anonymous referee for useful comments that improved this
paper.  A.P. acknowledges the support of NASA grant NNG04GE13G.

\appendix
\section{GOODS X-ray Sample Properties and Galaxy Selection}
Here we discuss our methodology for classifying the sources, and we
also discuss the statistical properties of the sample.  Note
that the relevant point of our classification is not whether the
entire SED is dominated by star-formation or an AGN, but which of these
is dominating the soft X-ray band.

\subsection{Bayesian Selection}
N04 selected normal/starburst galaxies from the full
CDF-N and CDF-S samples using a Bayesian classification procedure,
where priors were constructed from the a set of galaxies with
well-determined optical types, normal/starburst galaxy, type-1 AGN and
type-2 AGN (hereafter galaxy, AGN1 and AGN2).  The product of the
prior distributions for a class and the likelihood for the observed
parameters for a given source gave the probability that the source was
drawn from that class.
Here we follow the same procedure with several improvements.  First,
sources with small differences between the probabilities in each class should have been
considered uncertain since these conditions occur when the
separation of the sources parameter values and the parent
distribution means are small relative to the parameter errors.
Here we use the Bayesian ``odds ratio'', or the ratio of
posterior probabilities for the classes being compared. In Bayesian
model testing a model is considered to be favored only when the odds
ratio exceeds at least 3 (while odds ratios greater than 10 are
preferred). 
Second, the parameter likelihoods were handled somewhat simplistically,
with Gaussian errors assumed on each parameter and a constant error
in $\log L_X$ and $\log L_X - \log L_{opt}$ of 0.25.  The Gaussian
assumption is not correct for hardness ratio errors and
$\log L_X$ errors when the number of counts detected for the source is
  small.  However, it turns out that the Gaussian approximation to
  hardness ratio errors is {\it conservative} \citep{park2006}.  
  Here we use the larger of the asymmetric errors on count rate given
  in Alexander et al. (2003).  We assume that $\Delta F_X/F_X = \Delta
  C/C$ where $\Delta F_X$ is the error on X-ray flux and $\Delta C$ is
  the error on X-ray count rate C given in Alexander et al.  We then
  take the error on 
  $\log L_X$ to be given by $\frac{\Delta L_X}{L_X\log(10)}$ where
  $\Delta L_X = L_X \frac{\Delta F_X}{F_X}$.  This is valid only when
  $\Delta F_X/F_X \ll 1$, however this is the case for the 
  majority  of the X-ray sources. Finally, N04 did not
  account for k-corrections in the optical data when computing
  X-ray/optical flux ratios (k-corrections to the X-ray data were not
  necessary since an energy index of 1.0 was assumed for every
  source).  As discussed below, k-corrections are now included in
  the computation of the priors and in the galaxy classification.

\subsection{Priors}
  Szokoly et al. (2004) reported classifications based on the optical
  spectra alone, in the classes ``ABS'' (no or only absorption lines
  are present in the spectrum),
  ``LEX'' (a low-ionization emission line spectrum), ``HEX'' (a
  high-ionization emission line spectrum), and ``BLAGN'' (broad
  emission lines are found).  All of the BLAGN sources
  should correspond to AGN1 sources, by definition, however a broad
  line AGN may be present in the other classes where there was not
  sufficient signal to detect a broad-line component.  The HEX and LEX
  classes should be dominated by type-2 AGN and star-forming galaxies,
  respectively.  However the LEX classification includes some AGN
  where low signal-to-noise or dilution of the nuclear
  spectrum due to aperture effects \citep[][]{Moran2002} has precluded the
  detection of high-excitation emission lines. For several
  LEX sources the statistics are sufficient for an AGN component to be
  identified from line-ratio diagnostics. We therefore
  derive priors using only ABS sources as galaxies, HEX sources and
  LEX sources with AGN line ratios as
  AGN2, and BLAGN sources as AGN1.  We required that the X-ray sources
  have a corresponding entry in the Alexander et al. catalog since
  that catalog is used for the X-ray properties of the GOODS
  sample. 

We initial only selected sources 
  with z $<=$1.2 however this resulted in only 8 AGN2 sources.  Since
  AGN2 are known to have flat X-ray spectra, and hence a minimal
  k-correction, we relaxed the redshift constraint to z $<=$ 2 for
  that class.  The final tally was then 11 AGN2 sources, 11 AGN1
  sources and 15 (normal/starburst) galaxy sources.  
The mean offsets between the R and K band magnitudes reported in
  Szokoly et al. and the R and $\rm K_s$ bands used in the GOODS survey were
  computed in order to adopt the priors based on the Szokoly et
  al. source for use with GOODS data.  This
  was done regardless of spectral type and we found offsets of
  0.22 magnitudes in R and 1.9 magnitudes in K, with a standard
  deviations of 0.16 and 0.21 magnitudes.

\begin{figure}
\plotone{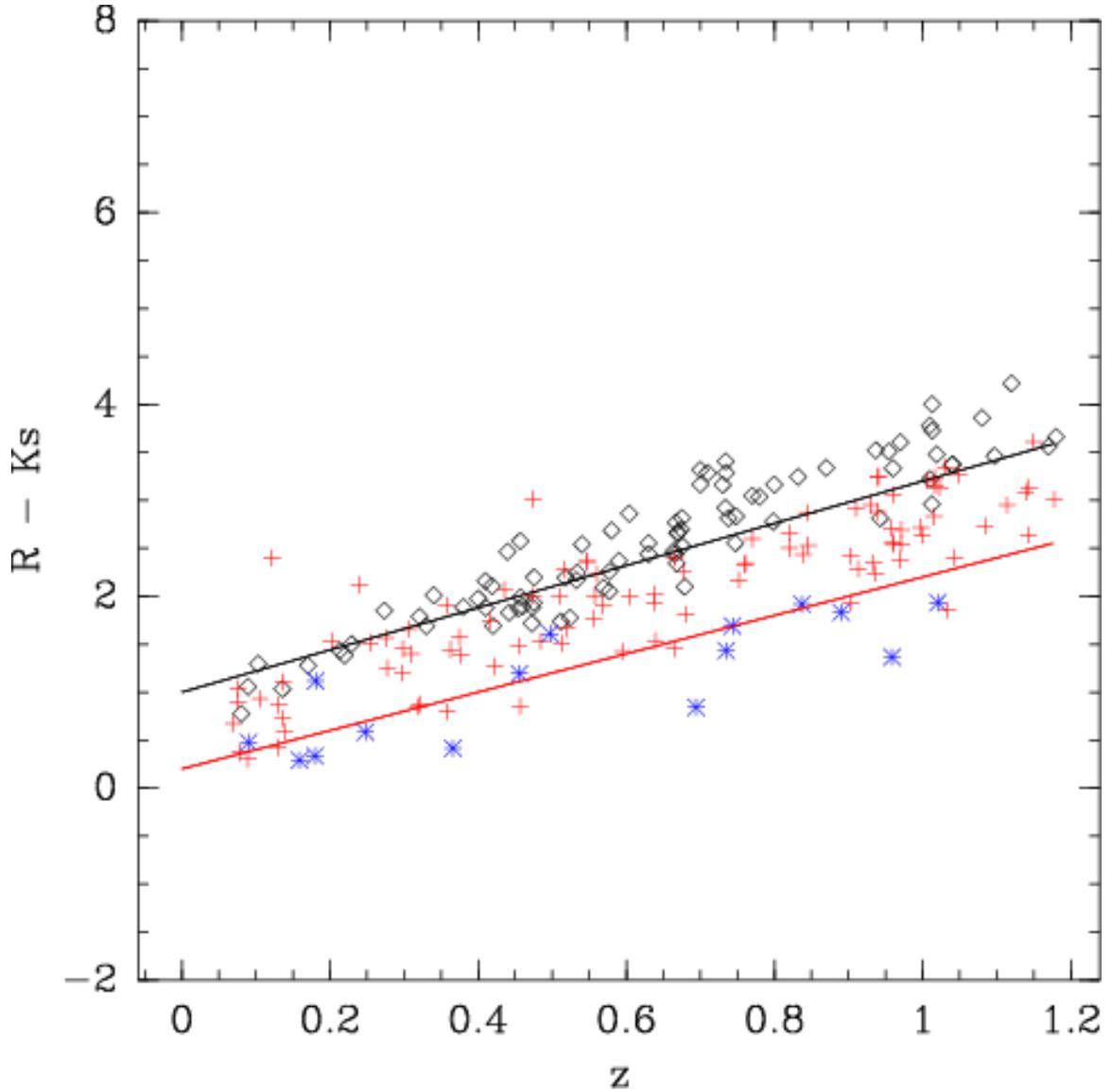}
\caption{R-$\rm K_s$ color plotted as a function of redshift for the
  early-type, late-type and irregular/starburst galaxies in the full
  sample.  Also plotted are the lines used to separate these spectral
  types in the computation of priors in the Szokoly
  sample. \label{f:r_k_full}}
\end{figure}

As shown in Figure \ref{f:kcor}, k-corrections to the X-ray/R and
X-ray/K band flux ratios can be significant.  Since only K and R band
magnitudes are listed in Szokoly et al., we cannot apply the same
interpolation procedure for k-corrections as was done with GOODS
data.  However, we can use the R-K color to estimate the spectral type
of the source, and apply a mean k-correction based on that type.  
We show in Figure \ref{f:r_k_full} the R-$\rm K_s$ colors for the full
sample, with early-type, late-type and irregular/starburst galaxies
marked separately.  While there is some overlap, we manually selected
linear functions to delineate the spectral type as a function of
redshift, as shown in the figure.  The early/late-type galaxy
separation is given by R-$\rm K_s$ = 1.0 + 2.2z and the late-type/starburst
galaxy separation is given by R-$\rm K_s$ = 0.2+2.0z.  

\begin{figure}
\plottwo{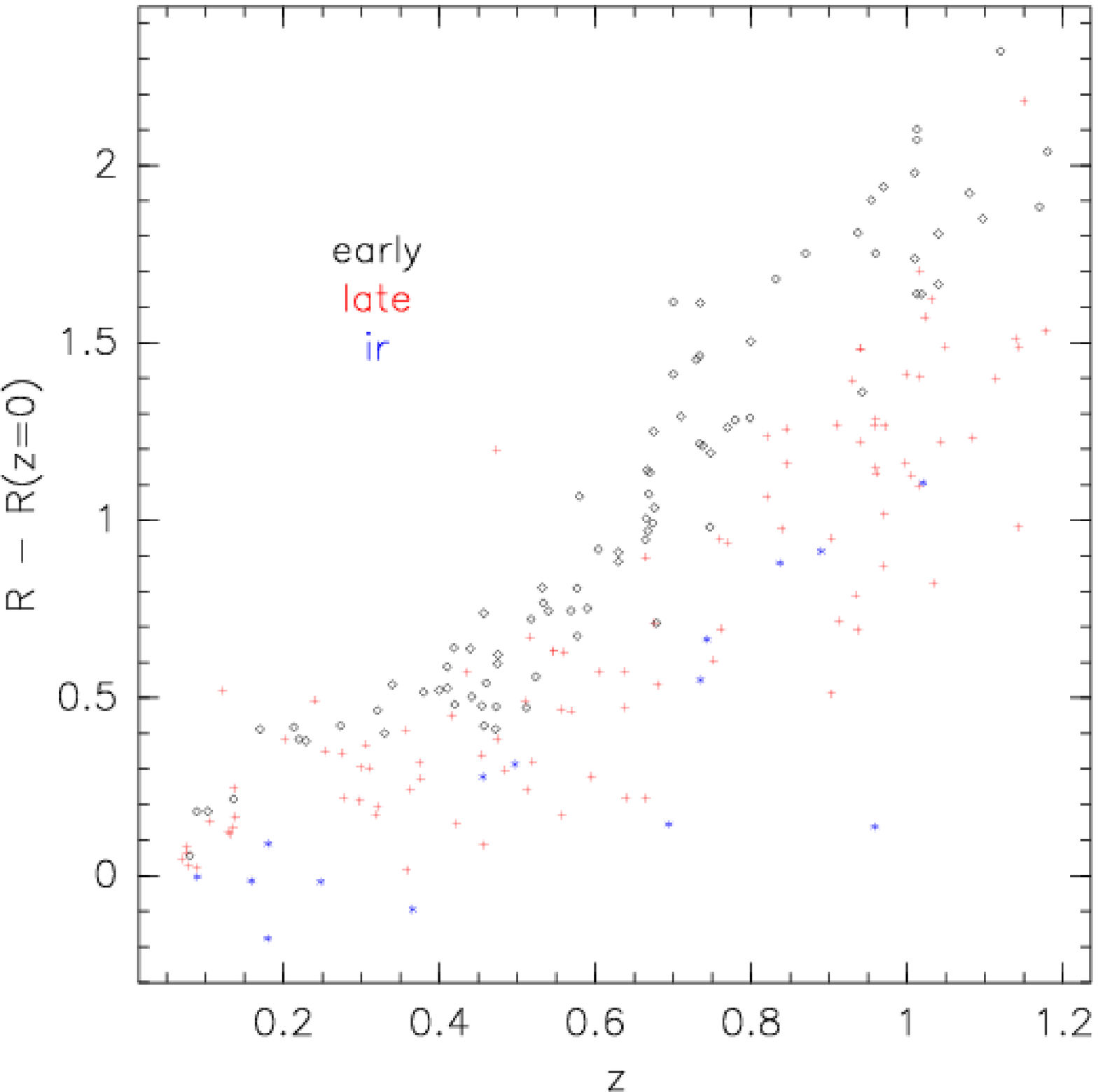}{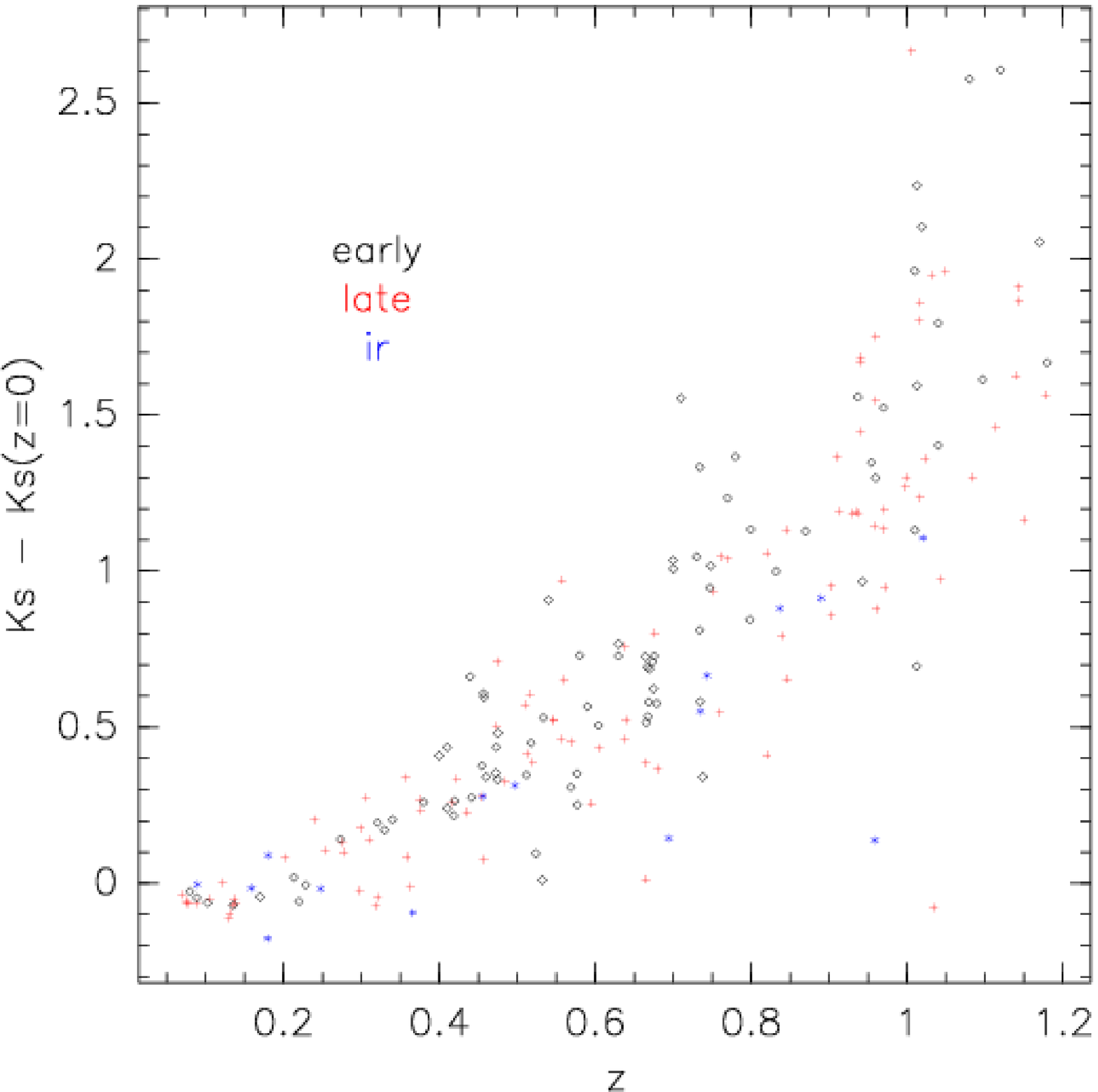}
\caption{K-corrections plotted as a function of redshift for R-band
  (left) and $\rm K_s$-band (right) magnitudes.  The quantity plotted is the
  difference between the observed and k-corrected 
  magnitude.  Early-type galaxies are plotted with diamonds , 
late-type galaxies are plotted with pluses, and
  irregular/starbursts are plotted with asterisks.  \label{f:r_k_kcor}}
\end{figure}

Having established a crude spectral type color selection, we proceeded
by plotting the R and $\rm K_s$ k-corrections as a function of redshift for
each spectral type, shown in Figure \ref{f:r_k_kcor}.  For each
spectral type we determined the k-correction/redshift
correlation. Again, while the dispersion in k-correction is large, 
this resulted in at least an approximate k-correction that was
applied to the sources when computing the priors.  

We list in Table \ref{t:priors} the mean and standard deviation for
the parameters $L_X$, $HR$ (X-ray hardness), X-ray/R-band flux ratio
and X-ray/$\rm K_s$-band flux ratio computed using the prior samples.  The
priors were then modeled as Gaussians with these values for the
Gaussian mean and standard deviation, as discussed in N04  
The full sample is
listed in Tables \ref{t:nsamp} and \ref{t:ssamp}, where we list the
final classification along with the Bayesian odds ratios for the
models.  For example, if the preferred class for a source is
``galaxy'', then the odds ratios are given for galaxy vs. AGN1 and
galaxy vs. AGN2.  For comparison, 30/43 of the OBXF sources from
\citet{HornOBXF} are in our GOODS-N sample, and all but one were
classified as (normal/starburst) galaxies.  
The remaining source
(XID=458) had ``galaxy'' as the highest probability, but only at a factor
of 2.1 higher than the AGN2 probability and hence is only in the
``optimistic'' sample.  
A spectral type was given for 28 of the 30 OBXF sources with matches,
which is also listed in Table \ref{t:nsamp}.  Overall there was good
agreement between our SED types and the spectral types, with 7/9 of the
absorption-line galaxies having an early-type SED classification and
15/17 emission-line galaxies having a late-type SED classification
(the remaining 2 sources were ``composite'' galaxies having both
absorption and emission lines).
Similarly we also list the redshift found in
Georgakakis et al. (2007), if present, for the GOODS-N sources, where
it can be seen there is good overall agreement between the redshifts.

\begin{deluxetable}{crrr}
\tablecaption{Bayesian Prior Parameters \label{t:priors}}
\tablehead{Class & Parameter & Mean & $\sigma$}
\startdata
Normal/Starburst Galaxies & HR & -0.19 & 0.46\\
& $\log L_X$ & 40.6 & 0.7 \\
& $\log F_X/F_{opt}$ & -3.2 & 0.7 \\
& $\log F_X/F_K$ & -3.4 & 0.7\\
AGN2 & HR & 0.16 & 0.37 \\
& $\log L_X$ & 41.1 & 1.1 \\
& $\log F_X/F_{opt}$ & -2.2 & 1.0 \\
& $\log F_X/F_K$ & -2.7 & 0.7\\
AGN1 & HR & -0.51 & 0.05 \\
& $\log L_X$ & 42.9 & 0.4 \\
& $\log F_X/F_{opt}$ & -1.2 & 0.4 \\
& $\log F_X/F_K$ & -1.4 & 0.5\\
\enddata
\end{deluxetable}


\section{Statistical Properties of the Samples}
\begin{figure*}
\plottwo{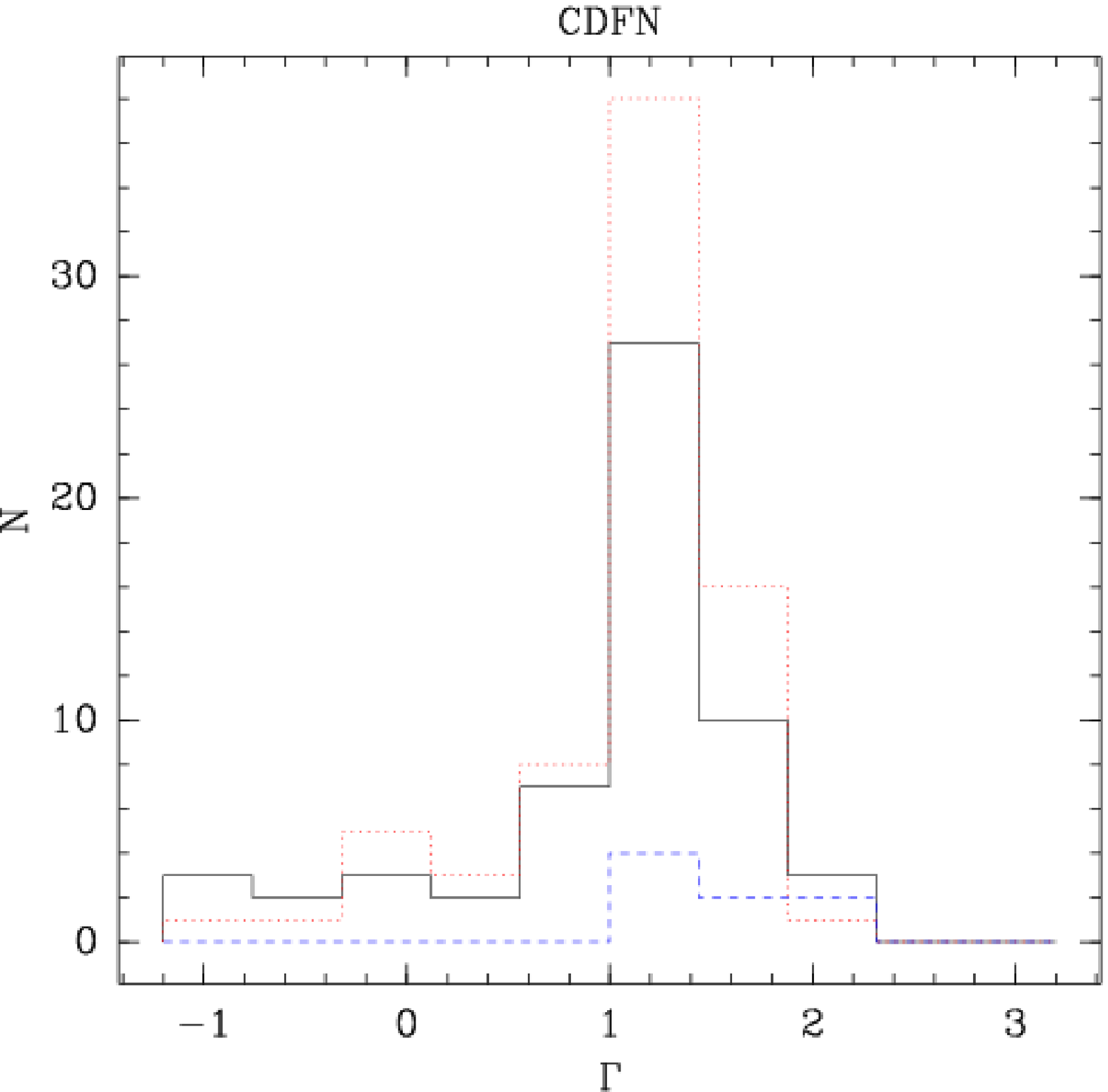}{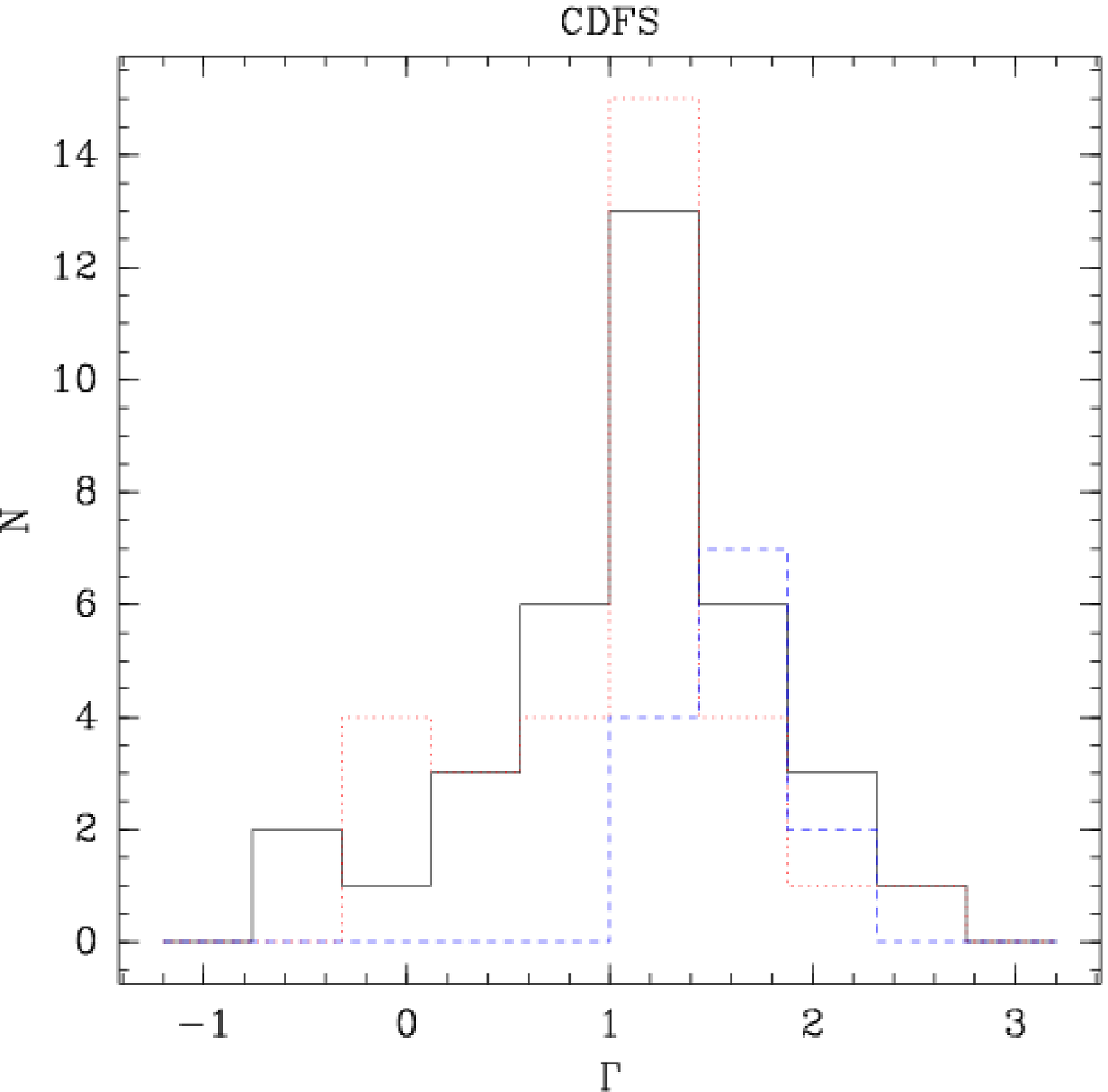}
\caption{Distribution of photon indices for the CDF-N (left) and CDF-S
  (right) sources, binned by galaxy spectral type (black, solid
  line=early-type, 
  red, dotted line=late-type, blue, dashed
  line=starburst/irregular).\label{f:gamma_hist}} 
\end{figure*}

In Figure \ref{f:gamma_hist} we show
histograms of the photon indices, binned separately by spectral
type. The photon index distributions are similar,
peaking at $\Gamma \sim 1.2-1.4$.  However note this is due in part to the
adoption of $\Gamma=1.4$ in Alexander et al. for sources with low
signal-to-noise.  The number of irregular/starburst sources is very
low and is likely to be similar to the rest of the sample.  We list in
Table \ref{t:stats}  the number of
sources in each sample (i.e., divided by field, host galaxy type and
bandpass) along with the mean, standard deviation, minimum and maximum
(k-corrected) magnitude.  We also give the similar statistics for
$\log F_X$ and $\log L_X$.  Note that $\rm K_s$ magnitudes were not always
available.  An immediate
conclusion is that the AGN contribution is most significant for the
irregular/starburst samples since the mean X-ray luminosities are $
\sim 10^{42} \rm \ erg s^{-1}$, the luminosity at which AGN emission
starts to dominate the X-ray band.  The corresponding
values after galaxy selection are listed in Table \ref{t:statsgal}.
\clearpage
\begin{deluxetable}{llllllll}
\tablecaption{X-ray and Optical Statistical Properties \label{t:stats}}
\tablehead{Field & Band & SED Type & N & Mean & $\sigma$ & Min. & Max.}
\startdata
CDFN & B & early & 57  & 22.26 & 1.07 & 19.13 & 24.27 \\
CDFN & R & early & 57  & 21.29 & 1.16 & 18.03 & 23.45 \\
CDFN & Ks & early & 57  & 18.90 & 0.94 & 17.09 & 21.32 \\
CDFN & J & early & 57  & 20.32 & 1.02 & 17.45 & 22.32 \\
CDFN & logFX & early & 57  & -16.16 & 0.48 & -17.38 & -14.64 \\
CDFN & logLX & early & 57  & 40.96 & 0.70 & 38.91 & 42.42 \\
CDFN & B & late & 73  & 21.94 & 1.16 & 19.64 & 24.66 \\
CDFN & R & late & 73  & 21.20 & 1.15 & 18.59 & 23.81 \\
CDFN & Ks & late & 73  & 19.18 & 0.85 & 17.62 & 21.69 \\
CDFN & J & late & 73  & 20.41 & 0.95 & 18.09 & 22.53 \\
CDFN & logFX & late & 73  & -16.03 & 0.79 & -17.39 & -13.64 \\
CDFN & logLX & late & 73  & 41.01 & 1.03 & 38.89 & 43.33 \\
CDFN & B & ir & 8  & 22.11 & 1.48 & 19.91 & 24.54 \\
CDFN & R & ir & 8  & 21.73 & 1.38 & 19.84 & 23.88 \\
CDFN & Ks & ir & 8  & 20.20 & 1.55 & 18.27 & 22.47 \\
CDFN & J & ir & 8  & 21.18 & 1.39 & 19.69 & 23.13 \\
CDFN & logFX & ir & 8  & -15.27 & 1.27 & -16.74 & -13.81 \\
CDFN & logLX & ir & 8  & 41.73 & 1.52 & 39.72 & 43.54 \\
CDFS & B & early & 36  & 21.62 & 1.48 & 18.11 & 25.95 \\
CDFS & R & early & 36  & 20.54 & 2.08 & 16.25 & 27.61 \\
CDFS & Ks & early & 29  & 18.73 & 1.22 & 16.07 & 21.59 \\
CDFS & J & early & 29  & 19.86 & 1.38 & 16.27 & 22.80 \\
CDFS & logFX & early & 36  & -15.80 & 0.56 & -16.88 & -14.51 \\
CDFS & logLX & early & 36  & 41.32 & 0.65 & 39.65 & 42.77 \\
CDFS & B & late & 32  & 22.03 & 1.68 & 17.34 & 25.95 \\
CDFS & R & late & 32  & 21.08 & 1.70 & 16.71 & 24.97 \\
CDFS & Ks & late & 26  & 19.08 & 1.27 & 15.94 & 21.82 \\
CDFS & J & late & 26  & 20.25 & 1.50 & 16.03 & 22.39 \\
CDFS & logFX & late & 32  & -15.83 & 0.63 & -16.74 & -14.39 \\
CDFS & logLX & late & 32  & 41.31 & 0.83 & 39.91 & 42.90 \\
CDFS & B & ir & 13  & 21.45 & 1.55 & 19.24 & 24.65 \\
CDFS & R & ir & 13  & 21.00 & 1.69 & 18.00 & 24.30 \\
CDFS & Ks & ir & 7  & 20.76 & 1.33 & 19.58 & 22.89 \\
CDFS & J & ir & 7  & 21.45 & 1.30 & 19.98 & 23.64 \\
CDFS & logFX & ir & 13  & -14.90 & 1.13 & -16.27 & -13.36 \\
CDFS & logLX & ir & 13  & 41.98 & 1.78 & 39.44 & 44.03 \\
\enddata
\end{deluxetable}

\begin{deluxetable}{llllllll}
\tablecaption{X-ray and Optical Statistical Properties , Galaxies only\label{t:statsgal}}
\tablehead{Field & Band & SED Type & N & Mean & $\sigma$ & Min. & Max.}
\startdata
CDFN & B & early & 27  & 21.54 & 0.91 & 19.13 & 22.91 \\
CDFN & R & early & 27  & 20.50 & 1.00 & 18.03 & 21.88 \\
CDFN & Ks & early & 27  & 18.31 & 0.61 & 17.09 & 19.56 \\
CDFN & J & early & 27  & 19.60 & 0.81 & 17.45 & 20.75 \\
CDFN & logFX & early & 27  & -16.32 & 0.31 & -16.93 & -15.60 \\
CDFN & logLX & early & 27  & 40.63 & 0.66 & 38.91 & 41.47 \\
CDFN & B & late & 36  & 21.45 & 1.01 & 19.64 & 23.65 \\
CDFN & R & late & 36  & 20.73 & 1.04 & 18.59 & 22.87 \\
CDFN & Ks & late & 36  & 18.95 & 0.66 & 17.78 & 20.59 \\
CDFN & J & late & 36  & 20.02 & 0.80 & 18.09 & 21.45 \\
CDFN & logFX & late & 36  & -16.39 & 0.32 & -17.12 & -15.72 \\
CDFN & logLX & late & 36  & 40.41 & 0.64 & 38.89 & 41.41 \\
CDFN & logFX & ir & 1  \\
CDFN & logLX & ir & 1  \\
CDFS & B & early & 13  & 20.47 & 1.01 & 18.11 & 21.52 \\
CDFS & R & early & 13  & 19.15 & 1.38 & 16.25 & 20.96 \\
CDFS & Ks & early & 11  & 17.63 & 0.76 & 16.07 & 18.74 \\
CDFS & J & early & 11  & 18.58 & 1.07 & 16.27 & 20.09 \\
CDFS & logFX & early & 13  & -15.97 & 0.46 & -16.59 & -15.18 \\
CDFS & logLX & early & 13  & 40.92 & 0.62 & 39.65 & 41.85 \\
CDFS & B & late & 10  & 20.52 & 1.61 & 17.34 & 21.83 \\
CDFS & R & late & 10  & 19.70 & 1.58 & 16.71 & 21.23 \\
CDFS & Ks & late & 9  & 18.08 & 1.21 & 15.94 & 19.54 \\
CDFS & J & late & 9  & 18.92 & 1.56 & 16.03 & 20.11 \\
CDFS & logFX & late & 10  & -15.94 & 0.43 & -16.36 & -15.07 \\
CDFS & logLX & late & 10  & 40.62 & 0.49 & 39.91 & 41.43 \\
CDFS & B & ir & 2  & 21.03 & 0.79 & 20.47 & 21.59 \\
CDFS & R & ir & 2  & 20.63 & 0.74 & 20.11 & 21.15 \\
CDFS & Ks & ir & 2  & 19.66 & 0.11 & 19.58 & 19.74 \\
CDFS & J & ir & 2  & 20.32 & 0.48 & 19.98 & 20.66 \\
CDFS & logFX & ir & 2  & -16.24 & 0.05 & -16.27 & -16.20 \\
CDFS & logLX & ir & 2  & 40.35 & 0.48 & 40.01 & 40.69 \\
\enddata
\end{deluxetable}

\clearpage
\section{X-ray Flux Ratios}
\begin{figure*}
\plottwo{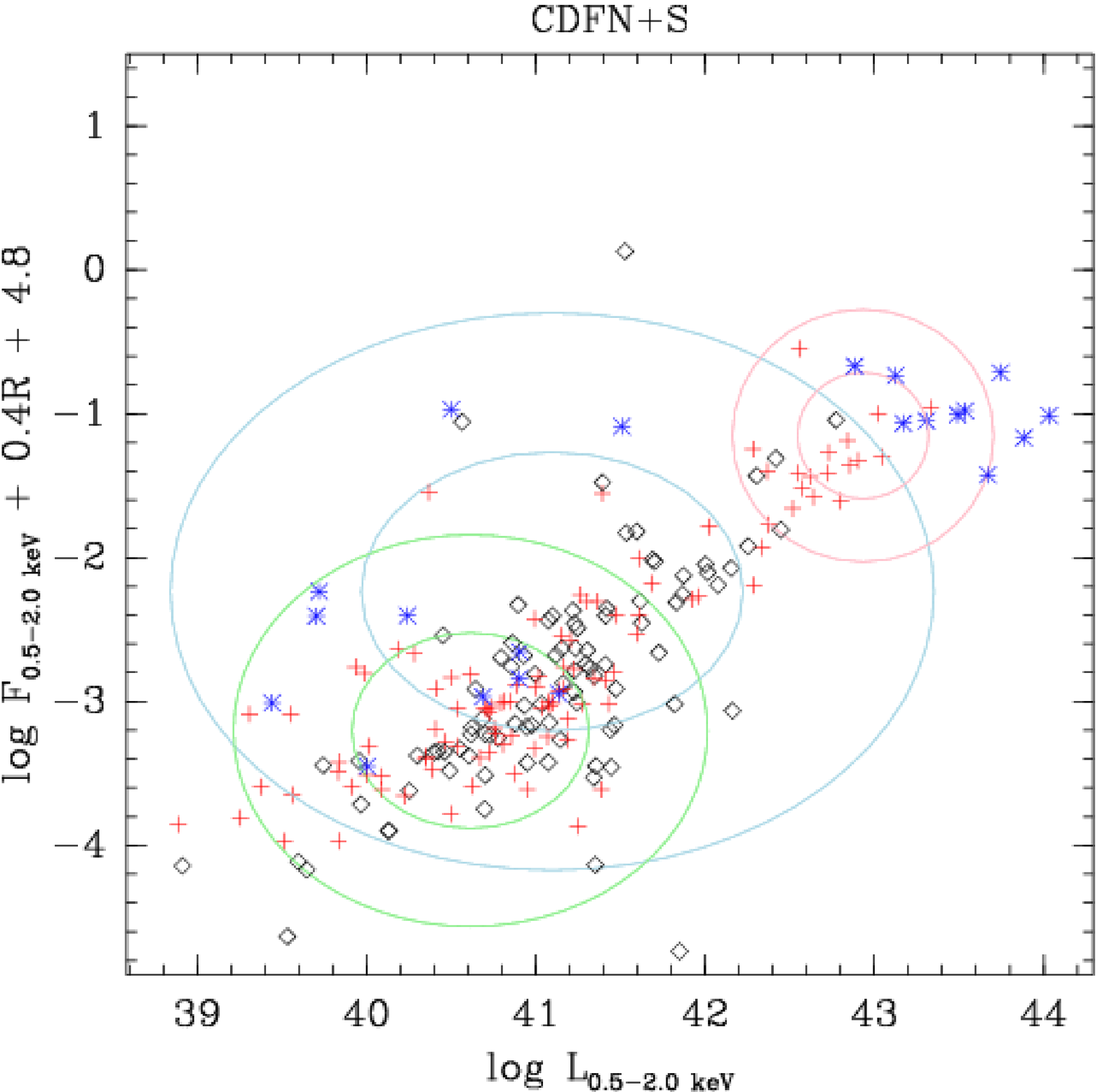}{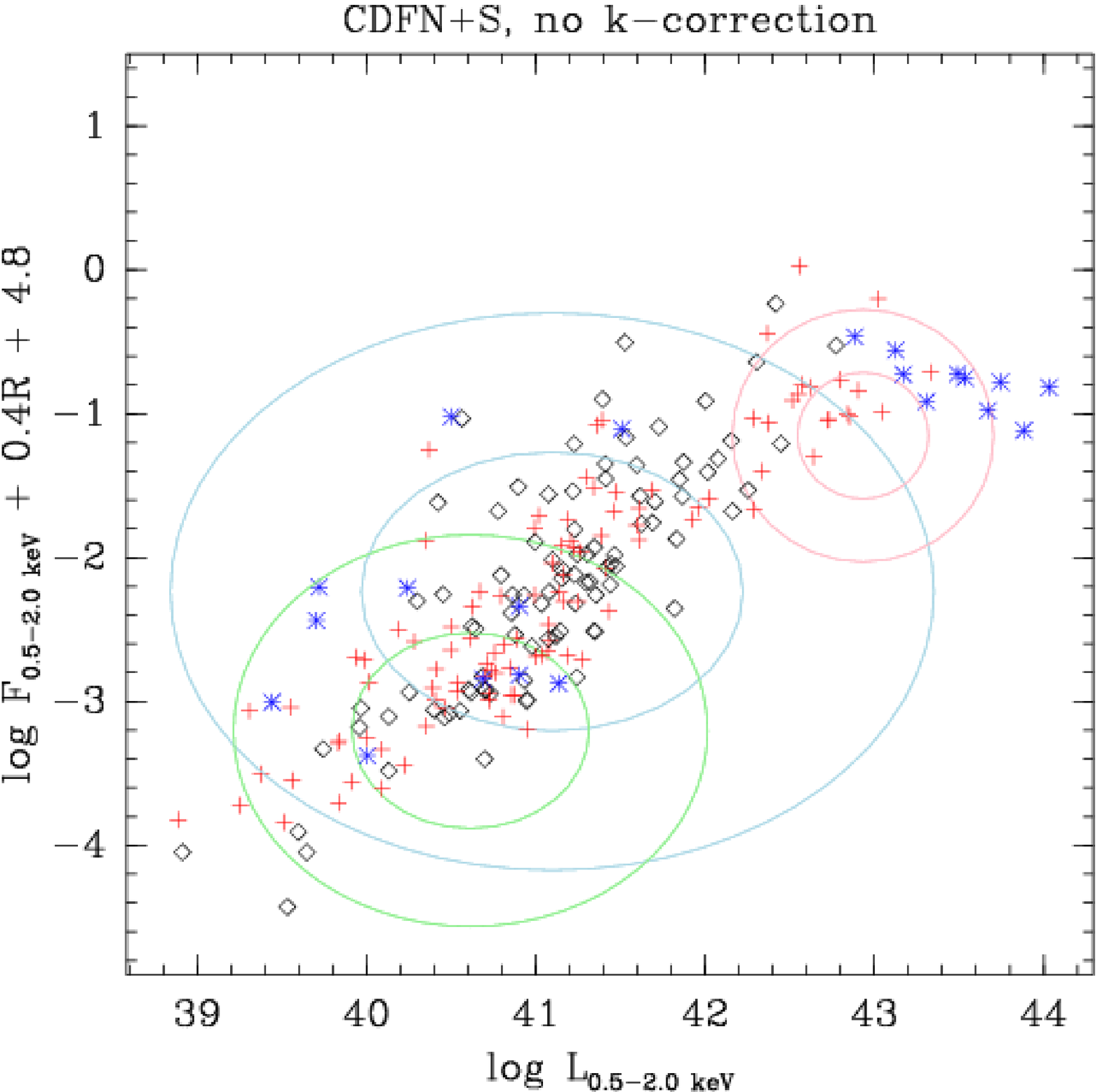}
\caption{The X-ray/R-band flux ratio plotted as a function
  of X-ray luminosity, including (left) and not including (right)
  k-corrections.   Early-type, late-type and irregular/starburst SED sources
  are marked with (black) diamonds, (red) pluses, and (blue) asterisks,
  respectively. The ellipses show the 1 and 2$\sigma$ intervals
  computed for the prior probability distributions for galaxies, AGN1
  and AGN2 for $\log L_X$ and $\log F_X/F_R$. Note that these
  intervals correspond to standard deviations computed separately for
  these parameters (i.e.,
  the widths of the ellipses give the 68\% and 95\% probability
  intervals for $\log L_X$ for each type), which assumes no
  correlation between $L_X$ and $F_X/F_R$.  \label{f:R_lx_vs_fluxrat}}
\end{figure*}

\begin{figure*}
\plottwo{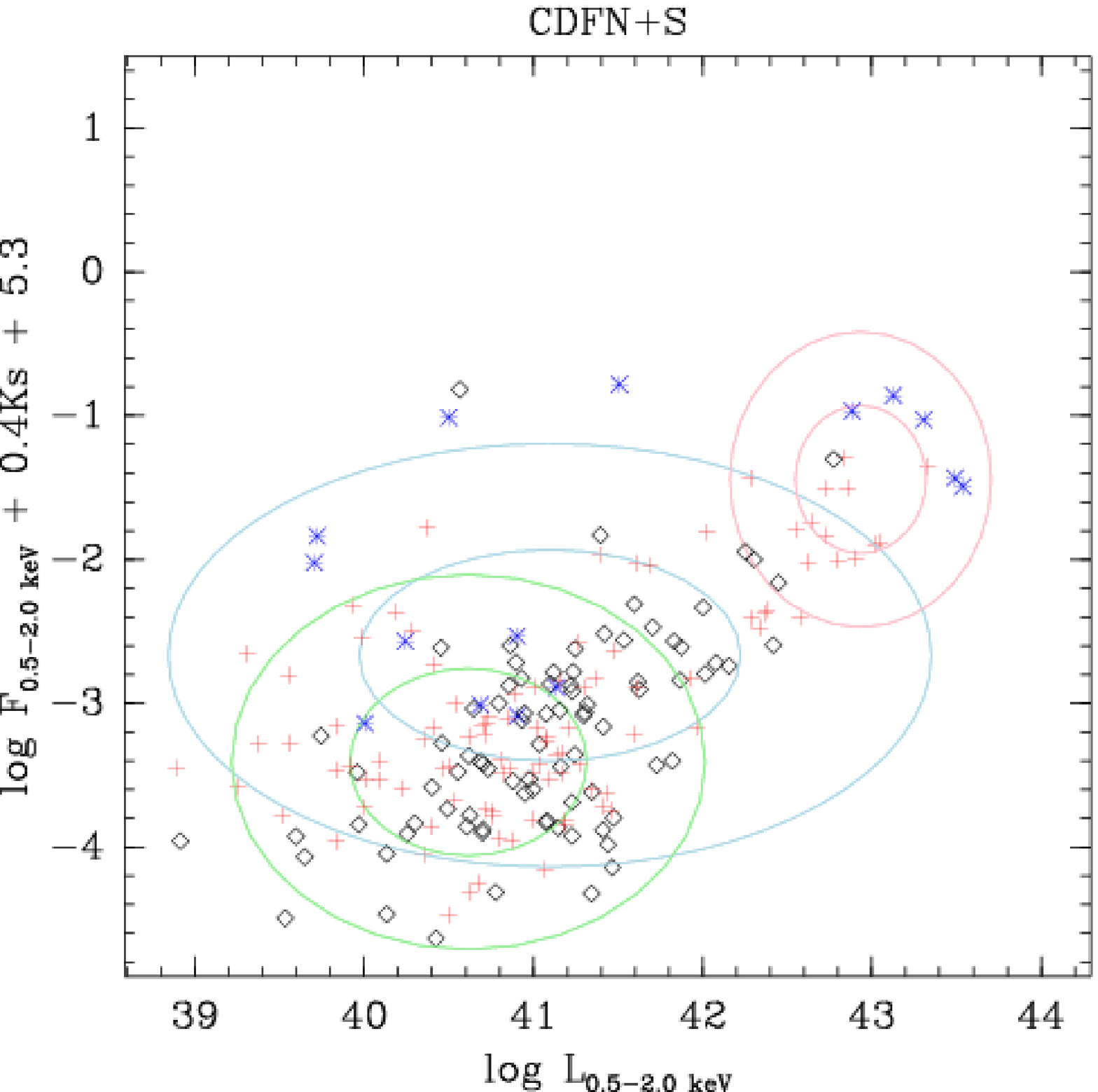}{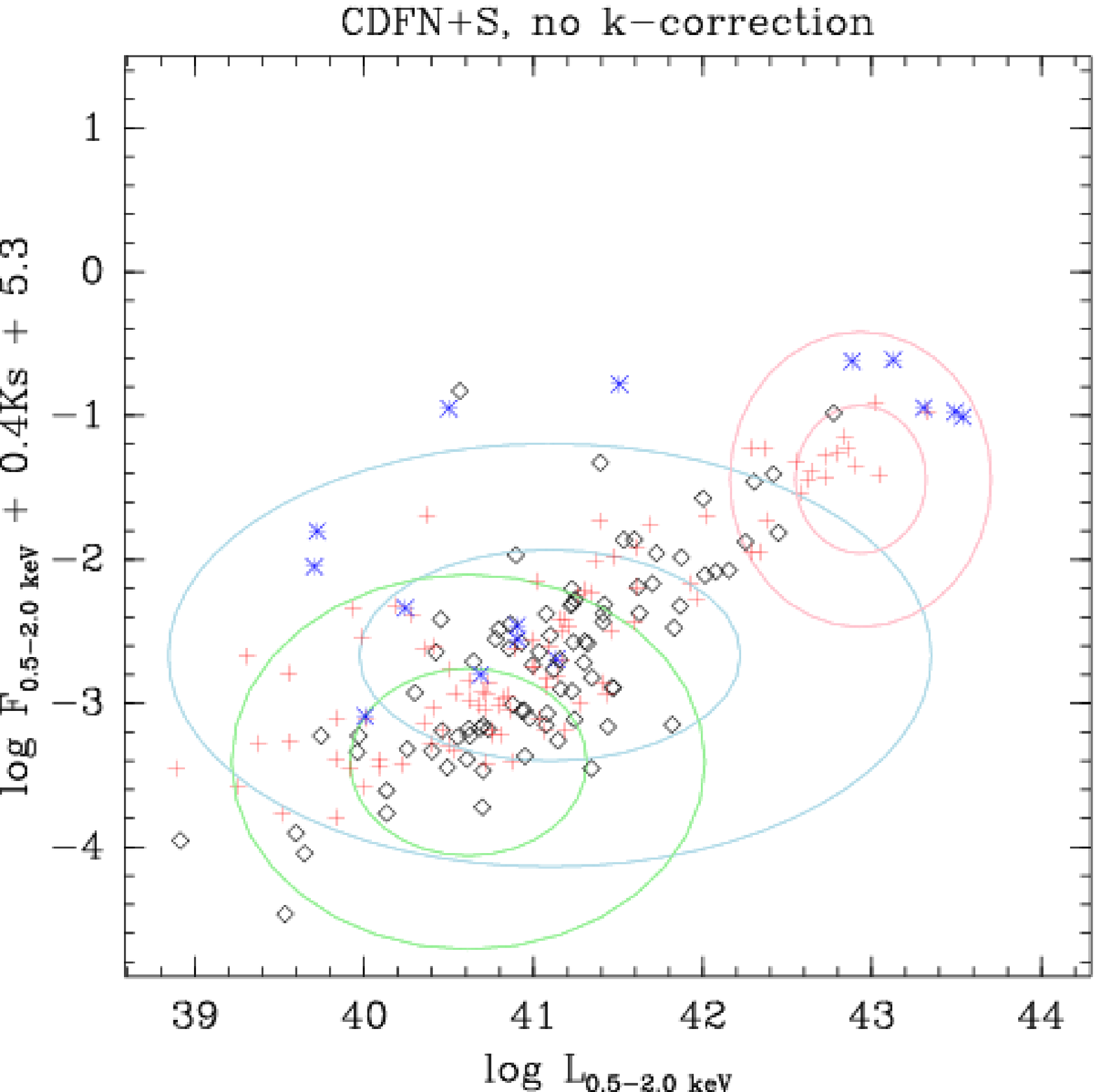}
\caption{The X-ray/$\rm K_s$-band flux ratio plotted as a function
  of X-ray luminosity, including (left) and not including (right)
  k-corrections.  Symbols and ellipses are as in Figure
  \ref{f:R_lx_vs_fluxrat}. \label{f:Ks_lx_vs_fluxrat}}
\end{figure*}

\begin{figure*}
\plottwo{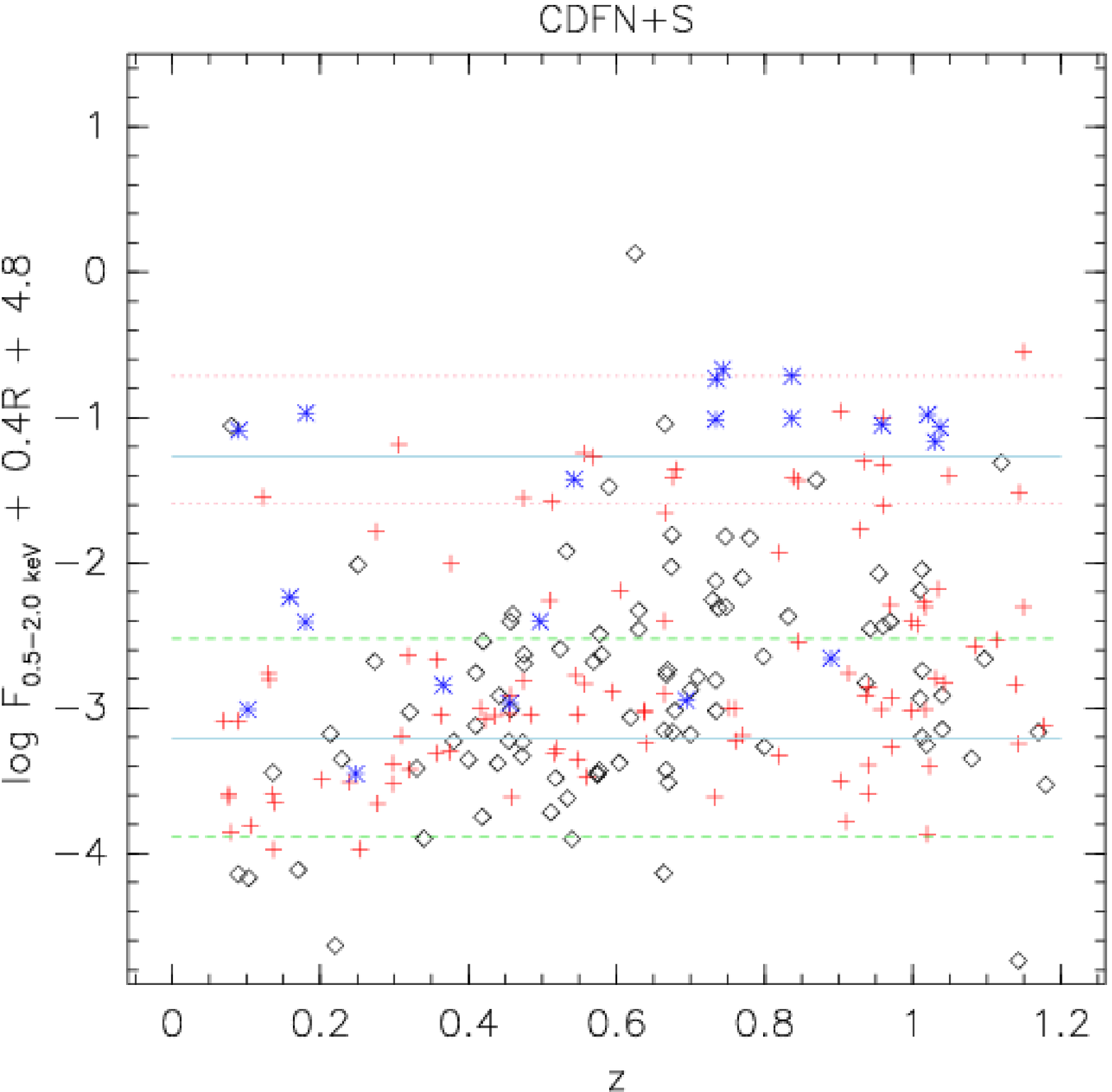}{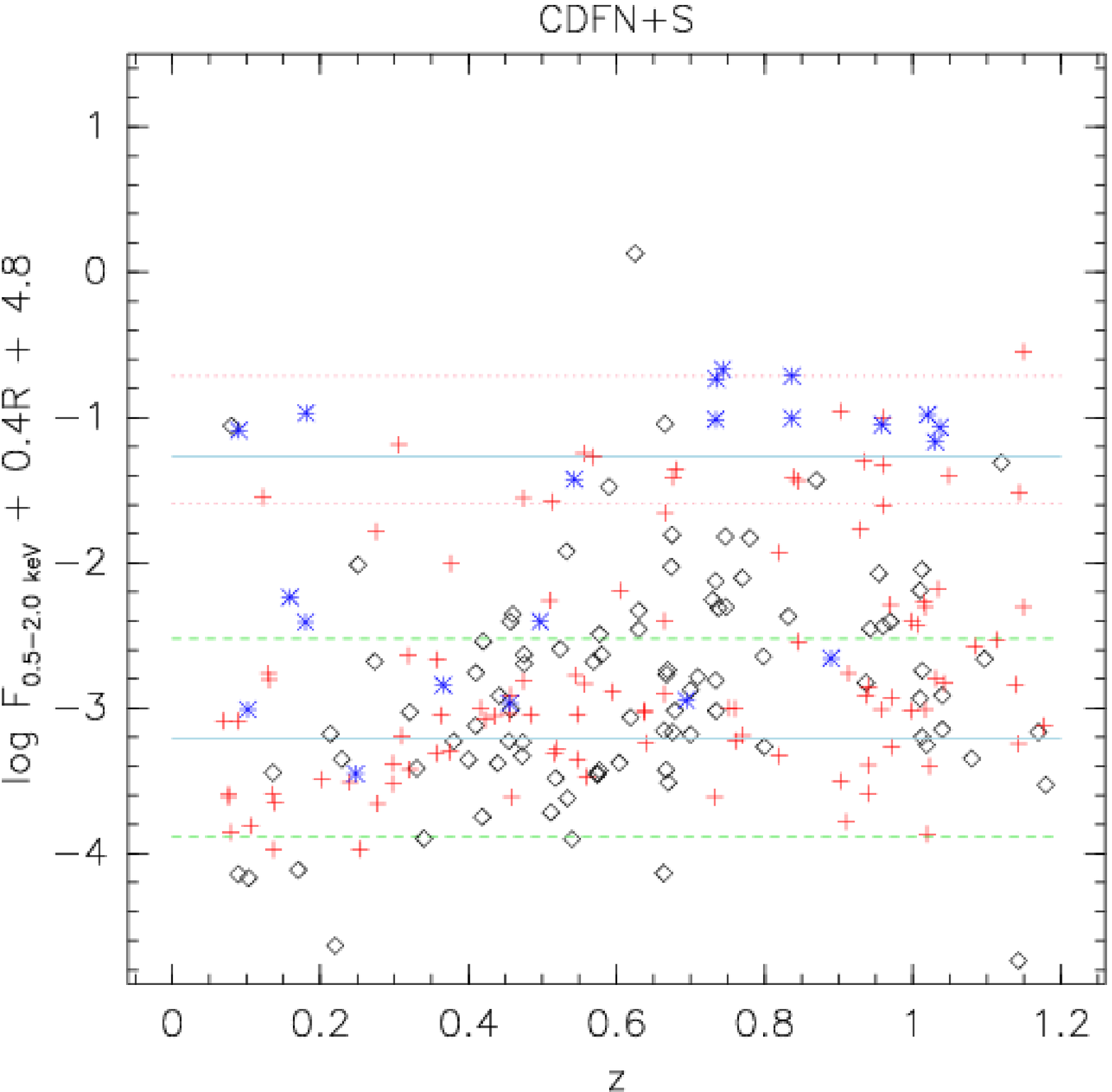}
\caption{The X-ray/R-band flux ratio plotted as a function of
  redshift, including (left) and not including (right)
  k-corrections.  Early-type, late-type and irregular/starburst SED sources
  are marked with (black) diamonds, (red) pluses, and (blue) asterisks,
  respectively.   Horizontal lines are drawn showing the 1$\sigma$
  probability intervals for the flux ratio for  
    galaxies (green dashed lines), AGN1 (red dotted lines) and AGN2
  (blue solid lines). \label{f:R_z_vs_fluxrat}}
\end{figure*}

\begin{figure*}
\plottwo{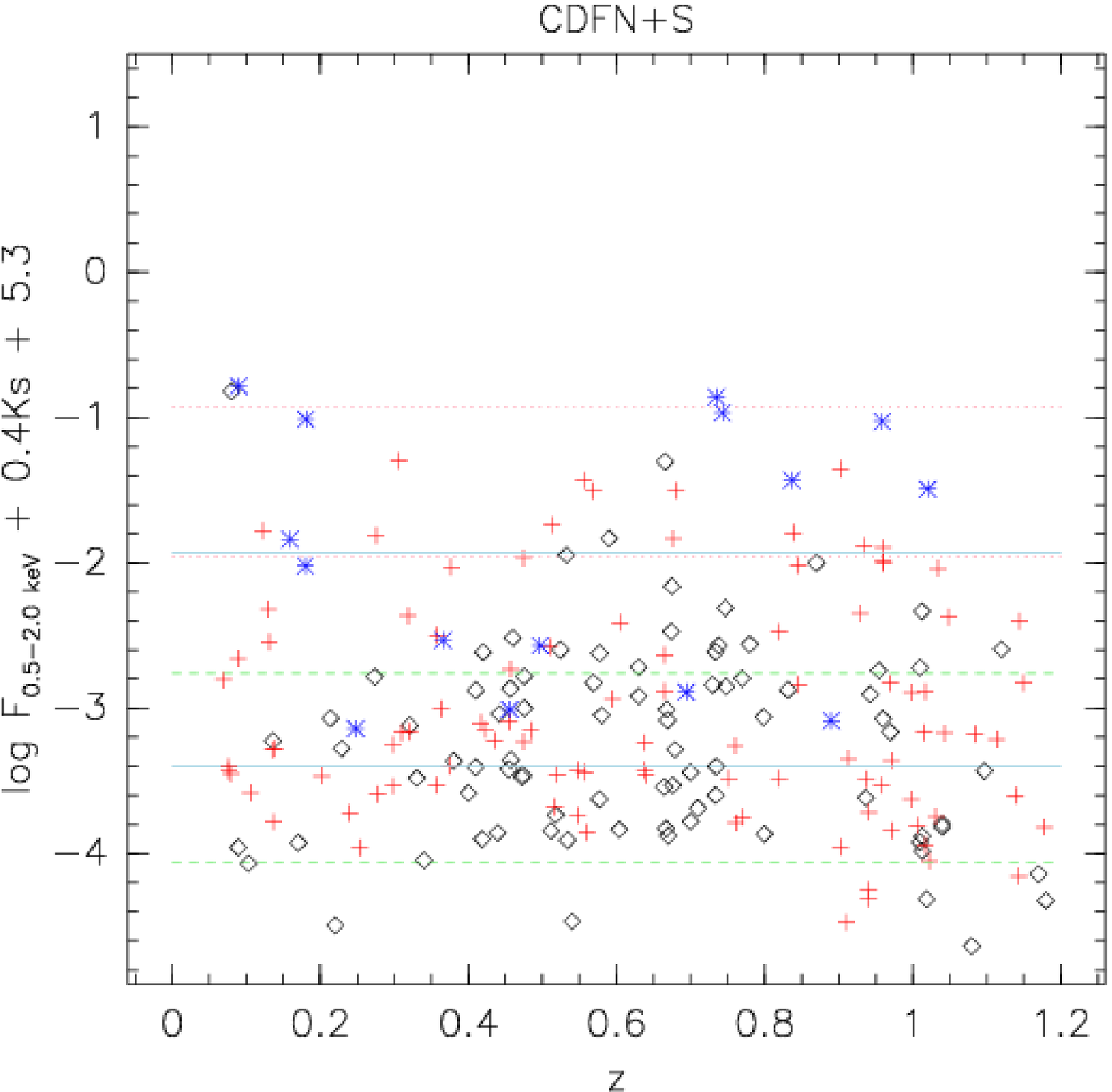}{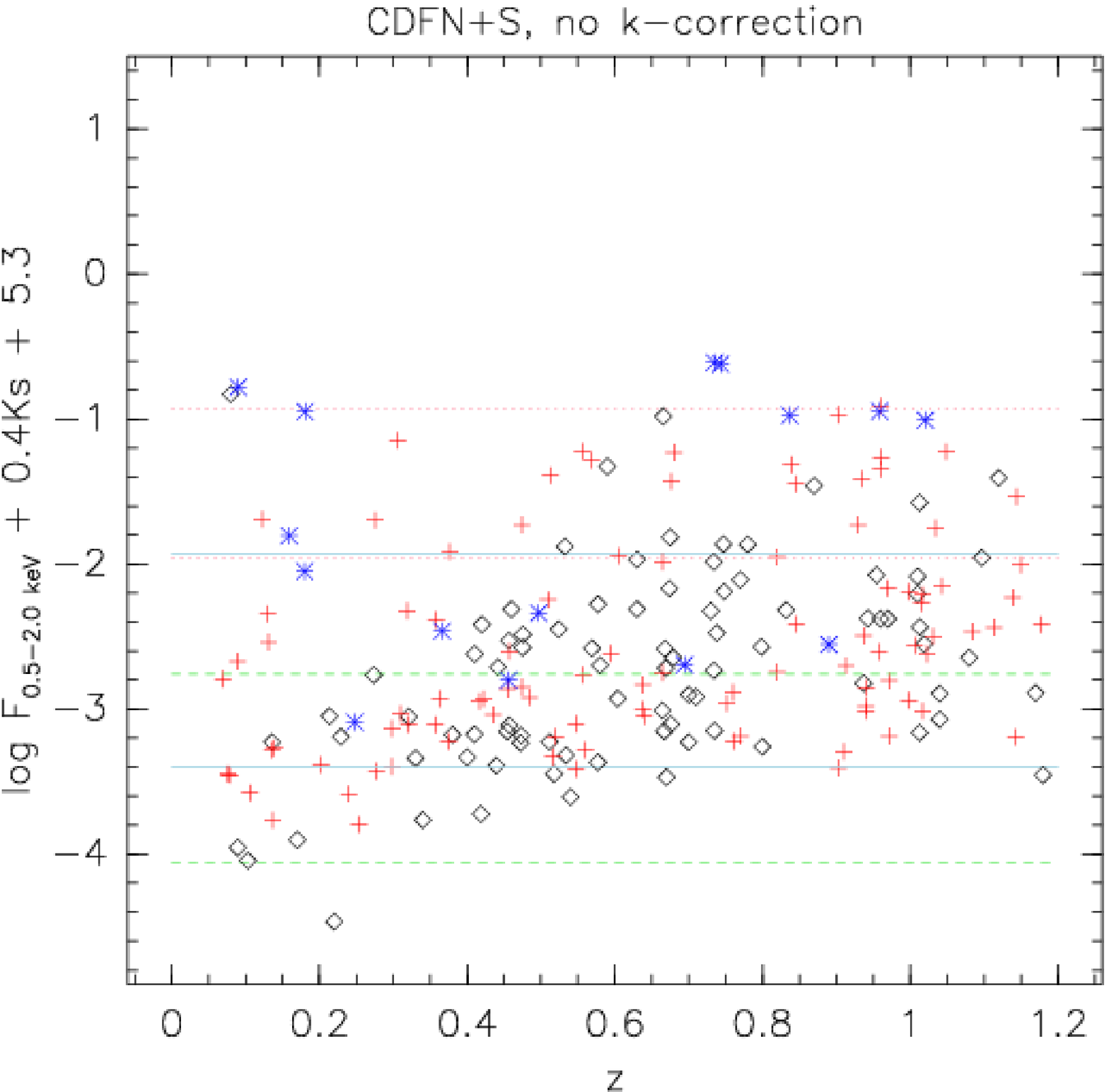}
\caption{The X-ray/$\rm K_s$-band flux ratio plotted as a function of
  redshift, including (left) and not including (right)
  k-corrections.  Early-type, late-type and irregular/starburst SED sources
  are marked with (black) diamonds, (red) pluses, and (blue) asterisks,
  respectively.   Horizontal lines are drawn showing the 1$\sigma$
  probability intervals for the flux ratio for  
  galaxies (green dashed lines), AGN1 (red dotted lines) and AGN2
  (blue solid lines). \label{f:Ks_z_vs_fluxrat}}
\end{figure*}

Since X-ray/R-band and X-ray/$\rm K_s$-band flux ratios are used in selection
criteria for this paper, we discuss here the potential
impact of any luminosity or redshift dependence of these quantities.
In Figures \ref{f:R_lx_vs_fluxrat} and \ref{f:Ks_lx_vs_fluxrat}
 we plot the X-ray/R-band and X-ray/$\rm K_s$-band flux
ratios as a function of luminosity, before and after applying
k-corrections. 
 Figures \ref{f:R_z_vs_fluxrat} and \ref{f:Ks_z_vs_fluxrat} 
show the flux ratios plotted as a
function of redshift.  In the luminosity versus flux ratio plots
we show the regions resulting from our Bayesian prior analysis.
These ellipses show the 1 and 2$\sigma$ regions where the $\sigma$ value
is based on the standard deviation of the parent populations.  In
other words, the width and height of the 1$\sigma$ regions are the
standard deviations of the X-ray luminosity and given flux ratio for
the corresponding class (galaxy, AGN1 or AGN2), and correspond to the
68\% and 95\% probability intervals for that parameter.  However, this
assumes that the X-ray luminosity and flux ratios are not 
correlated,  and since this is not the case, these regions should {\it
  not} be interpreted as joint confidence regions.  The intention here
is to simply show the regions in the $\log L_X - \log F_X/F_R$ and
$\log L_X - \log F_X/F_{K_s}$ planes from which our normal/starburst
samples are being selected.

In the redshift versus flux ratio plots we show the 1$\sigma$
regions for the flux ratios, as in luminosity/flux ratio plots, except
in this case the regions are simply marked with horizontal lines
(i.e., since redshift is not a selection criterion).  From these plots
we conclude that any evolution in the flux ratios with redshift is
dominated by the k-corrections and the scatter in the flux ratios.  In
other words, after k-correction, the flux ratios are consistent with
no redshift dependence.  This emphasizes the importance of
k-correcting the data prior to utilizing flux ratios as a selection
criterion.  On the other hand there is clearly a luminosity dependence
in the flux ratios.  This is due in large part to the increased
prevalence of AGN activity at higher X-ray luminosities 
\citep[e.g.,][]{Fiore03, Barger2005}.
The correlation between
X-ray luminosity and the flux ratios mitigates the benefit of
including both luminosity and the ratios as selection criteria.
However, in practice there are a significant number of sources in
the luminosity range $10^{41-42}$ ergs s$^{-1}$, consistent with
either AGN or normal/starburst galaxies, where the X-ray/R-band
and/or X-ray/$\rm K_s$-band flux ratio is within the 1$\sigma$ region for
AGN2 galaxies.  Thus, including the X-ray/optical and X-ray/K-band
flux ratios as a selection criteria should improve the separation, and
hence selection, of normal/starburst galaxies from type II AGN.

\bibliography{goods_xlf_emulateapj}



\clearpage
\LongTables 
\begin{landscape}
\pagestyle{empty}
\begin{deluxetable}{rrrrrrrrrrrrl}
\tablewidth{0pt}
\tabletypesize{\footnotesize}
\tablecaption{CDF-N X-ray Sample}
\tablehead{XID & z & Flag & $F_{X}$ & Class &
  log(odds1) & log(odds2) & $\log(L_{X})$ & 
  log($\rm F_X/F_R$) &  log($\rm F_X/F_K$) & 
  Type & OBXF ID& $z_{\rm G07}$}
\startdata
48& 1.01& 4& 0.27& agn2& 38.53& 0.90& 42.16 & -1.97 & -2.99 & early &\\
55& 0.64& 1& 0.12& & \ldots& \ldots& 41.30 & -2.78 & -3.02 & late &\\
56& 0.13& 4& 0.20& gal& 14.16& 0.51& 39.94 & -2.72 & -2.30 & late & & 0.08\tablenotemark{a$\dagger$}\\
57& 0.38& 2& 0.15& gal& 12.78& 1.36& 40.87 & -3.21 & -3.33 & late & E (0.38) & 0.33\tablenotemark{a$\dagger$}\\
60& 0.42& 4& 0.05& & \ldots& \ldots& 40.54 & -2.42 & -2.52 & early &\\
62& 0.22& 4& 0.03& gal& 18.12& 1.21& 39.58 & -4.56 & -4.44 & early & A (0.09)\\
67& 0.64& 1& 0.14& gal& 8.68& 0.98& 41.39 & -2.87 & -3.30 & late & & 0.64\tablenotemark{a}\\
72& 0.94& 1& 0.11& & \ldots& \ldots& 41.68 & -2.36 & -2.97 & late &\\
78& 0.75& 1& 0.18& agn2& 1.39& 1.14& 41.66 & -1.72 & -2.24 & early & & 0.75\tablenotemark{a}\\
81& 0.38& 4& 0.10& gal& 10.92& 0.84& 40.70 & -3.12 & -3.28 & early &\\
82& 0.68& 1& 0.14& & \ldots& \ldots& 41.44 & -2.58 & -2.87 & early &\\
87& 0.14& 1& 0.12& gal& 20.57& 1.31& 39.76 & -3.39 & -3.20 & early & A (0.14)\\
90& 1.14& 1& 0.16& agn2& 13.13& 0.84& 42.06 & -2.09 & -2.88 & late &\\
93& 0.28& 1& 5.08& agn2& 1.80& 2.12& 42.08 & -1.70 & -1.75 & late &\\
101& 0.45& 1& 0.08& gal& 12.60& 1.19& 40.81 & -3.08 & -3.30 & early & A (0.45) & 0.45\tablenotemark{a}\\
103& 0.97& 2& 0.28& agn2& 1.51& 0.66& 42.13 & -2.05 & -2.62 & late & & 0.81\tablenotemark{a$\dagger$}\\
105& 0.33& 4& 0.03& gal& 14.66& 0.96& 40.03 & -3.31 & -3.41 & early &\\
110& 1.01& 4& 0.34& agn2& 0.89& 1.05& 42.26 & -1.98 & -2.54 & early &\\
111& 0.52& 1& 0.04& gal& 13.33& 1.15& 40.64 & -3.17 & -3.57 & late & E (0.52) & 0.52\tablenotemark{a}\\
113& 0.84& 2& 1.59& agn1& 4.98& 1.93& 42.74 & -1.29 & -1.90 & late & & 0.84\tablenotemark{a}\\
114& 0.53& 1& 0.04& gal& 17.54& 0.65& 40.63 & -3.22 & -3.53 & early &\\
115& 0.68& 1& 4.67& agn1& 5.48& 1.43& 42.98 & -1.21 & -1.38 & late & & 0.68\tablenotemark{a}\\
119& 0.47& 1& 0.08& gal& 12.31& 1.14& 40.83 & -3.11 & -3.35 & early & E (0.47) & 0.47\tablenotemark{a}\\
120& 0.69& 1& 0.09& gal& 7.17& 0.51& 41.27 & -2.78 & -2.75 & ir &\\
126& 0.77& 4& 0.03& gal& 10.68& 0.93& 40.90 & -3.01 & -3.60 & late & & 0.84\tablenotemark{a}\\
132& 0.64& 1& 0.07& gal& 11.36& 0.67& 41.06 & -3.01 & -3.26 & late & E (0.65) & 0.64\tablenotemark{a}\\
136& 0.47& 1& 0.05& gal& 13.55& 1.16& 40.65 & -3.20 & -3.37 & early & A (0.47) & 0.47\tablenotemark{a}\\
138& 0.48& 1& 0.07& gal& 10.81& 0.80& 40.80 & -2.92 & -3.05 & late & E (0.48) & 0.48\tablenotemark{a}\\
139& 0.93& 4& 0.76& agn1& 2.87& 0.75& 42.52 & -1.60 & -2.20 & late & & 1.01\tablenotemark{a}\\
142& 0.75& 1& 0.29& agn2& 7.05& 0.97& 41.87 & -2.02 & -2.60 & early &\\
150& 0.63& 4& 0.13& agn2& 23.99& 1.38& 41.35 & -1.85 & -2.26 & early &\\
158& 1.01& 2& 0.17& agn2& 20.44& 0.49& 41.96 & -2.16 & -3.33 & early &\\
160& 0.82& 4& 0.12& & \ldots& \ldots& 41.57 & -2.73 & -2.91 & late &\\
166& 0.46& 1& 0.04& & \ldots& \ldots& 40.50 & -2.78 & -2.63 & late & & 0.46\tablenotemark{a}\\
169& 0.31& 4& 0.10& gal& 13.84& 1.10& 40.48 & -3.08 & -3.09 & late & A (0.84)\\
170& 0.63& 4& 0.20& agn2& 11.44& 0.77& 41.53 & -2.13 & -2.62 & early &\\
177& 1.02& 1& 0.25& agn2& 2.73& 0.66& 42.14 & -2.07 & -2.99 & late & & 1.01\tablenotemark{a}\\
180& 0.46& 1& 0.23& gal& 9.37& 1.05& 41.25 & -2.97 & -3.35 & early & & 0.46\tablenotemark{a}\\
187& 0.94& 4& 0.04& gal& 14.53& 0.49& 41.23 & -2.80 & -3.68 & late &\\
188& 1.15& 5& 0.07& agn2& 4.05& 0.88& 41.71 & -1.93 & -2.47 & late &\\
189& 0.41& 1& 0.17& & \ldots& \ldots& 41.01 & -2.57 & -2.71 & early &\\
194& 0.56& 1& 1.62& agn1& 6.01& 2.94& 42.30 & -1.19 & -1.41 & late & & 0.56\tablenotemark{a}\\
197& 0.08& 1& 0.05& gal& 28.79& 0.90& 38.90 & -3.81 & -3.43 & late & E (0.08) & 0.08\tablenotemark{a}\\
200& 0.97& 1& 0.04& & \ldots& \ldots& 41.31 & -2.72 & -3.18 & late &\\
203& 1.14& 2& 0.03& gal& 10.14& 0.98& 41.27 & -3.02 & -3.96 & late &\\
209& 0.51& 2& 0.23& & \ldots& \ldots& 41.36 & -2.13 & -2.47 & late & & 0.51\tablenotemark{a}\\
210& 0.70& 4& 0.09& gal& 8.43& 0.87& 41.30 & -2.71 & -3.30 & early &\\
211& 0.76& 4& 0.06& gal& 8.76& 0.51& 41.23 & -2.82 & -3.10 & late & & 0.85\tablenotemark{a}\\
212& 0.94& 1& 0.13& & \ldots& \ldots& 41.77 & -2.28 & -2.76 & early &\\
214& 1.04& 4& 0.03& gal& 8.63& 0.75& 41.27 & -2.93 & -3.63 & early &\\
215& 1.01& 2& 0.03& & \ldots& \ldots& 41.17 & -2.22 & -3.63 & late &\\
217& 0.54& 4& 0.04& gal& 24.31& 0.79& 40.63 & -3.38 & -3.97 & early &\\
218& 0.09& 1& 0.10& gal& 21.17& 0.65& 39.31 & -3.05 & -2.65 & late & E (0.09)\\
219& 0.70& 4& 0.03& gal& 10.65& 0.85& 40.76 & -3.02 & -3.64 & early &\\
222& 0.77& 4& 0.59& agn2& 3.17& 1.29& 42.21 & -1.88 & -2.60 & early & & 0.86\tablenotemark{a}\\
227& 0.52& 4& 0.04& gal& 13.17& 1.03& 40.57 & -3.14 & -3.34 & late & E (0.56) & 0.56\tablenotemark{a}\\
230& 1.01& 1& 0.08& gal& 9.67& 1.11& 41.62 & -2.99 & -3.80 & early & & 1.01\tablenotemark{a}\\
234& 0.45& 1& 0.09& gal& 10.56& 0.84& 40.85 & -2.89 & -2.97 & late & & 0.45\tablenotemark{a}\\
244& 0.97& 1& 0.05& gal& 9.90& 0.91& 41.36 & -3.06 & -3.67 & late & & 0.97\tablenotemark{a}\\
245& 0.32& 1& 0.02& gal& 15.27& 0.81& 39.90 & -3.32 & -3.09 & late & E (0.32) & 0.32\tablenotemark{a}\\
249& 0.47& 2& 0.24& & \ldots& \ldots& 41.30 & -2.53 & -2.71 & early &\\
251& 0.14& 1& 0.07& gal& 22.92& 1.31& 39.59 & -3.59 & -3.24 & late & E (0.14) & 0.14\tablenotemark{a}\\
256& 0.60& 1& 0.09& & \ldots& \ldots& 41.10 & -2.65 & -2.72 & late &\\
257& 0.09& 1& 0.04& gal& 30.11& 1.08& 38.92 & -4.09 & -3.93 & early & A (0.09)\\
258& 0.75& 1& 0.04& gal& 9.97& 0.79& 40.96 & -2.83 & -3.34 & late & & 0.75\tablenotemark{a}\\
260& 0.47& 1& 0.06& gal& 10.82& 0.86& 40.71 & -2.69 & -3.13 & late & & 0.47\tablenotemark{b}\\
262& 0.87& 4& 0.67& agn1& 4.47& 1.93& 42.39 & -1.31 & -1.91 & early & & 0.81\tablenotemark{b$\dagger$}\\
264& 0.32& 1& 0.05& & \ldots& \ldots& 40.25 & -2.54 & -2.29 & late & & 0.32\tablenotemark{b}\\
265& 0.41& 1& 0.09& gal& 12.74& 1.30& 40.74 & -3.03 & -3.35 & early & C (0.41) & 0.41\tablenotemark{b}\\
266& 1.08& 4& 0.04& & \ldots& \ldots& 41.38 & -2.36 & -3.67 & early &\\
269& 0.36& 2& 0.05& & \ldots& \ldots& 40.36 & -2.56 & -2.42 & late & & 0.36\tablenotemark{b}\\
274& 0.32& 1& 0.24& gal& 10.66& 1.13& 40.91 & -3.01 & -3.13 & early & A (0.32)\\
278& 1.02& 2& 0.03& & \ldots& \ldots& 41.24 & -2.49 & -3.16 & late &\\
279& 0.89& 2& 0.03& & \ldots& \ldots& 41.07 & -2.46 & -2.92 & ir &\\
280& 0.96& 4& 0.04& & \ldots& \ldots& 41.25 & -2.23 & -2.90 & early &\\
282& 0.20& 2& 0.06& gal& 18.94& 1.10& 39.87 & -3.42 & -3.42 & late & E (0.20) & 0.08\tablenotemark{b$\dagger$}\\
286& 0.95& 1& 0.41& agn2& 0.71& 1.16& 42.28 & -1.92 & -2.61 & early & & 0.95\tablenotemark{b}\\
288& 0.71& 4& 0.11& gal& 8.51& 0.88& 41.38 & -2.61 & -3.54 & early & & 0.79\tablenotemark{b}\\
291& 0.52& 1& 0.04& gal& 14.27& 1.20& 40.60 & -3.35 & -3.62 & early &\\
292& 0.50& 2& 0.02& & \ldots& \ldots& 40.34 & -2.27 & -2.46 & ir &\\
294& 0.47& 2& 0.31& agn2& 1.29& 1.50& 41.42 & -1.50 & -1.93 & late & & 0.47\tablenotemark{b}\\
295& 0.85& 1& 0.06& & \ldots& \ldots& 41.31 & -2.36 & -2.68 & late &\\
296& 0.66& 1& 0.07& gal& 11.67& 0.55& 41.12 & -2.89 & -3.30 & early &\\
300& 0.14& 2& 0.07& gal& 25.94& 1.48& 39.54 & -3.91 & -3.75 & late & E (0.14) & 0.14\tablenotemark{b}\\
304& 0.68& 1& 3.22& agn1& 5.87& 2.60& 42.81 & -1.30 & -1.75 & late &\\
305& 0.30& 1& 0.05& gal& 17.69& 1.30& 40.15 & -3.43 & -3.47 & late & E (0.30) & 0.30\tablenotemark{b}\\
309& 1.14& 2& 0.68& agn1& 4.25& 1.77& 42.70 & -1.37 & -2.28 & late & & 1.14\tablenotemark{b}\\
310& 0.76& 1& 0.03& gal& 11.44& 1.02& 40.91 & -3.05 & -3.64 & late & & 0.76\tablenotemark{b}\\
311& 0.91& 1& 0.05& gal& 7.15& 0.49& 41.33 & -2.56 & -3.18 & late & & 0.91\tablenotemark{b}\\
313& 0.80& 1& 0.07& gal& 9.70& 0.88& 41.30 & -3.08 & -3.71 & early &\\
320& 0.14& 1& 0.05& gal& 23.11& 1.04& 39.40 & -3.53 & -3.25 & late & E (0.14) & 0.14\tablenotemark{b}\\
323& 0.51& 1& 6.48& agn2& 11.69& 3.65& 42.82 & -1.37 & -1.55 & late &\\
326& 0.36& 2& 0.05& & \ldots& \ldots& 40.29 & -3.00 & -3.24 & late &\\
332& 0.56& 1& 0.04& gal& 16.42& 0.71& 40.70 & -3.13 & -3.54 & late &\\
333& 0.38& 1& 0.86& agn2& 0.95& 0.97& 41.62 & -1.96 & -2.01 & late & & 0.38\tablenotemark{b}\\
337& 0.90& 1& 0.03& gal& 12.22& 0.90& 41.04 & -3.31 & -3.79 & late &\\
339& 0.25& 1& 0.05& gal& 22.38& 1.14& 39.95 & -3.82 & -3.84 & late & E (0.25) & 0.25\tablenotemark{b}\\
344& 0.09& 4& 15.70& agn1& 4.26& 0.99& 41.50 & -1.06 & -0.78 & ir &\\
346& 1.02& 2& 0.02& gal& 7.37& 0.69& 40.98 & -2.80 & -3.76 & late & & 1.02\tablenotemark{b}\\
349& 1.12& 4& 0.53& agn1& 3.71& 1.23& 42.56 & -1.14 & -2.45 & early & & 1.25\tablenotemark{b$\dagger$}\\
351& 0.94& 1& 0.09& gal& 7.85& 0.50& 41.59 & -2.64 & -3.53 & late & & 0.94\tablenotemark{b}\\
352& 0.91& 4& 0.03& gal& 17.62& 0.69& 41.13 & -3.12 & -3.84 & late &\\
353& 0.42& 1& 0.10& gal& 11.41& 1.05& 40.80 & -2.97 & -3.06 & late & C (0.42) & 0.42\tablenotemark{b}\\
354& 0.57& 2& 0.09& & \ldots& \ldots& 41.05 & -2.53 & -2.70 & early &\\
367& 0.78& 4& 0.17& agn2& 2.27& 1.19& 41.68 & -1.65 & -2.41 & early &\\
373& 0.48& 1& 0.15& agn2& 22.40& 0.52& 41.11 & -2.35 & -2.68 & early &\\
378& 1.08& 2& 0.04& & \ldots& \ldots& 41.40 & -2.35 & -2.98 & late & & 1.08\tablenotemark{b}\\
383& 0.17& 4& 0.05& gal& 25.18& 1.47& 39.63 & -4.05 & -3.88 & early & A (0.11) & 0.11\tablenotemark{b}\\
384& 1.02& 2& 0.09& & \ldots& \ldots& 41.70 & -2.30 & -3.39 & early &\\
387& 0.97& 4& 0.08& & \ldots& \ldots& 41.59 & -2.20 & -2.99 & early & & 1.01\tablenotemark{b$\dagger$}\\
389& 0.56& 1& 0.05& & \ldots& \ldots& 40.79 & -2.52 & -3.15 & late &\\
392& 0.40& 4& 0.05& gal& 14.47& 1.13& 40.48 & -3.24 & -3.50 & early & E (0.41) & 0.41\tablenotemark{b}\\
401& 0.94& 1& 0.07& gal& 7.52& 0.63& 41.52 & -2.63 & -3.44 & early &\\
404& 0.11& 1& 0.07& gal& 25.62& 1.11& 39.27 & -3.76 & -3.56 & late & A (0.11)\\
405& 0.94& 4& 0.05& & \ldots& \ldots& 41.38 & -2.80 & -3.56 & late &\\
414& 0.80& 1& 0.09& & \ldots& \ldots& 41.45 & -2.46 & -2.91 & early &\\
418& 0.28& 1& 0.09& gal& 17.04& 1.26& 40.34 & -3.50 & -3.47 & late & & 0.28\tablenotemark{b}\\
426& 0.16& 2& 0.08& agn2& 12.68& 0.75& 39.75 & -2.17 & -1.80 & ir &\\
428& 0.30& 2& 0.10& gal& 14.84& 1.05& 40.47 & -3.23 & -3.13 & late & E (0.30)\\
433& 1.04& 4& 0.08& gal& 7.72& 0.69& 41.66 & -2.70 & -3.61 & early & & 1.02\tablenotemark{b}\\
437& 0.82& 4& 0.85& agn1& 2.18& 0.54& 42.44 & -1.80 & -2.37 & late & & 0.84\tablenotemark{b}\\
453& 0.73& 4& 0.38& agn2& 2.17& 0.59& 41.96 & -2.13 & -2.74 & early & & 0.84\tablenotemark{b}\\
454& 0.46& 4& 0.39& & \ldots& \ldots& 41.48 & -2.26 & -2.45 & early & & 0.46\tablenotemark{b}\\
458& 0.07& 1& 0.31& & \ldots& \ldots& 39.57 & -3.03 & -2.78 & late &   (0.07)\\
462& 0.51& 1& 0.03& & \ldots& \ldots& 40.44 & -3.21 & -3.36 & early &\\
466& 0.44& 1& 0.09& gal& 13.55& 1.16& 40.81 & -3.15 & -3.65 & early &\\
471& 1.17& 2& 0.10& gal& 7.89& 0.64& 41.90 & -2.71 & -3.71 & early &\\
473& 0.31& 1& 24.90& agn1& 7.71& 3.56& 42.88 & -1.12 & -1.25 & late &\\
477& 0.44& 1& 0.10& & \ldots& \ldots& 40.86 & -2.67 & -2.82 & early &\\
478& 0.08& 4& 2.27& agn2& 4.40& 2.84& 40.55 & -1.03 & -0.82 & early &\\
480& 0.46& 1& 0.20& & \ldots& \ldots& 41.19 & -2.28 & -2.77 & early &\\

\enddata
\tablecomments{XID is ID number from Alexander et al. (2003).  z gives
  the adopted redshift.  Flag describes the source and quality of the
  redshift: (1 = high-quality spectroscopic redshift,
  2=unknown-quality spectroscopic redshift, 3=low-quality spectroscopic
  redshift, 4=high-quality photometric redshift, 5=low-quality
  photometric redshift). $F_{X}$ gives the 0.5-2.0 keV X-ray flux from
  Alexander et al. in units of $10^{-15}$~\ergcms.  Class gives the results of
  the Bayesian classification, if any.  log(odds1) and log(odds2) give
  the logarithms of the Bayesian odds ratios for the alternate models
  (agn1 and agn2 in the case of galaxies, galaxies and agn2 in the
  case of agn1, and agn1 and galaxies in the case of agn2). 
  $L_X$ is the 0.5-2.0 keV luminosity of the source in~\ergs.
  log($F_X/F_R$) and log($F_X/F_K$) give the X-ray/R-band and
  X-ray/K-band flux ratios.
Type gives the SED type computed in the photometric analysis.
  OBXF ID gives the spectral type and redshift of the sources found in \citet{HornOBXF}.
  $z_{\rm G07}$ gives the redshift from Georgakakis et al. (2007), if
  present.}
\tablenotetext{a}{source is in the G07 infrared-faint sample}
\tablenotetext{b}{source is in the G07 infrared-bright sample}
\tablenotetext{$\dagger$}{the G07 redshift is a photometric redshift}
\label{t:nsamp}
\end{deluxetable}
\clearpage
\end{landscape}
\clearpage
\LongTables
\begin{landscape}
\begin{deluxetable}{rrrrrrrrrrr}
\tablewidth{0pt}
\tablecaption{CDF-S X-ray Sample}
\tablehead{XID & z & Flag & $F_{X}$ & Class &
  log(odds1) & log(odds2) & $\log(L_{X})$ & 
  log($\rm F_X/F_R$) &  log($\rm F_X/F_K$) & 
Type}
\startdata
29& 0.57& 1& 0.24& gal& 7.12& 0.81& 41.51 & -3.25 & \ldots & early\\
44& 0.57& 1& 0.60& & \ldots& \ldots& 41.91 & -2.96 & \ldots & early\\
53& 0.67& 1& 0.14& & \ldots& \ldots& 41.43 & -2.57 & -2.95 & early\\
60& 0.54& 1& 0.23& & \ldots& \ldots& 41.44 & -2.53 & \ldots & late\\
73& 0.42& 1& 0.10& gal& 13.41& 1.22& 40.78 & -3.62 & -3.81 & early\\
75& 1.00& 1& 0.08& gal& 6.74& 0.48& 41.61 & -2.81 & -3.45 & late\\
80& 0.58& 1& 0.23& agn2& 6.17& 0.77& 41.49 & -2.21 & -2.38 & early\\
83& 0.68& 1& 0.06& gal& 9.35& 0.76& 41.11 & -3.00 & -3.39 & early\\
84& 1.03& 1& 0.15& & \ldots& \ldots& 41.93 & -2.30 & -3.28 & late\\
88& 0.60& 1& 2.48& agn2& 21.15& 2.29& 42.58 & -1.87 & -2.12 & late\\
94& 0.12& 1& 0.73& agn2& 30.05& 2.26& 40.45 & -1.43 & -1.69 & late\\
103& 0.68& 1& 1.76& agn1& 3.64& 1.47& 42.54 & -1.68 & -2.06 & early\\
106& 0.67& 1& 0.27& agn2& 1.67& 0.50& 41.72 & -1.98 & -2.45 & early\\
113& 0.52& 1& 0.09& & \ldots& \ldots& 40.97 & -2.45 & -2.49 & early\\
115& 0.34& 1& 0.06& gal& 18.63& 1.06& 40.33 & -3.66 & -3.84 & early\\
117& 0.57& 1& 4.50& agn1& 7.26& 3.69& 42.77 & -1.19 & -1.46 & late\\
118& 1.10& 1& 0.55& agn2& 38.53& 1.68& 42.56 & -1.80 & -2.60 & early\\
121& 0.73& 1& 0.28& gal& 6.78& 0.80& 41.84 & -2.97 & -3.38 & early\\
122& 0.18& 1& 0.36& agn2& 5.20& 2.97& 40.52 & -0.91 & -0.98 & ir\\
124& 0.96& 1& 0.08& & \ldots& \ldots& 41.56 & -2.51 & -3.06 & late\\
126& 0.59& 4& 0.33& agn2& 5.84& 2.85& 41.67 & -1.17 & -1.55 & early\\
129& 0.23& 1& 0.22& gal& 14.37& 1.15& 40.54 & -3.23 & -3.19 & early\\
134& 0.60& 1& 0.07& & \ldots& \ldots& 41.00 & -2.64 & -3.13 & early\\
146& 0.73& 1& 0.09& & \ldots& \ldots& 41.33 & -2.44 & -3.26 & early\\
149& 0.13& 1& 0.23& & \ldots& \ldots& 40.02 & -2.73 & -2.50 & late\\
152& 0.25& 1& 0.06& gal& 15.29& 0.90& 40.06 & -3.36 & -3.08 & ir\\
155& 0.53& 1& 1.85& agn1& 3.26& 1.37& 42.32 & -1.82 & -1.88 & early\\
158& 0.36& 1& 0.09& gal& 11.29& 0.73& 40.61 & -2.94 & -2.92 & late\\
159& 1.05& 1& 0.91& agn2& 4.15& 3.43& 42.73 & -1.00 & -2.01 & late\\
161& 0.73& 1& 0.31& & \ldots& \ldots& 41.87 & -3.09 & \ldots & late\\
162& 1.04& 1& 0.08& agn2& 8.43& 0.78& 41.65 & -2.17 & -2.54 & late\\
167& 0.58& 1& 0.09& gal& 11.41& 1.00& 41.06 & -3.28 & -3.51 & early\\
169& 1.02& 4& 0.05& & \ldots& \ldots& 41.43 & -3.65 & \ldots & late\\
171& 1.03& 1& 0.18& agn2& 0.64& 1.56& 42.00 & -1.83 & -1.72 & late\\
176& 0.73& 1& 0.61& agn2& 6.11& 1.71& 42.17 & -1.79 & -2.31 & early\\
179& 0.66& 1& 0.49& agn2& 30.04& 1.77& 41.97 & -1.87 & -2.13 & late\\
181& 0.74& 1& 0.25& & \ldots& \ldots& 41.79 & -2.32 & -2.61 & early\\
182& 0.21& 1& 0.67& gal& 10.83& 1.08& 40.95 & -3.13 & -3.06 & early\\
184& 0.62& 1& 1.32& & \ldots& \ldots& 42.33 & -2.86 & \ldots & early\\
185& 0.18& 4& 0.06& agn2& 11.38& 0.60& 39.74 & -2.33 & -1.98 & ir\\
189& 0.08& 1& 0.81& gal& 21.22& 2.08& 40.06 & -3.60 & -3.42 & late\\
190& 1.02& 1& 0.06& & \ldots& \ldots& 41.48 & -2.09 & -2.70 & late\\
192& 0.08& 1& 0.57& gal& 21.21& 1.53& 39.91 & -3.56 & -3.43 & late\\
193& 0.96& 1& 1.81& agn1& 6.23& 3.07& 42.93 & -1.27 & -1.96 & late\\
196& 0.67& 1& 0.07& & \ldots& \ldots& 41.13 & -2.74 & -2.75 & late\\
203& 1.18& 4& 0.04& gal& 6.08& 0.72& 41.54 & -3.29 & -4.12 & early\\
207& 0.10& 1& 0.18& gal& 27.50& 1.25& 39.69 & -4.09 & -4.02 & early\\
210& 0.83& 1& 0.07& & \ldots& \ldots& 41.38 & -2.18 & -2.71 & early\\
212& 0.67& 1& 0.17& & \ldots& \ldots& 41.53 & -2.52 & -2.80 & early\\
214& 0.84& 1& 1.51& agn1& 5.35& 1.98& 42.71 & -1.22 & -1.63 & late\\
224& 0.55& 1& 0.12& gal& 9.46& 0.88& 41.14 & -2.90 & -3.31 & late\\
225& 0.55& 1& 0.06& gal& 11.56& 1.05& 40.83 & -3.21 & -3.62 & late\\
226& 1.00& 1& 0.12& & \ldots& \ldots& 41.79 & -2.19 & -2.71 & late\\
227& 0.67& 1& 4.01& agn1& 7.49& 2.95& 42.89 & -0.90 & -1.19 & early\\
229& 0.67& 1& 1.86& agn1& 3.40& 1.91& 42.55 & -1.59 & \ldots & late\\
236& 0.46& 1& 0.08& gal& 9.70& 0.59& 40.78 & -2.84 & -2.91 & ir\\
238& 0.24& 1& 0.06& gal& 17.46& 1.25& 40.05 & -3.42 & -3.66 & late\\
244& 0.58& 1& 0.12& & \ldots& \ldots& 41.21 & -2.54 & -2.99 & early\\
247& 0.62& 2& 0.54& agn2& 22.40& 0.76& 41.95 & 0.59 & \ldots & early\\
248& 0.67& 4& 0.03& gal& 10.45& 1.02& 40.83 & -3.35 & -3.74 & early\\
256& 0.74& 1& 3.66& agn1& 9.56& 4.09& 42.96 & -0.55 & -0.88 & ir\\
260& 0.27& 1& 0.51& gal& 7.85& 0.95& 41.08 & -2.68 & -2.82 & early\\
262& 0.42& 1& 0.13& gal& 10.01& 0.71& 40.89 & -2.91 & -3.05 & late\\
263& 1.01& 1& 0.56& agn2& 12.02& 2.67& 42.48 & -1.53 & -1.85 & early\\
265& 0.46& 1& 0.14& gal& 9.69& 0.88& 41.04 & -3.48 & \ldots & late\\
266& 0.44& 1& 0.09& gal& 10.56& 0.81& 40.80 & -2.93 & -3.13 & late\\
267& 0.10& 1& 0.11& & \ldots& \ldots& 39.45 & -2.96 & \ldots & ir\\
269& 0.66& 1& 0.22& & \ldots& \ldots& 41.63 & -3.82 & \ldots & early\\
271& 1.18& 1& 0.12& & \ldots& \ldots& 41.96 & -2.32 & -3.04 & late\\
273& 1.11& 1& 0.09& & \ldots& \ldots& 41.79 & -2.30 & -3.02 & late\\
276& 0.67& 1& 0.15& & \ldots& \ldots& 41.47 & -2.99 & -3.43 & early\\
277& 1.15& 4& 0.78& agn1& 2.04& 1.16& 42.76 & -0.31 & \ldots & late\\
292& 0.37& 1& 0.22& & \ldots& \ldots& 41.01 & -2.70 & -2.42 & ir\\
293& 1.14& 1& 0.18& gal& 8.46& 0.49& 42.13 & -4.43 & -2.47 & early\\
303& 0.25& 1& 2.70& & \ldots& \ldots& 41.71 & -1.95 & \ldots & early\\

\enddata
\tablecomments{Columns are as in Table \ref{t:nsamp}. \label{t:ssamp}}
\end{deluxetable}
\clearpage
\end{landscape}

\end{document}